\PassOptionsToPackage{pdfpagelabels=false}{hyperref}
\documentclass[usenatbib]{mnras}
\usepackage{graphicx}
\usepackage{amsmath}
\usepackage{longtable}
\usepackage{hyperref}
\usepackage{aas_macros}
\usepackage{epsfig,color}
\usepackage{cleveref}
\usepackage{url}
\usepackage[caption=false]{subfig}
\DeclareSymbolFont{matha}{OML}{txmi}{m}{it}% txfonts
\DeclareMathSymbol{\varv}{\mathord}{matha}{118}

\long\def\symbolfootnote[#1]#2{\begingroup%
\def\thefootnote{\fnsymbol{footnote}}\footnote[#1]{#2}\endgroup} 

\def\HII {H{\sc ii}}

\title{Millimeter Spectral Line Mapping Observations Toward Four Massive Star Forming \HII\ Regions}

\author[Shanghuo Li et al.]
{
        \parbox{\textwidth}{Shanghuo Li$^{1,2,3}$\thanks{E-mail: \texttt{shanghuo.li@shao.ac.cn; jzwang@shao.ac.cn}},  Junzhi Wang$^{1,4}$, Zhi-Yu Zhang$^{5,6}$,  Min Fang$^7$, Juan Li$^{1,4}$, Jiangshui Zhang$^3$, Junhui Fan$^3$, Qingfeng Zhu$^8$, Fei Li$^{1,2,9}$}\vspace{0.4cm}\\
 \parbox{\textwidth}{
  $^1$Shanghai Astronomical Observatory, Chinese Academy of Sciences, 80 Nandan Road, Shanghai 200030, China\\
  $^2$University of Chinese Academy of Sciences, 19A Yuquanlu, Beijing 100049, China\\
  $^3$Center for Astrophysics, GuangZhou University, 230 Wai Huan Xi Road, GuangZhou Higher Education Mega Center, GuangZhou 510006, China\\
  $^4$ Key Laboratory of Radio Astronomy, Chinese Academy of Sciences, 210008, Nanjing, PR China  \\
  $^5$ Institute for Astronomy, University of Edinburgh, Royal Observatory, Blackford Hill, Edinburgh EH9 3HJ, UK  \\
  $^6$ European Southern Observatory, Karl-Schwarzschild-Str.~2, D-85748 Garching,  Germany \\ 
  $^7$Department of  Astronomy and Steward Observatory, University of Arizona, 933 N Cherry Ave., Tucson, AZ 85721, USA  \\
  $^8$Astronomy Department, University of Science and Technology, Chinese Academy of Sciences, Hefei 210008, China \\ 
  $^9$ College of Physics and Information Engineering, Hebei Normal University, Shijiazhuang, 050024, China
 \\  
  }}

\date{Received xxx}

\pagerange{\pageref{firstpage}--\pageref{lastpage}} \pubyear{2016}

\def\LaTeX{L\kern-.36em\raise.3ex\hbox{a}\kern-.15em
    T\kern-.1667em\lower.7ex\hbox{E}\kern-.125emX}

\begin{document}

\label{firstpage}

\maketitle

\begin{abstract}

We present spectral line mapping observations toward four massive 
star-forming regions (Cepheus A, DR21S, S76E and  G34.26+0.15), 
with the IRAM 30 meter telescope at 2\,mm and 3\,mm bands. Totally 396 
spectral lines from 51 molecules, one helium recombination line, ten 
hydrogen recombination lines, and 16 unidentified lines were detected 
in these four sources. An emission line of nitrosyl cyanide (ONCN,
14$_{0,14}$-13$_{0,13}$) was detected in G34.26+0.15,  as first detection
in massive star-forming regions. We found that the $c$-C$_{3}$H$_{2}$ and 
NH$_{2}$D show enhancement in shocked regions as suggested by 
evidences of SiO and/or SO emission. Column density and rotational 
temperature of CH$_{3}$CN were estimated with the rotational diagram 
method  for all four sources.  Isotope abundance ratios of $^{12}$C/$^{13}$C 
were derived  using HC$_{3}$N and its $^{13}$C  isotopologue, which were  
around 40 in all four massive star-forming regions and  slightly lower than the 
local interstellar value ($\sim$65).   $^{14}$N/$^{15}$N and $^{16}$O/$^{18}$O 
abundance ratios in these sources were also derived using double isotopic  
method, which were slightly lower than that in local interstellar medium.  
Except for Cep A,  $^{33}$S/$^{34}$S ratio in the other three targets were 
derived, which were similar  to that in  the local interstellar medium.  The 
column density ratios of N(DCN)/N(HCN) and N(DCO$^{+}$)/N(HCO$^{+}$) 
in these sources were more than two orders of magnitude higher than the 
elemental [D]/[H] ratio, which is 1.5$\times$10$ ^{-5}$.
Our results show the later stage sources, G34.26+0.15 in particular, present 
more molecular species than earlier stage ones. Evidence of shock activity is 
seen in all stages studied.

\end{abstract}

\section{Introduction}

Massive stars play an important role in galaxy evolution. The formation of
massive stars is difficult to study due to large distances, complex cluster
environments and rapid time scales in evolution \citep[e.g.,][]{Zinnecker07,
Herbst09}.  Massive star formation involves different physical and
astro-chemical properties  in various evolutionary stages, i.e., starless
cores, Hot Molecular Cores (HMCs), Hyper-Compact \HII\, (HC\HII\,), Ultra
Compact \HII\ (UC\HII\,), Compact \HII\ (C\HII\,) and evolved \HII\ regions
\citep{Kurtz00,Churchwell02}.  Feedback from massive stars in turn affects
surrounding molecular clouds through energetic stellar winds, powerful outflows,
UV radiation, expansion of \HII\, regions, and supernova explosions
\citep{Kennicutt98}. Emissions of radio continuum and recombination  lines have
been used to investigate  HC\HII,\ UC\HII\ and \HII\ regions \citep{Churchwell02}.
Molecular probes are also essential for investigating  the physical and
chemical conditions of massive star-forming regions (e.g, temperature, density,
radiation field, etc.). They are also utilized to study chemical 
evolution in massive star formation regions \citep{van98}.

Millimetre wavelength window is important for studying massive star formation,
because it contains many useful molecular probes, such as shock tracers: SiO,
SO, HNCO and H$ _{2}$CO \citep{Schilke97,Esplugues13,Rizzo05}, long  carbon
chain molecules: $c$-C$ _{3}$H$ _{2}$, HC$ _{3}$N, and  HC$ _{5}$N
\citep{Morris77,Chung91,Bergin96}, deuterated molecules: DCN, DCO$ ^{+}$, C$
_{2}$D, and  NH$ _{2}$D \citep{Shah01,Pety07,Parise09}, and dense gas traces:
HCN, HCO$^+$, CS and C$ _{2}$H \citep{Padovani09}, etc.

The various molecules can probe different physical and chemical components. 
Thus, molecular tracers offer a powerful tool for investigating the physical and 
chemical conditions of star-forming regions. Studying the physical 
and chemical evolutions in various scales are essential to 
developing a picture of massive star formation, and are the main focus of this 
study. To achieve these main goals, we mapping four massive star formation 
regions, at different evolutionary stages from HC\HII\ to \HII, 
with total 11.8 GHz bandwidth coverage.

In this paper we present observational results of four massive star-forming
regions, which were carried out by the IRAM 30m telescope. In Section 2, we
describe the observations and the data reduction.  Observational results,
including molecular lines, velocity-integrated maps of molecular lines, isotope
abundance ratios and long carbon chain molecules, which are introduced in
Section 3. In Section 4, we derive the rotational temperature and column density 
of CH$_{3}$CN for each source, and investigating the properties of shocked region 
with shock tracers and deuterated molecules. A brief summary is given in Section 5.

\section{Source selection and Observations}
\subsection{Sources}

We select four well studied massive star formation regions from
\cite{Wu10}, which are representative of different evolutionary stages.  
Cepheus A (hereafter Cep A), DR21S, G34.26+0.15 
(hereafter G34) and S76E lie at different distances from 0.7 kpc to 3.3 kpc, 
and associated with HC\HII\ \citep{Hughes84},  cometary shaped \HII\ \citep{Immer14}, 
C\HII\  \citep{Reid85,Mookerjea07} and an extended \HII\ 
regions \citep{Li12}, respectively, according to radio observations. 
These regions provide ideal laboratories to study the properties of massive 
star-forming regions with mapping observations, since they are high mass and luminosity.

Cep A molecular cloud complex is the second nearest star formation region after
the Orion complex, located at a distance of 700$\pm$40 pc \citep{Moscadelli09}.
Herbig-Haro objects \citep{Hartigan86}, HC\HII\ regions
\citep{Hughes84}, and two bipolar molecular outflow aligned east-west and
northeast-southwest (NE-SW) \citep{Rodriguez80,Gomez99} have been detected in
Cep A. The bolometric luminosity of this region is 2.5 $\times$ 10$^{4}$
L$_{\odot}$ with a mass of $\sim$ 330 M$_{\odot}$ \citep{Evans81,1996ApJ...466..844N}. 
Our mapping centre of Cep A locates in the radio
source HW2, which is a massive protostar \citep{Cunningham09,Dzib11}, with a
mass of $\sim$ 18 M$_{\odot}$ \citep{Torrelles11}. There is a fast, bipolar,
highly collimated radio jet ejected from the HW2 young stellar object (YSO),
with well aligned NE-SW direction and scale of $\sim$ 1500 AU \citep{Curiel06}.
Furthermore, a rotating disk around the HW2 have been identified from
sub-millimetre observations \citep{Patel05}.  The infrared observations toward
HW2 show that there are strong shocks, implying that the object is at the early
evolutionary stage.

DR21S is a massive star-forming region within the Cygnus X molecular cloud at a
distance of 1.5$\pm$0.08 kpc \citep{Rygl12}. This region has been studied in
multiple wavelength observations
\citep{Wilson90,Liechti97,Jakob07,Schneider06,Schneider10}. It contains
molecular gas mass of about 4.7$\times$10$^{6} M_{\bigodot}$ and a far-infrared
luminosity of 1.5 $\times$ 10$^{5}$ L$_{\odot}$ \citep{Schneider06,Campbell82}.
The cloud also hosts one of the strong energetic outflow, with a mass of
about 3000 $M_{\bigodot}$ \citep{Cruz-Gonzalez10}.  Our mapping observations
have covered two cometary \HII\, regions \citep{Immer14} and several regions
where ongoing shock activity \citep{Davis07} toward DR21S.

G34 molecular cloud complex is a massive star-forming region at a distance of
3.3 kpc \citep{Kuchar94}, and its envelope scale is 8.1 $\times$ 10$^{4}$ AU.
The cloud contains 1.1$\times$ 10$^{4}$ $M_{\bigodot}$ of dense gas \citep{Hill05}, 
and has a luminosity of 3.2 $\times$ 10$^{5}$ L$_{\odot}$ \citep{vanderTak13}, with an
age of a few $\times 10^{4}$ yr \citep{Paron09}. From radio continuum
observations, this source shows two very condensed UC \HII\, regions (named as
A and B),  a more evolved  C\HII\, region with a cometary shape (named as C)
and an extended ringlike \HII\, region (named as D) \citep{Reid85,Mookerjea07}. 
There are several outflows in G34, as evidenced by SiO emission extends to the northwest, 
southeast and northeast of the compact \HII\, regions \citep{Hatchell01}.

S76E is a massive star-forming region containing an extended \HII\, region
\citep{Li12},  with an emitting region size of about 2$'$, i.e., about 1.2 pc
given a distance of 2.11 kpc \citep{Plume92}. This region has an IRAS
total IR (TIR) luminosity of 2.3$\times$10$^{4}$ L$_{\odot}$
\citep{Zinchenko97}, with a molecular mass of 1.64 $\times$ 10$^{3}$
$M_{\bigodot}$ \citep{Zinchenko94}.

\subsection{IRAM 30m observations and VLA archive data}

We carried out mapping observations with the IRAM 30m telescope at Pico Veleta,
Spain\footnote{Based on observations carried out with the IRAM 30m Telescope.
IRAM is supported by INSU/CNRS (France), MPG (Germany) and IGN (Spain).}, in
January 2012.  On-the-fly (OTF) mapping mode was used for all four objects,
with a mapping size of 120$'' \times$ 120$''$. The mapping centers are listed
in Table \ref{tab:coordination}. The Eight MIxer Receiver (EMIR) and Fourier
Transform Spectrometers (FTS) backends were used.  
These observations cover the frequency range between 84.5GHz and 92.3 GHz 
in the 3 mm atmospheric window, and between 143.7GHz and 147.7 GHz 
in the 2 mm atmospheric window. A variety of molecular lines are covered
simultaneously, which can improves observational efficiency and avoid deviation
of relative position and flux calibration. The reasonable  frequency resolution
(V.V. Guzm$\acute{a}$n et al. 2013) of FTS with 195 kHz sampling 
correspond  to  $\sim$ 0.67 km s$^{-1}$ and  $\sim$ 0.4 km s$^{-1}$ at 87 and
145 GHz, respectively. The data were converted to the main beam temperature
$T_{mb}$ using $T_{mb} = \frac{F_{eff}}{B_{eff}} T^{\ast}_{A}$, where $B_{eff}$
is main beam efficiency and $F_{eff}$ is forward efficiency.  The main beam
efficiencies are 0.81 for 86 GHz and  0.73 for 145 GHz. The forward
efficiencies are 95\% at 3 mm and 93\% at 2 mm.  The beam sizes are about
17$''$ and 28$''$, and the $\tau$ are about 0.05 and 0.1, while  the typical
system temperatures are about 125 K and 95 K at 2 mm and 3 mm, respectively.
All data were re-gridded to the step of 15$''$, and assumed that the 
beam-filling factor is unity for following analysis.

Free-free emission at centimetre wavelength is an useful tool to study  HC\HII,
UC\HII, and \HII\, regions. Radio continuum data at $X$ band (3.6 cm) observed
with Very Large Array (VLA) were obtained from the \cite{Li12}.  
In Table \ref{tab:coordination}, we list the peak intensity $S_{p}$ in Column 7, 
the flux density $S_{\nu}$ in Column 8, the deconvolved size in Column 9.

%8 GHz (84.5-88.5, 88.3-92.3 GHz) 
%%%%%%%%%%%%%%%%%%%%%%%%%%      RESULTS     %%%%%%%%%%%%%%%%%%%%%%%%%%%%%%%%%%%%
\section{Results}

\subsection{Line identification} \label{sec:lines}

To increase sensitivity of weak lines, we averaged spectrum over the map of
$30'' \times 30''$ for each source (figure \ref{fig:1}). The molecular lines
were identified using the Splatalogue, Cologne and CDMS spectroscopic
databases. Most of the molecular lines have similar line widths and central
velocities, while several lines are more broad than the typical line widths,
such as radio recombination lines (RRLs) and SiO lines. RRLs were detected in
DR21S, G34 and S76E, but not detected in Cep A, where the radio
continuum emission was dominated by high-velocity ionized jet \citep{Jim11}.
The line parameters derived from Gaussian fitting are listed in
table \ref{tab:CEPA}, \ref{tab:DR21S}, \ref{tab:G34}, \ref{tab:S76E},
including velocity integrated intensity  (\textit{A}), local standard of rest
velocity $v_{LSR}$, line width (full width at half maximum FWHM), and emission
peak temperature (in $T_{mb}$). Detailed information of the molecular lines for
each source are described below.

\textbf{Cep A}. In total, 61 lines from 35 molecules were identified (see table
\ref{tab:CEPA}), including long carbon chain molecules (e.g, HC$ _{2n-1} $N,
CH$ _{3} $C$ _{2} $H,  H$ ^{13} $C$ _{2n-1}$N, $c$-C$ _{3} $H$ _{2}$,  C$
_{4}$H, n=2,3),  deuterated molecules (DCN, DCO$^{+}$, C$ _{2} $D, NH$ _{2}
$D.) and shock tracers (SiO and SO, etc.), etc. Most of these lines
velocities are around   $v_{LSR} \sim$ -11 km s$ ^{-1}$, while the line width
(FWHM) range from 3.3 to 4.5 km s$ ^{-1}$.

\textbf{DR21S}. We identified 78 spectral features corresponding to 43
different molecular species, including long carbon chain molecules, deuterated
molecules, PDR and shock tracers, etc. In addition, one helium recombination
(He42$\alpha$), six RRLs and one unidentified line were detected (see  table
\ref{tab:DR21S}).  A complex oxygen-bearing molecule of acetone (CH$ _{3}$)$
_{2}$CO 23$_{11,13}$-22$_{12,10}$ AE was detected for the first time in this
massive star-forming region. These  lines velocities are around $v_{LSR} \sim$
-2 km s$ ^{-1}$, while the typical FWHM range from 3 to 5 km s$ ^{-1}$. 

\textbf{G34}.  A total of 169 spectral features were detected  (table
\ref{tab:G34}).  We identified 78 molecular species and 3 RRLs, including long
carbon chain molecules, deuterated molecules, PDR and shock tracers, etc.  In
addition, 9 unidentified lines were detected. Most of these molecular lines
velocities are around $v_{LSR} \sim$ 58 km s$ ^{-1}$ and the typical FWHM range
from 3.5 to 4.5  km s$ ^{-1}$. Several molecule were detected in this source,
but absent in the other three sources, e.g, SiC$ _{2}$, H$ _{2}$CN, HDCS, N$
^{15} $NH$ ^{+} $, $ ^{15} $NNH$ ^{+} $, $ ^{13} $C$ ^{34} $S, SO$ _{2} $, ONCN
and HCO. The  line width  of SiC$ _{2}$ is 6.23 km s$ ^{-1}$ (FWHM)  which is
significantly broader than the typical FWHM.

\textbf{S76E}. We detected 71 spectral lines from 44 molecules, and one RRL
(table \ref{tab:S76E}), including long carbon chain molecules, deuterated
molecules, several well known PDR and shock tracers, etc. In addition, we found 6
unidentified lines. These lines are around $v_{LSR} \sim$ 33 km s$ ^{-1}$,
while the typical FWHM range from 2.8 to 3.5 km s$ ^{-1}$.
 
In summary, many fundamental molecules were detected in all four sources, such
as long carbon chain molecules, deuterated molecules, sulfur-bearing molecules,
PDR and shock tracers, etc. Both C$ _{2}$S and HCO  were detected in DR21S and
G34, while SO$_{2}$, ONCN, SiC$_{2}$, C$_{3}$S, NH$_{2}$CO and
H$_{2}$NCH$_{2}$CN were only detected in G34,  and (CH$ _{3}$)$ _{2}$CO is only
detected in DR21S.  In addition, the number of molecules in G34 is
significantly more than other sources. This indicates that the chemical
environment in G34 is more complicated than other three sources.  Molecules were
detected in all four sources might be formed or existed at the early stage of
massive star formation. The molecules were only detected in DR21S and/or G34, but
absent in Cep A and S76E, likely only form in specific environments. 
The physical and chemical environments can be affected by
interaction between two compact \HII\, regions in DR21S. The G34 molecular
cloud complex is interacting with a SNR, which could affect the chemical
environments of molecular cloud, accelerate chemical reaction and produce
complicated molecules. These sources exhibit different radio continuum specific
intensities and chemical complexities, indicating that they may be in different
evolutionary stage. G34 and DR21S show more chemical complexity than 
Cep A and S76E,  which is consistent  with the evolutionary stages as suggested 
by evidences of radio continuum and RRLs. The comparison between  
our results and previous line survey, toward infrared dark clouds (IRDCs), 
high-mass protostellar objects (HMPOs), HMCs, and UC\HII\, regions at 3
mm, show that the molecular abundances in our sources are significantly
higher than in IRDCs and HMPOs \citep{Sanhueza12,Marseille08}, but generally in
agreement with the HMCs and UC\HII\ \citep{Gerner14}.

\subsection{ Velocity-integrated intensity maps} \label{sec:maps}

In each target, most molecular lines show similar spatial distributions with
their emission peaking at the core position, while some other lines exhibit
distinct morphologies. For example, both $c$-C$ _{3}$H$_{2}$ and DCO$ ^{+}$ 
have multiple emission peaks in all four sources, while few cases 
(HCO,  ONCN and (CH$ _{3}$)$_{2}$CO) show anti-correlated with the RRLs 
(or free-free continuum emission) in the velocity-integrated intensity maps.

\textbf{Cep A.} Our mapping centre locates at the radio source HW2. The
velocity-integrated intensity map of DCO$ ^{+}$ 2-1 shows three main
condensations, as well as $c$-C$ _{3}$H$_{2}$ 2-1 (figure \ref{fig:2}), which
will be discussed in Sect.\ref{sec:C3H2} and Sect.\ref{sec:deuterium}.  Figure
\ref{fig:3} shows the velocity-integrated intensity maps of the deuterated
molecules and shock tracers.  The SiO 2-1 emission is stretched along the 
northeast-southwest (NE-SW) direction, which is consistent with what was found
in literature \cite{Codella06}. 

\textbf{DR21S.} Velocity-integrated intensity maps and spectra of deuterated
molecules are shown in figure \ref{fig:4}.  Some of them show slightly
different morphologies comparing with its corresponding hydrogenated molecules.
Detail discussion about the deuterated molecules will be presented in
Sect.\ref{sec:deuterium}. Large oxygen-bearing molecule of acetone (CH$ _{3}$)$
_{2}$CO shows strong emission in several positions, with a cavity
around the emission peak of H41$\alpha$ (figure \ref{fig:7}), which is similar to the 
results of \cite{Friedel08}. The $c$-C$_{3}$H$_{2}$ 2-1 emission also shows 
a cavity around the radio continuum emission peak, which will be discussed 
in Sect.\ref{sec:C3H2}. He42$\alpha$ emission was  detected toward the core region (figure
\ref{fig:7}). 

\textbf{G34.}  Figure \ref{fig:8} shows different spatial distributions between
H41$ \alpha$ and HC$ _{3}$N. H41$\alpha$ emission is extended in the southeast,
while HC$ _{3}$N is extended in the west. HCO and $c$-C$ _{3} $H$ _{2}$ show
significantly different morphologies compared to RRLs. HCO shows
anti-correlated with H41$ \alpha$ (figure \ref{fig:8}), while $c$-C$ _{3} $H$
_{2}$ shows a cavity feature in the emission peak of RRL, which will be
discussed in Sect.\ref{sec:C3H2}.

\textbf{S76E}.  As shown in figure \ref{fig:2}, $c$-C$ _{3}$H$_{2}$ 2-1 and
$c$-C$ _{3}$H$_{2}$ 3-2 transitions exhibit  slightly different morphologies.  The
velocity-integrated intensity maps and spectra of deuterated molecules are
presented in figure \ref{fig:6}, which will be discussed in
Sect.\ref{sec:deuterium}.

\subsection{Isotopic abundances} 
\label{sec:isotope}

\textbf{$^{12}$C/$ ^{13}$C.} Since the millimeter HC$ _{3}$N lines are
optically thin in most molecular clouds \citep{Li12}, the line intensity ($I =
\int T\, d \varv\,$) ratios between HC$ _{3}$N 10-9 and its isotopologues
(HCC$^{13}$CN 10-9 and HC$^{13}$CCN 10-9) can be used to determine the 
abundance ratios of these molecules. The $^{12}$C/$^{13}$C  isotope abundance 
ratios were derived from the line ratio between HC$ _{3}$N and its isotopologues 
assuming similar astro-chemical processes. The $^{12}$C/$ ^{13}$C isotope ratios 
are about 40 in all four sources (table \ref{tab:ratio}), which are slightly lower
than the ratio of 60--70 found in the local interstellar medium (ISM)
\citep{Milam05}, but agree well with the result in Orion ($\sim$40)
\citep{Persson07,Tercero10}.  The rarer isotope $^{13}$C is a secondary element
which can be produced from $^{12}$C in the carbon-nitrogen-oxygen (CNO) cycle
in low- and intermediate-mass stars \citep{Prantzos96}. While the main isotope of carbon
$^{12}$C, as a primary element, is mainly produced by the triple alpha reaction
\citep{Timmes95}.  A positive gradient along the Galactic radius is apparent in
the $^{12}$C/$ ^{13}$C ratios, which are $\sim$25 to $>$100 from the Galactic
centre to the outer disk, following the relationship $^{12}$C/$ ^{13}$C =
6.21D$_{GC}$ + 18.71, where D$_{GC}$ is the distance from the Galactic center
\citep{Milam05}.  The $^{12}$C/$ ^{13}$C ratios that are calculated from HC$
_{3}$N and its isotopologues seem exhibiting a gradient with D$_{GC}$, 
except for DR21S and S76E, since the ratio is lower than the prediction of 
 \cite{Milam05} for the former and the Galactic latitude is much higher 
 than 2$^{\circ}$ for the later.
The discrepancy of isotope abundance ratios between our sources and the local 
ISM may arise from Galactic chemical evolution, 
since the solar value represents the ISM 4.5 Gyr ago. The
$^{12}$C/$ ^{13}$C ratios also can be effected by isotope selective
photodissociation from UV photon because both $^{12}$C and $^{13}$C have
different self-shielding \citep{Wilson94,Milam05}. The $^{13}$C is more
easily to be photodissociated than $^{12}$C, which can result in a lower
$^{12}$C/$ ^{13}$C ratios in star-forming region comparing with local ISM
\citep[e.g.,][]{2014MNRAS.445.4055S}.

\textbf{$^{14}$N/$ ^{15}$N.} The HCN molecule and its isotopologues have large
Einstein co-efficiencies, making their line emission easily optically thick.
To accurately estimate the abundance ratios from lines with high optical
depths, we used the double isotope method, which has one of the isotope
abundances is either measured from different species or fixed with the local
ISM value. We estimated the isotope abundance ratios of $^{14}$N/$ ^{15}$N,
which were calculated from the observed line ratios of H$^{13}$CN(1-0) and HC$
^{15}$N(1-0), after adopting the ratio of $ ^{12}$C/$ ^{13}$C derived from HC$
_{3}$N 10-9 and its isotopologues. 

The $^{14}$N/$ ^{15}$N isotope ratios in these sources range from 109 to 304,
which are lower than $\sim$400 found in local ISM \citep{Terzieva00}, and 1000
$\pm$ 200 found in the prototypical starless cloud core L1544
\citep{Bizzocchi13}. The $^{14}$N/$ ^{15}$N ratio in S76E is 304 $\pm$ 15,
which is roughly consistent with 300 $\pm$ 50 in Barnard 1 \citep{Daniel13}.
The HC$^{15}$N abundance could be enhanced in the early stages of
molecular dense cores \citep{Rodgers08}, while the effect can be neglected in
our sources because they are all \HII\ regions. The $^{15}$N element is
believed to be formed in the synthesis of massive stars and potential can be destroyed
in low mass stars \citep{2009ApJ...690..580W}. This indicates that a part of the $^{15}$N 
element may be destroyed in local ISM.  \cite{Adande12} found that the $^{14}$N/$ ^{15}$N 
ratio shows a positive gradient with distance of D$_{GC}$, 
$^{14}$N/$ ^{15}$N = 21.1(5.2) kpc$^{-1}$ D$_{GC}$ + 123.8(37.1), which likely
supports the Galactic chemical evolution model that $^{15}$N has a secondary
origin in novae. The results of Cep A and G34 are consistent with the above
scenario. The Galactic chemical evolution can be used to explain the discrepancy of 
$^{14}$N/$ ^{15}$N ratios between our sources and the local ISM.

\textbf{$^{16}$O/$ ^{18}$O.} With the double isotope method, we derived the
$^{16}$O/$ ^{18}$O isotopic ratios with H$ ^{13}$CO$ ^{+}$(1-0), HC$ ^{18}$O$
^{+}$(1-0) and the ratio of $^{12}$C/$^{13}$C.  The local ISM value
(557$\pm$30) is slightly higher than our results \citep{Wilson99}, which are
about 422, 415, 532 and 288 in Cep A, S76E DR21S and G34, respectively. The
ratio of $\sim$288 found in G34 is consistent with 287 $\pm$ 27 found in Orion
derived from OCS and its isotopologues \citep{Tercero10}.  The fractionation of
oxygen can be ignored due to its high first ionization potential
\citep{1984ApJ...277..581L}.

\textbf{$^{34}$S/$ ^{33}$S.}  C$^{33}$S 3-2 and C$^{34}$S 3-2  lines are 
optically thin \citep{Goicoechea06}, thus their flux ratio can be used to derive the
ratios of $ ^{34}$S/$ ^{33}$S. The ratios of  $ ^{34}$S/$ ^{33}$S are 5.1, 4.7
and 5.5 in DR21S, G34 and S76E, respectively. These values are in agreement
with the local ISM ($\sim$ 6) and Orion KL ($\sim$ 5) isotope abundance ratio
\citep{Chin96,Persson07,Tercero10}.  For Cep A, the $ ^{34}$S/$ ^{33}$S isotope
ratio was not derived, because the C$^{33}$S 3-2 line was not detected.

\textbf{D/H.} Several deuterated molecules were found in all four sources.  DCN,
DCO$ ^{+}$, H$^{13}$CN and H$^{13}$CO$^{+}$ are adopted to derive the abundance
ratios of N(DCN)/N(HCN) and N(DCO$^{+}$)/N(HCO$^{+}$), and the
$^{12}$C/$^{13}$C ratios are obtained with HC$_{3}$N and its isotopologues. To
derive the abundance ratio of {D/H}, we adopt several assumptions.  Firstly we
assume local thermal equilibrium (LTE) conditions and optically thin for all these lines. 
In addition, these lines have similar excitation temperatures, emitting regions and filling
factors. With the above assumptions, we derived the abundance (column density)
ratio using the formula from \citet{Sakai12}.
 
\begin{multline}
\label{ratio}
\frac{N({\rm DCN})}{N({\rm H^{13}CN})}=(\frac{\mu_{0},_{\rm H^{13}CN}}{\mu_{0},_{\rm DCN}})^{2} (\frac{\nu_{\rm H^{13}CN}}{\nu_{\rm DCN}})^{2} \\
 \times exp(\frac{E_{\rm u,DCN}-E_{\rm u,H^{13}CN}}{kT_{\rm ex}}) \frac{1-\frac{J_{\rm H^{13}CN}(T_{\rm BB})}{J_{\rm H^{13}CN}(T_{\rm ex})}}{1-\frac{J_{\rm DCN}(T_{\rm BB})}{J_{\rm DCN}(T_{\rm ex})}} \frac{I_{\rm DCN}}{I_{\rm H^{13}CN}},
\end{multline}

where $N$ is the column density, $\mu_{0}$ is the dipole moment, $\nu$ is the
rest frequency of the line, $E_{\rm u}$ is the upper state energy, $T_{\rm ex}$
is the excitation temperature, $T_{\rm BB}$ is the background radiation
temperature (2.7K), and $I$ is the velocity-integrated in $T_{\rm mb}$. $J(T)$
is the Planck function as:

\begin{equation}
J(T)=\frac{\frac{h\nu}{k}}{exp(\frac{h\nu}{kT})-1},
\end{equation}
  
With the values of $\mu_{0}$ , $\nu$ and $E_{u}$ shown in Table
\ref{Deuterated}. The above equation can be reduced as
 
\begin{equation}
        \frac{N({\rm DCN})}{\rm N(H^{13}CN)} \simeq 0.42  exp(\frac{6.28}{T_{\rm ex}}) \frac{I_{\rm DCN}}{I_{\rm H^{13}CN}},
\end{equation}
 
We can also derived the N(DCO$^{+}$)/N(H$^{13}$CO$^{+}$) ratio with the same method, 

\begin{equation}
        \frac{N({\rm DCO^{+}})}{N({\rm H^{13}CO^{+}})} \simeq 0.29  exp(\frac{6.21}{T_{\rm ex}}) \frac{I_{\rm DCO^{+}}}{I_{\rm H^{13}CO^{+}}},
\end{equation}
 
We take $T_{\rm ex}$ from the highest value of the rotational temperature
 of CH$_{3}$CN (table \ref{tab:excitation}), which will be discussed in
Sect.\ref{sec:excitation}. 
 
The column density ratios of N(DCN)/N(HCN) and N(DCO$ ^{+}$)/N(HCO$^{+}$) are
converted by N(DCN)/N(H$^{13}$CN) and N(DCO$^{+}$)/N(H$^{13}$CO$^{+}$), with the
$^{12}$C/$ ^{13}$C ratios which derived from the aforementioned method for each sources. 
The ratios of N(DCN)/N(HCN) are slightly higher than the N(DCO$
^{+}$)/N(HCO$^{+}$), while these values are two order of magnitude  higher than
the elemental [D]/[H] $\sim$1.5$\times$10$ ^{-5}$ ratio \citep{Linsky06}.
These ratios  are consistent with  0.018 - 0.08 from a study of DNC and HNC
toward UC\HII\, regions, which are an order of magnitude lower than the
high-mass starless core candidates (HMSCs) \citep{Fontani14}. This result
confirms the theoretic prediction that the deuterated fraction (column density
ratio between a deuterated molecules and its hydrogenated  counterpart, $D_{\rm
frac}$) decreases with the core evolution  after the young stellar object
formed \citep{Caselli02}.  The high $D_{\rm frac}$ found in our sources is
likely because the abundances of deuterated molecules were enhanced in cold 
molecular clouds before the formation of the proto-star or warm deuterium
chemistry drived by CH$_{2}$D$^{+}$ and/or C$_{2}$HD$^{+}$ \citep{Parise09}.
These ratios are summarized in table \ref{tab:ratio}.

\subsection{Long carbon chain molecules} \label{sec:} 

\subsubsection{$c$-C$ _{3}$H$_{2}$}
\label{sec:C3H2}

Two lines of $c$-C$ _{3}$H$_{2}$ (2$_{1,2}$-1$_{0,1}$, 3$_{1,2}$-2$_{2,1}$) were
detected in all four sources.  As a small cyclic hydrocarbon,  $c$-C$
_{3}$H$_{2}$ was first detected towards Sgr B2 \citep{Thaddeus85}, and then was
detected in a variety of sources \citep{Vrtilek87,Madden89,Lucas00,Qi13}.
$c$-C$ _{3}$H$_{2}$ is  easily dissociated by radiation from the central star \citep{Qi13}. 
Acetylene (C$_2$H$_2$), which is released from polycyclic aromatic hydrocarbons (PAHs),
has been used to explain the $c$-C$ _{3}$H$_{2}$ enhancement in UV exposed
regions  \citep{Fuente03}.   

The velocity-integrated intensity map of $c$-C$ _{3}$H$_{2}$ 2-1 shows a cavity
around the emission peak of  the 8.4 GHz radio continuum in DR21S (figure
\ref{fig:2}).  However, the $c$-C$ _{3}$H$_{2}$ 3-2 emission peaks at the radio
continuum emission peak, which implies that the abundance of $c$-C$
_{3}$H$_{2}$ in the radio continuum emission peak region should not be lower
than its surrounding region. In addition, the spectra of $c$-C$ _{3}$H$_{2}$
2-1 have an absorption feature toward the cavity region as shown in figure
\ref{fig:2}. Therefore, the cavity feature of $c$-C$ _{3}$H$_{2}$ 2-1 toward
the radio continuum emission peak should be caused by absorption. The
velocity-integrated intensity map of $c$-C$ _{3}$H$_{2}$ 2-1 in G34 also shows an
absorption feature similar to that in DR21S (figure \ref{fig:2}).  In figure
\ref{fig:2}, we also show the velocity-integrated intensity map of $c$-C$ _{3}$H$_{2}$ in
S76E, with two emission peaks. The slightly different spatial distributions
between $c$-C$ _{3}$H$_{2}$ 2-1  and 3-2 might be due to  both transitions 
have different excitation conditions.

As shown in figure \ref{fig:2}, the $c$-C$ _{3}$H$_{2}$ 2-1 emission in CepA is
resolved into three main condensations, which are labeled as  P1, P2 and P3 in
the order of clockwise rotation. Similar to DR21S and G34, the $c$-C$
_{3}$H$_{2}$  cavity on top of the radio continuum peak is caused by
absorption. The $c$-C$ _{3}$H$_{2}$ emission is  associated with CS 3-2
emission in P1 and P3, while CS emission is weaker in P2. The $c$-C$
_{3}$H$_{2}$ line emission in P1 is stronger and narrower than
those in P2 and P3.  Since $c$-C$ _{3}$H$_{2}$ has similar spatial distribution
and line profile compared to SO, which is considered as a low-velocity shock
tracer in massive star-forming regions \citep{Podio15}, we suggest that the
main mechanism to enhance abundance of $c$-C$ _{3}$H$_{2}$ in the P2 region is
likely low-velocity shock activity.

\subsubsection{Other long carbon chain molecules}
\label{sec:}

We also detected some other long carbon chain molecules,
including the cyanopolyynes (HC$_{2n+1}$N), such as cyanoacetylene (HC$_3$N) 
and cyanodiacetylene (HC$_5$N), and  C$_{4}$H. Three high-$J$ excitation 
lines of HC$ _{5}$N (J=34-33, J=33-32, and  J=32-31), two transitions of HC$_{3}$N 
(10-9 and  16-15) and their isotopoogues (HCC$^{13} $CN 10-9 and HC$ ^{13}$CCN 
10-9) were detected  in these targets. The HC$_{3}$N isomeride molecule of HCCNC 
and  C$_{3}$S 15-14 was only detected in DR21S and G34, respectively.  
CH$_3$C$_5$N was only detected in G34, which is the largest 
molecule in this observation.  We present these spectra in figure \ref{fig:10}.

\subsection{ONCN} 
\label{sec:ONCN}

Nitrosyl cyanide ONCN (14$_{0,14}$-13$_{0,13}$, 145397.59 MHz) emission, which
has an upper level energy $E_{u}$ of 52.4 K and an Einstein A coefficient
A$_{ij}$ of 2.66$\times 10^{-5}$,  was detected in G34. It is the first
detection of ONCN in massive star formation regions.  Figure \ref{fig:9}
presents the spectra and the velocity-integrated intensity map of ONCN, which
shows an anti-correlated with the ionized gas tracer H41$\alpha$. ONCN emission
is neither associated with the core, nor associated with the shock activity.
The ONCN line was not detected in previous observations probably because its emission not
focused on the center of source,  and previous observations did not cover this frequency.

%%%%%%%%%%%%%%%%%%%%%%%%%%      DISCUSSION     %%%%%%%%%%%%%%%%%%%%%%%%%%%%%%%%%%%%
\section{Discussions}

\subsection{Excitation and column densities}
\label{sec:excitation}

In all the four sources, we have detected several transitions of CH$_{3}$CN
(methyl cyanide), which is an excellent probe of gas temperature in warm dense
environments \citep{Goldsmith99}. We estimated the excitation conditions of CH$_{3}$CN
using the population diagram method introduced by \cite{Goldsmith99}:

\begin{equation}
\label{population}
\ln{\left( \frac{N_u}{g_u} \right)}=\ln{ \left( \frac{N_{\rm{tot}}}{Q_{\rm{rot}}} \right)}-\frac{E_u}{k T_{\rm{ex}}}-\ln{\left(\frac{\Omega_{source}}{\Omega_{beam}}\right)}-\ln{(C_\tau)},
\end{equation}

where $f_{beam}$ = $\Omega_{source}/\Omega_{beam}$, is the beam filling
factor. C$_{\tau}$ = $\tau$/(1-e$^{-\tau}$), $\tau$ is the optical depth. For
an uniform beam filling ($\Omega_{beam}\sim \Omega_{source}$) and low optical
depth, the population diagram reduces to a rotational diagram with a rotational
temperature $T_{rot}$ and total column density $N_{tot}$. Column density
($N_{tot}$) and rotational temperature ($T_{rot}$) can be estimated based on 

\begin{equation}
\label{population}
\ln{\left( \frac{N_u}{g_u} \right)}=\ln{ \left( \frac{N_{\rm{tot}}}{Q_{\rm{rot}}} \right)}-\frac{E_u}{k T_{\rm{rot}}},
\end{equation}

where $N_{u}$ is the observed upper state column density of the molecule
including line opacity and beam-source coupling effects; $g_{u}$ is the
degeneracy of the upper state; $Q_{rot}$ is the rotational partition function;
$k$ is the Boltzmann constant; $E_{u}$ is the upper level energy. 

The rotation diagrams are shown in figure \ref{fig:11}. Under the LTE
assumption, the excitation temperature is the same as the rotational
temperature.  The excitation temperature of CH$_{3}$CN in G34 ($\sim$ 68K) and
Cep A ($\sim$ 71K) are higher than DR21S  ($\sim$ 47K) and S76E ($\sim$ 47K).
While the total column density of CH$_{3}$CN  in G34 ($\sim$ 1.5$\times
10^{14}cm^{-2}$) and S76E ($\sim$5.4$\times 10^{14}cm^{-2}$) are higher than
that in  Cep A ($\sim$ 1.1$\times 10^{13}cm^{-2}$) and 
DR21S ($\sim$ 4.0$\times 10^{13}cm^{-2}$).  CH$_{3}$CN in  G34 
has a higher excitation temperature and
higher column density than those in the other three sources, indicating that
G34 is in an more evolved stage. The high excitation temperature of CH$_{3}$CN
found in Cep A might be affected by the radio jet from HW2. We summarize the
diagnostic results of CH$_{3}$CN in table \ref{tab:excitation}.

\subsection{Shock related molecules}
\label{sec:shock}

\textbf{Cep A}. Figure \ref{fig:3} shows the velocity-integrated intensity map
of SiO 2-1, which is a fast shock tracer \citep[e.g.,][]{Qiu07}.  The SiO emission stretchs 
along the direction of NE-SW, indicating that shock is in the NE-SW direction, 
which is consistent with the results in literature \cite{Codella06}.  
SiO emission peaks are located
at two positions: offset (0$''$, 0$''$) with $v_{LSR}$ of -9.77 km s$ ^{-1}$,
and offset (30$''$, 30$''$) with  $v_{LSR}$ of  -8.38 km s$ ^{-1}$, which give
an velocity gradient of $ \sim $0.66 km s$^{-1}$kAU$^{-1}$ (1arcsec =
700\, AU at the given distance of 0.7 kpc).  The line width of the second peak
(8.57 km s$^{-1}$) is broader than the first one (5.55 km s$^{-1}$).

SO is an important shock tracer in star formation regions
\citep{Podio15}, and it can be significantly enhanced by low-velocity shocks
\citep{Chernin94}. From the channel map of SO (figure \ref{fig:3}), it appears
that the -5.36 km s$ ^{-1}$ component of SO shows similar spatial distribution as SiO.
This might be due to both emissions are affected by central star, since they are emissions 
associated with radio continuum. SO also shows strong
emission in northern region (P2) of the core at a velocity of -9.44 km s$
^{-1}$, and its spectra profile shows a blue-shifted line wing with high
T$_{mb}^{*}$ and a narrow line width. The HCO$^{+}$ 1-0 emission in P2 also 
shows a blue profile, the spectrum presents a blue-shifted emission and red-shifted absorption, 
which are considered as a signature of infalling gas \citep{Liu13}. 
Thus, low-velocity shocks likely  dominate gas properties here,  due to the gas infall.

\textbf{DR21S}. Figure \ref{fig:4}  shows the velocity-integrated intensity maps
of shock tracers (SiO 2-1, HNCO 4-3, SO 2-1). They show strong emission 
 on the south side of the core with narrow line width.  
This region likely has a shock activity, which is consistent
with previous observations (see the figure A1: A 4-2 in \cite{Davis07}).  
As shown in figure \ref{fig:4}, a blue-shifted line wing with an inverse P-Cygni profile of HCO$^{+}$ 
can be seen  in the western region of the core, while SiO,
HNCO  and SO  without obvious emission in this region. 
Class I CH$ _{3} $OH maser emission at 84.5 GHz  was also detected in this region, 
and it was considered to be associated with outflows/inflows in star-forming region \citep{Sobolev93}.
These results are consistent with previous high spatial resolution observations
that there are detected several shock activities (see the figure A2: B 1-1, B
1-2 and B 1-3 in \cite{Davis07}).

\textbf{G34}. We detected several shock gas tracers in G34, including SO 2-1, HNCO 4-3, 
SiO 2-1 and its isotopologues ($ ^{29}$SiO 2-1 and $ ^{30}$SiO 2-1).
They show  strong emission in the west region of the core (figure \ref{fig:5}), 
this shocked region and the tracer velocity are in agreement with the results of study SiO in \cite{Hatchell01}. 
Their emitting region correspond to a cometary compact HII region, where is named ``C" in literature 
\cite{Mookerjea07}. Figure \ref{fig:5} shows that the spectra of HCO$^{+}$ present a a blue-shifted 
line wing with an inverse P-Cygni profile  in the western region.  
CS 3-2 emission also shows a redshifted absorption feature in the same region.
These features may be due to the gas infalling towards cometary HII region ``C'' \citep[e.g.,][]{Liu13}.

\textbf{S76E}.  Figure \ref{fig:6} shows the velocity-integrated intensity maps of SiO 2-1 and SO 2-1. 
Their emission only associate with a part of radio continuum emission, where the south-west side of the core.
The velocity-integrated intensity map of SiO redshifted emission, integrated from 33.5 to 47 km s$^{-1}$, 
shows similar spatial distribution to the blueshifted emission, integrated from  28.2 to 33.5 km s$^{-1}$. 
In figure \ref{fig:6} we show the position-velocity (pv) diagram of SiO, which presents a strong emission in the blueshifted.

In all the four sources, we have detected several shock gas tracers, 
including SO, HNCO, SiO and its isotopologues, at a level of $>$ 5$\sigma$. 
The SiO emission usually associate with strong shocks, 
it is a ideal probe for revealing the processes of jet/outflow activity associated with young protostars, 
such as in the case of Cep A.  
SO not only can be enhanced in hot cores, but also in shocked region. 
This emission likely can be enhanced by low-velocity shocks, due to the gas infall, 
such as in the P2 region of Cep A. 
These tracers provide a powerful tool to investigate the shock processes in star formation regions.

\subsection{Deuterium-bearing molecules}
\label{sec:deuterium}

The deuterated molecules are widespread in massive star-forming regions \citep{Case12}.   
High $D_{frac}$ toward these regions provide a chemical ``fingerprint" of physical conditions \citep{Shah01}.  
We have detected DCN 2-1, DCO$^{+}$ 2-1 and NH$_{2}$D 1-1  in all four sources, 
C$ _{2}$D 2-1 in three sources except for G34, and  HDCS (3-2) in G34.

\textbf{Cep A}. The velocity-integrated intensity maps of DCN 2-1 and DCO$^{+}$ 2-1 show  different spatial 
distributions compared to the radio continuum emission (figure \ref{fig:3}). 
DCO$^{+}$ emissions are resolved into three main condensations surrounding the emission peak 
of radio continuum with different line central velocities, 
which are labeled as  M1, M2 and M3 in order of clockwise rotation.  
The velocity is -12.0 km s$ ^{-1}$ in M1, -10.1 km s$ ^{-1}$  in M2, and -9.7 km s$ ^{-1}$ in M3. 
The DCO$^{+}$ emission in M1 and M3 have comparable line widths (FWHM) of $\sim$1.4 km s$ ^{-1}$, 
which is broader than 1.0 km s$ ^{-1}$ in M2.  
DCN and NH$ _{2}$D also show strong emission around the M1 region, 
and DCN presents similar spatial distribution to the DCO$^{+}$ in M1. 
The deuterated molecules likely have been enhanced around the M1 region, 
since there are strong emission of DCO$^{+}$, DCN and NH$ _{2}$D.
As discussed in Section \ref{sec:shock},  
there are strong shock activity in  M1 region and low-velocity shock activity in M2 region.  Moreover, 
weak emission of SiO toward M3 indicates that there is likely associated with an intermediate shock activity.

 Two main pathways are considered for forming deuterated molecules in different temperature ranges:  
 at low temperature of $T\sim$ 10 - 30 K, the dominant pathway is via H$_{3}^{+}$ isotopologues; 
 at slightly high of $T\sim$ 30-80 K, the dominant pathway is via light hydrocarbons 
 (CH$_{2}$D$^{+}$ and C$_{2}$HD$^{+}$) \citep{Albertsson13}.
The DCO$^{+}$ is considered mainly forming at low temperated through H$_{3}^{+}$ isotopologues, 
 thus it is expected to be formed during the prestellar phase and sensitive to freeze-out \citep{Albertsson13}.  
 One possible explanation of DCO$^{+}$ shows strong emission around M2 and M3 regions is that 
 it's abundance has been enhanced by shock, since it is likely to be injected into gas phase from 
 the grain mantles through shock activity \citep{Mangum91}.   
 The DCO$^{+}$ emission around M1 not only associated with radio continuum emission, 
 but also in the vicinity of the shocked region with the same velocity compared to SiO (-9.7 km s$ ^{-1}$), 
 as well as the DCN emission.
 The DCN is considered to be formed by light hydrocarbon at relatively high temperature 
 of $T\sim$ 30 - 80 K \citep{Albertsson13}, and it is not mainly formed on grain-surface.
The gas, in the surrounding areas of central young protostar, 
will be warming with the shock activity and  evolution of dense cores. 
This may be responsible for enhancement of DCO$^{+}$ and DCN emission around M1 region, 
since DCO$^{+}$ can be released into gas phase from dust grain 
and DCN can be formed with an increase in temperature. 
DCO$^{+}$ shows a cavity in the vicinity of the emission peak of radio continuum, 
which may be due to there are low abundances or it was destroyed by high temperature (heated by central star), 
since DCO$^{+}$ can be destroyed  in high temperature environment \citep{Pety07}. 
NH$_{2}$D is considered formed at the low temperatures ($<$30 K) and high densities ($\simeq10^{6} cm^{3}$) 
regions (e.g., starless cores, protostars), 
and it not only can be formed via gas-phase reactions but also via grain-surface reactions with D atoms \citep{Tielens83}. 
The NH$ _{2}$D shows strong emission around the shocked region (M1),  
implying that it likely has been released into gas phase from dust grain through shock activity  \citep[e.g.,][]{Mookerjea09}.

\textbf{DR21S}. As shown in figure \ref{fig:4},  the emission peak of DCN 2-1, 
DCO$^{+}$ 2-1 and NH$_{2}$D 1-1 are away from the emission peak of radio continuum. 
In addition, both DCO$^{+}$ and NH$_{2}$D show obvious emission on the south side of the core, 
 where likely dominated by shock activity considering the aforementioned shock tracers. 
The shock interacts with ambient gas/dust, which can affects the chemical composition of 
the grain mantles and of the molecular gas, is possibly responsible for the enhancement of  
DCO$^{+}$ and NH$ _{2}$D emission on the south side of the core. 
For C$_{2}$D(2-1), the  signal-to-noise ratio is not good  enough to make  a velocity-integrated map, 
and its spectrum is shown in figure \ref{fig:4}.

\textbf{G34}. The velocity-integrated intensity maps and spectra of deuterated molecules are presented 
in figure \ref{fig:5}. Both DCN 2-1 and NH$_{2}$D 1-1 show obvious emission on the west side of the core, 
where is associated with shock activity as suggested by evidences of SiO and SO.
This indicates that the shock activity likely responsible for the enhancement of DCO$^{+}$ and NH$_{2}$D 
emission in this region, since the gas temperature can be increased in shocked region, 
and both molecules can be evaporated to gas phase from the dust grain with an increase in temperature. 
DCO$^{+}$ 2-1 shows obvious emission on the north side of the core, as well as NH$_{2}$D.
The gas temperature in this region is around 30 K derived from CH$_{3}$CN. 
Both DCO$^{+}$ and NH$_{2}$D lines can be excited in such high temperature environments. 
We present  the spectrum of HDCS 3-2 in figure \ref{fig:5} instead of velocity-integrated intensity map, 
due to the low signal-to-noise ratio.

\textbf{S76E}. Figure \ref{fig:6} shows the velocity-integrated intensity maps of DCN 2-1 and NH$_{2}$D 1-1.  
Both NH$_{2}$D and DCN emissions are in the vicinity of the radio continuum emission, 
while looks more extended in NH$_{2}$D. 
NH$_{2}$D shows obvious emission in the south-western region of the radio continuum emission, 
where also found obvious SiO emission. 
This indicates that the NH$_{2}$D emission likely associate with shock activity in this region. 
We show the spectra of DCO$^{+}$  2-1 and C$ _{2}$D 2-1 in figure \ref{fig:6} instead of velocity-integrated intensity map, due to the low signal-to-noise ratio.

From above analysis, we noted that the NH$_{2}$D emission has been enhanced by shock activity toward all sources, while DCO$^{+}$ and DCN emission have been enhanced by shock activity only in some cases, e.g., Cep A and DR21S for DCO$^{+}$, Cep A and G34 for DCN.

%%%%%%%%%%%%%%%%%%%%%%%%%%%%%%%%%%%%%%%%%%%%%%%%%%%%%%%%%%%%%%%%%%%%%%%%%%%%

\section{Summary and future prospectus}
\label{sec:summary}

We  carried out  spectral line mapping observations toward four
massive star-forming regions (CepA, DR21S, G34 and S76E) with the IRAM 30 meter
telescope. Our main results are summarized as: 

\begin{enumerate}

\item 
A total of  395 spectral features were detected in four massive star-forming
regions. In total we identified 61 lines from 25 molecules in Cep A; 78 lines
from 43 molecules, an unidentified line, a helium combination line and six RRLs
in DR21S; and 169 lines from 86 molecules,  three RRLs, and nine unidentified
lines in G34; 78 lines from 44 molecules, one RRLs and six unidentified lines in S76E.

\item 
 We detected a large oxygen-bearing molecules of acetone (CH$_{3}$)$_{2}$CO in DR21S. 
For all four sources, we detected three transitions of HC$ _{5}$N,  
two transitions of HC$_{3}$N and its  isotopologues  of HCC$ ^{13} $CN and  HC$ ^{13}$CCN 
in all four sources. The HC$_{3}$N isomeride  molecule, HCCNC  was only detected in DR21S, 
while C$ _{3}$S  was detected in G34, and  C$_{4}$H was detected in DR21S and S76E.
The nitrosyl cyanide ONCN(14$_{0,14}$-13$_{0,13}$) emission, 
as the first detection of ONCN in massive star-forming region, 
was detected in G34, and its emission shows anti-correlated with the RRLs.

\item 
 We derived several isotope abundance ratios in each source: For example,  
 $^{12}$C/$^{13}$C, $^{14}$N/$^{15}$N, $^{16}$O/$^{18}$O and  $^{33}$S/$^{34}$S.  
 The $^{12}$C/$^{13}$C ratio are about 40, 
 which are slightly lower than the ratio of 60-70 found in the local ISM \citep{Milam05}.
 The $^{14}$N/$^{15}$N ratios are from 109 to 304, 
 and the $^{16}$O/$^{18}$O  ratios are  from 287 to 532.  
 The $^{33}$S/$^{34}$S ratios are about 5, except for CepA, in which C$^{33}$S 3-2 line is not detected. 
The ratios of N(DCN)/N(HCN) and N(DCO$^{+}$)/N(HCO$^{+}$) in each source are two orders 
of magnitude higher than the elemental [D]/[H] ratio (1.5$\times$10$^{-5}$), which is consistent 
with results in literature \cite{Parise09}.

\item 
Several shock tracers (SiO, SO and HNCO) were detected and used to investigate the shock 
properties in these star-forming regions. Comparing these shock traces and deuterated 
molecules, we found that NH$_{2}$D emission can be enhanced in shocked region, and 
both DCO$^{+}$ and DCN emissions also can be enhanced toward shocked region in some 
cases (e.g., Cep A and DR21S/G34).

\item
The CH$_{3}$CN lines are used to derive rotational temperature and column density of this 
line in each source, which show the highest excitation temperature and column density in G34. 
                                                                                                                                                                                                                                                                                                                                                                                                                                                                                                                                                                                                                                                                                                                                                                                                                                                                                                                                                                                                                                                                                                                                                                                                                                                                                                                                                                                                                                                                                                                                                                                                                                                                                                                                                                                                                                                                                                                                                                                                                                                                                                                                                                                                                                                                                                                                                                                                                                                                                                                                                                                                                                                                                                                                                                                                                                                                                                                                                                                                                                                                                                                                                                                                                                                                                                                                                                                                                                                                                                                                                                                                                                                                                                                                                                                                                                                                                                                                                                                                                                                                                                                                                                                                                                                                                                                                                                                                                                                                                                                                                                                                                                                                                                                                                                                                                                                                                                                                                                                                                                                                                                                                                                                                                                                                                                                                                                                                                                                                                                                                                                                                                                                                                                                                                                                                                                                                                                                                                                                                                                                                                                                                                                                                                                                                                                                                                                                                                                                                                                                                                                                                                                                                                                                                                                                                                                                                                                                                                                                                                                                                                                                                                                                                                                                                                                                                                                                                                                                                                                                                                                                                                                                                                                                                                                                                                                                                                                                                                                                                                                                                                                                                                                                                                                                                                                                                                                                                                                                                                                                                                                                                                                                                                                                                                                                                                                                                                                                                                                                                                                                                                                         
\end{enumerate}

Mapping observations of multiple lines toward massive star-forming regions at different 
evolutionary stages allow detailed studies of chemical properties, shock activities and 
physical conditions. Sensitive high-angular resolution observations with (sub-)millimeter 
interferometers, such as ALMA, will be a powerful tool to constrain  chemical and physical 
properties during massive star formation process.

%%%%%%%%%%%%%%%%%%%%%%%%%%%%%%%%%%%%%%%%%%%%%%%%%%%%%%%%%%%%%%%%%%%%%%%%%%%%

\section*{\textit{Acknowledgements}}

We thank the  support  of  the China Ministry of Science and Technology under
the State Key Development Program for Basic Research (2012CB821800), the
Natural Science Foundation of China under grants of 11173013.  Z-Y. Z  acknowledges 
support from the European Research Council in the form of the Advanced 
Investigator Programme, 321302, {\sc cosmicism}. We thank helpful
discussions and comments  from Professors   Keping Qiu, Di Li, and Qizhou Zhang
to improve the manuscript.  
We thank the anonymous referee for her/his valuable comments that helped to improve the paper.

\bibliographystyle{mn2e}
\bibliography{bibtex}
\label{lastpage}

\begin{table*}
%\begin{minipage}{200mm}
 \centering
  \begin{minipage}{160mm}
     \begin{center}
	 \caption{3.6 cm radio continuum emission of these four sources}
	  \label{tab:coordination}
		\begin{tabular}{c c c c  c c c c c c}
 \hline
 \hline        
  Source 		& R.A. 			& Decl. 		&Gal Lat	& D 					& D$ _{g}$ & $v_{LSR}$&	$S_{p}$ 	&$S_{\nu}$&Synthesized Beam\\
 			&	(J2000)  		&   (J2000) 	& degree	& (kpc) 				& (kpc)   & (km s$ ^{-1}$)&	mJy beam$^{-1}$ &mJy&($^{''}\times ^{''}$, $^{\circ}$)\\
	         \hline
	          \hline        
Cep A		&	22:56:18.1	 &	62:01:46.2	& 2.1	& 0.7$\pm$0.04(ref 1)   	& 8.7  	 &  -11 	&4.81$\pm$0.05	&10.3$\pm$0.1&8.8$\times$6.6, -22\\
DR21S		&	20:39:00.8	&	42:19:29.8 	& 0.5 	&  1.5$\pm$0.08(ref 2) 	&  8.6  	 & -2 	&5781$\pm$6	&13932$\pm$19&8.8$\times$7.2,89\\
G34  		&	18:53:18.5 	&	01:14:56.7	& 0.1	&  3.3  (ref 3)			& 5.8  	 & 58		&2051$\pm$3	&2512$\pm$5&12$\times$9,-23\\
S76E		&	18:46:10.4 	&	07:53:14.1 	& 4.7	&  2.11 (ref 4)			&  7.0   	 & 33	&133$\pm$1		&2437$\pm$28&14.4$\times$8.7,50\\
	         \hline	  
\end{tabular}
\end{center}
Notes. Columns are (1) source name, (2) right ascension, (3) declination, (4)Galactic latitude, (5)distance D and the reference, (6) galactocentric distance D$ _{g}$, (7) Local standard of rest velocity. (8) peak intensities $S_{p}$ of radio continuum emission, (9) flux densities $S_{\nu}$ of radio continuum emission, (10) Synthesized beam of radio continuum emission. \\
References. (1) \cite{Moscadelli09}, (2) \cite{Rygl12}, (3) \cite{Kuchar94}, (4) \cite{Plume92}
\end{minipage}
\end{table*}

\begin{table*}
 \centering
   \begin{minipage}{160mm}
    \begin{center}
		\caption{Isotope abundance ratios}
			\label{tab:ratio}
				\begin{tabular}{c c c c c c c}
 \hline
 \hline
Source & $^{12}$C/$^{13}$C & $^{14}$N/$^{15}$N 	& $^{16}$O/$ ^{18}$O & $ ^{34}$S/$^{33}$S & N(DCN)/N(HCN) 	& N(DCO$^{+}$)/N(HCO$^{+}$)\\
	         \hline
	          \hline
Cep A	&	49.0$\pm$8.6	&	214.1$\pm$41.3	&	422.1$\pm$87.9	&	 				&(2.4$\pm$0.5)$\times10^{-3}$	&(7.9$\pm$1.6)$\times10^{-4}$	\\
	         \hline
DR21S	&	 32.8$\pm$2.4	&	109.9$\pm$11.2	&	532.9$\pm$113.1	&5.1$\pm$0.8	&(3.6$\pm$0.4)$\times10^{-3}$	&(1.5$\pm$0.2)$\times10^{-3}$	   \\
	         \hline
G34  	&	38.7$\pm$2.7	&	289.2$\pm$23.6	&	287.9$\pm$27.4		&4.7$\pm$0.3	&(1.3$\pm$0.1)$\times10^{-3}$	&(3.5$\pm$0.7)$\times10^{-4}$	\\
	         \hline
S76E	&	46.4$\pm$4.2		&	304.2$\pm$31.9	&	 415.9$\pm$60.1	&5.5$\pm$0.4	&(1.1$\pm$0.1)$\times10^{-3}$	&(9.8$\pm$4.8)$\times10^{-4}$    \\
	         \hline	
 Local interstellar medium	&	60-70(ref 1)	&	400(ref 2)  &557$\pm$30(ref 3) 	 & 6.3$\pm$1(ref 4)	&		&	   \\	 
 \hline         
\end{tabular}
\end{center}
Notes. References. (1) \cite{Milam05}, (2) \cite{Terzieva00}, (3) \cite{Wilson99}, (4) \cite{Chin96}.\\
The errors of isotope ratios were calculated by the error propagation equation in \cite{2003drea.book.....B}.
\end{minipage}
\end{table*}

\begin{table*}
 \centering
   \begin{minipage}{160mm}
    \begin{center}
		\caption{Results of the rotation diagram analysis for CH$_{3}$CN with multiple transitions}
			\label{tab:excitation}
				\begin{tabular}{c c c c c c  c c c c c}
 \hline
 \hline
Source	&Cep A&		&	DR21S	&		&	G34	&		&	S76E	&	\\
\hline
					&T$_{rot}$	&N$_{tot}$&T$_{rot}$		&N$_{tot}$&T$_{rot}$		&N$_{tot}$&T$_{rot}$	&N$_{tot}$\\
Position		&K	&$\times 10^{13}cm^{-2}$&K			&$\times 10^{13}cm^{-2}$&K		&$\times 10^{13}cm^{-2}$&K		&$\times 10^{13}cm^{-2}$\\
\hline
(0$''$,0$''$)		&39$\pm$9&1.0$\pm0.3$&47$\pm$4&4.0$\pm0.5$	&		&			&		&			\\
(0$''$,15$''$)		&		&			&45$\pm$9&2.6$\pm0.8$	&		&			&40$\pm$4&53.7$\pm8$	\\
(0$''$,30$''$)		&		&			&33$\pm$6&1.6$\pm0.4$	&		&			&		&			\\
(0$''$,-15$''$)		&		&			&20$\pm$2&1.9$\pm0.3$	&		&			&47$\pm$13&1.9$\pm0.7$\\
(0$''$,-30$''$)		&		&			&24$\pm$4&1.4$\pm0.3$	&		&			&		&			\\	
(15$''$,0$''$)		&		&			&		&			&		&			&35$\pm$4&3.0$\pm$0.4	\\
(15$''$,15$''$)		&71$\pm$27&1.1$\pm0.6$	&		&			&		&			&		&			\\
(-15$''$,15$''$)		&		&			&		&			&68$\pm$2&15.4$\pm$0.1&		&			\\	
(-15$''$,30$''$)		&		&			&		&			&38$\pm$3&3.8$\pm0.4$	&		&			\\	
(-15$''$,45$''$)		&		&			&		&			&26$\pm$3&1.0$\pm0.2$	&		&			\\	
(-15$''$,60$''$)		&		&			&		&			&33$\pm$7&1.1$\pm0.3$	&		&			\\	
(-30$''$,15$''$)		&45$\pm$15&0.7$\pm0.3$&		&			&		&			&		&			\\	
(-30$''$,-15$''$)		&		&			&		&			&		&			&44$\pm$13&0.9$\pm0.3$\\	
	         \hline	 
\end{tabular}
\end{center}
Notes: From the mapping results,  we only selected the positions that detected three transitions of CH$_{3}$CN for calculating the T$_{rot}$ and N$_{tot}$.
\end{minipage}
\end{table*}

\clearpage
%%%%%%%%%%%%%%%%%%%%%%%%%%%%%%   Cep A  %%%%%%%%%%%%%%%%%%%%%%%%%%%%%%%%%%%%%%%%%%%%%%%

\begin{figure*}
	\centering
\scalebox{0.44}{\includegraphics{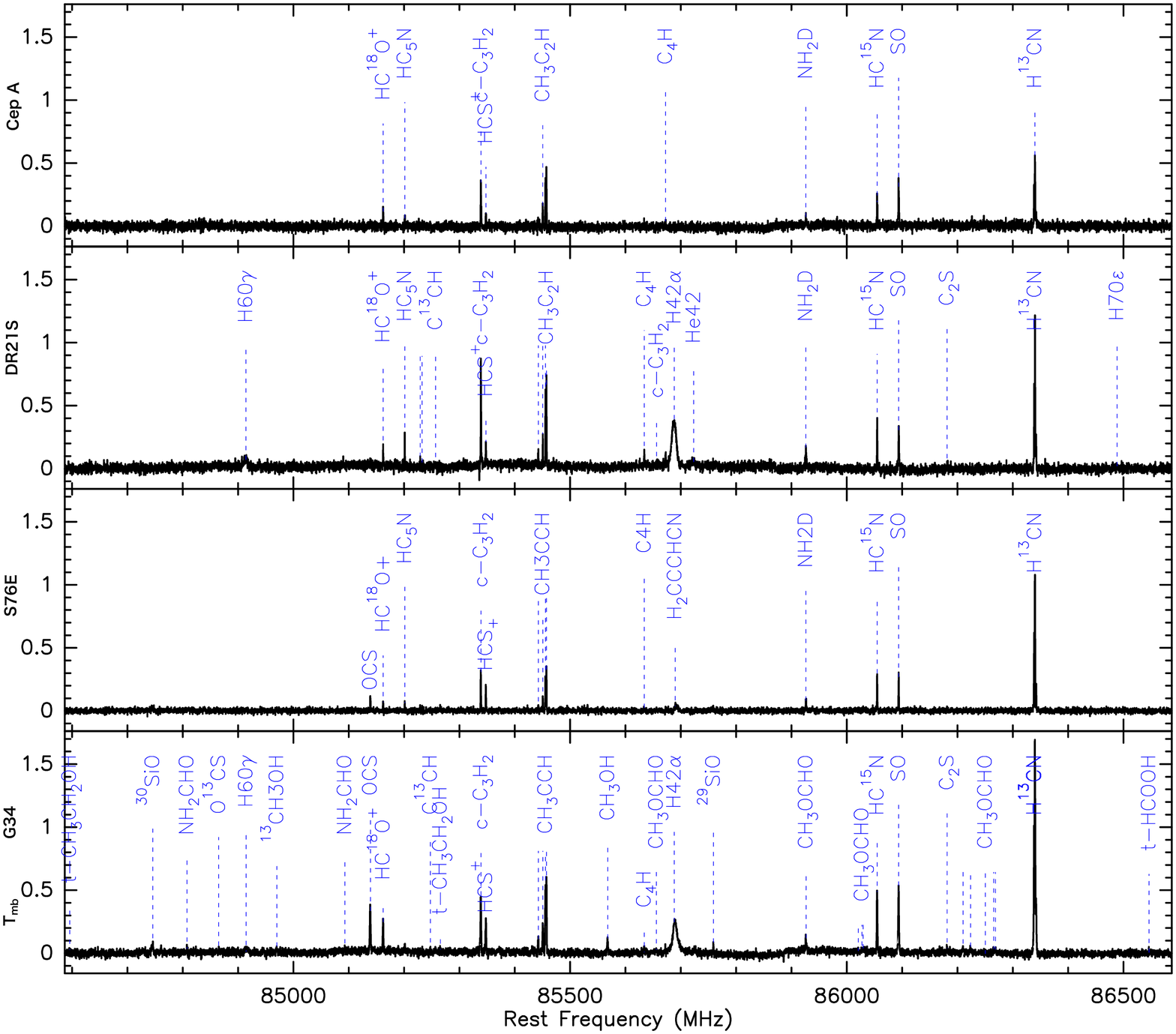}}
\scalebox{0.44}{\includegraphics{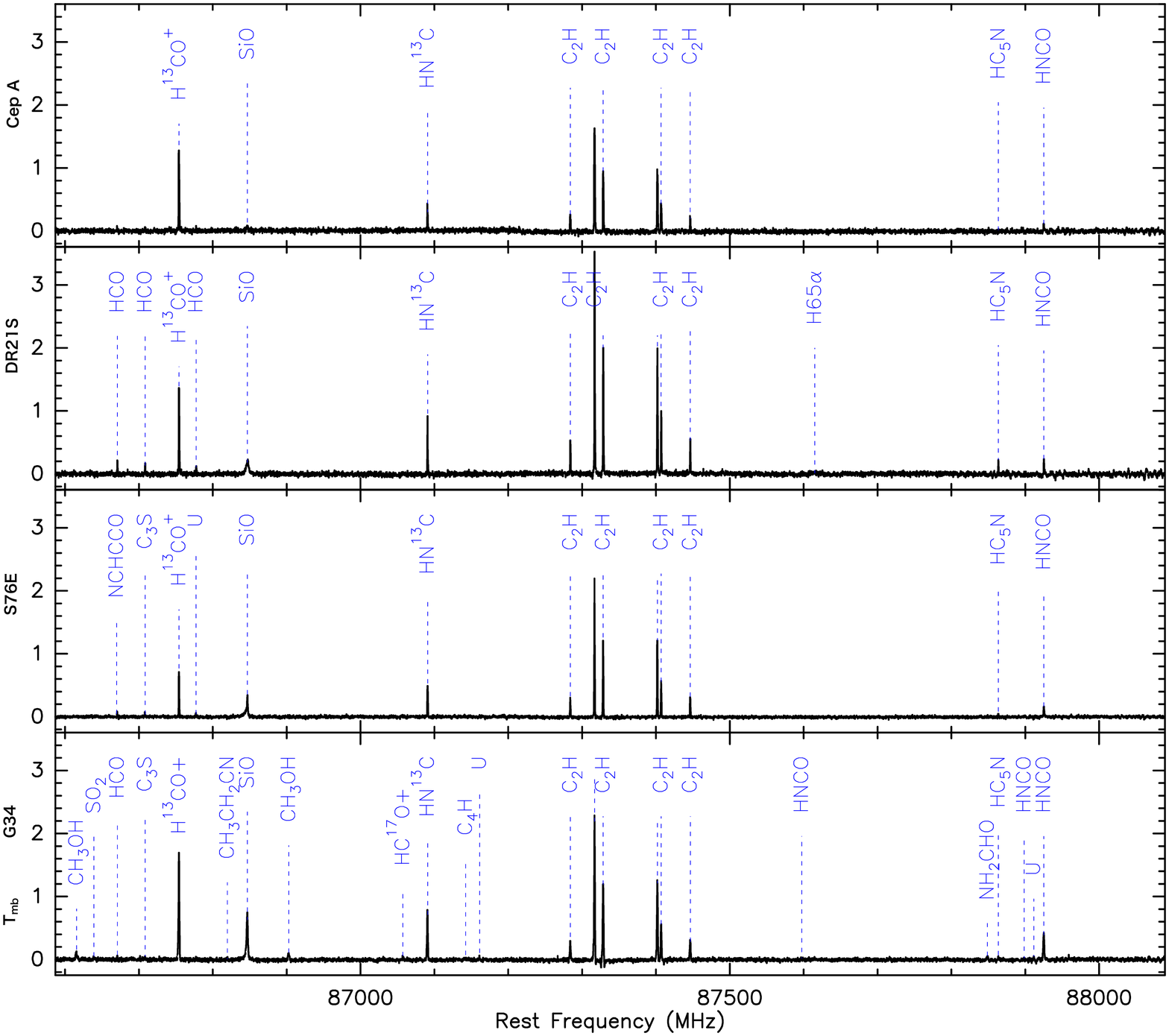}}
%\vspace*{-0.2 cm}
\caption{Comparison of the Cep A, DR21S, S76E and G34 survey at 3mm and 2mm wavelengths. The spectrum is averaged over 
 	\label{fig:1}}
%    \vskip-10pt
\end{figure*}

\addtocounter{figure}{-1} 

\begin{figure*}
  \addtocounter{subfigure}{1}
	\centering
\scalebox{0.44}{ \includegraphics{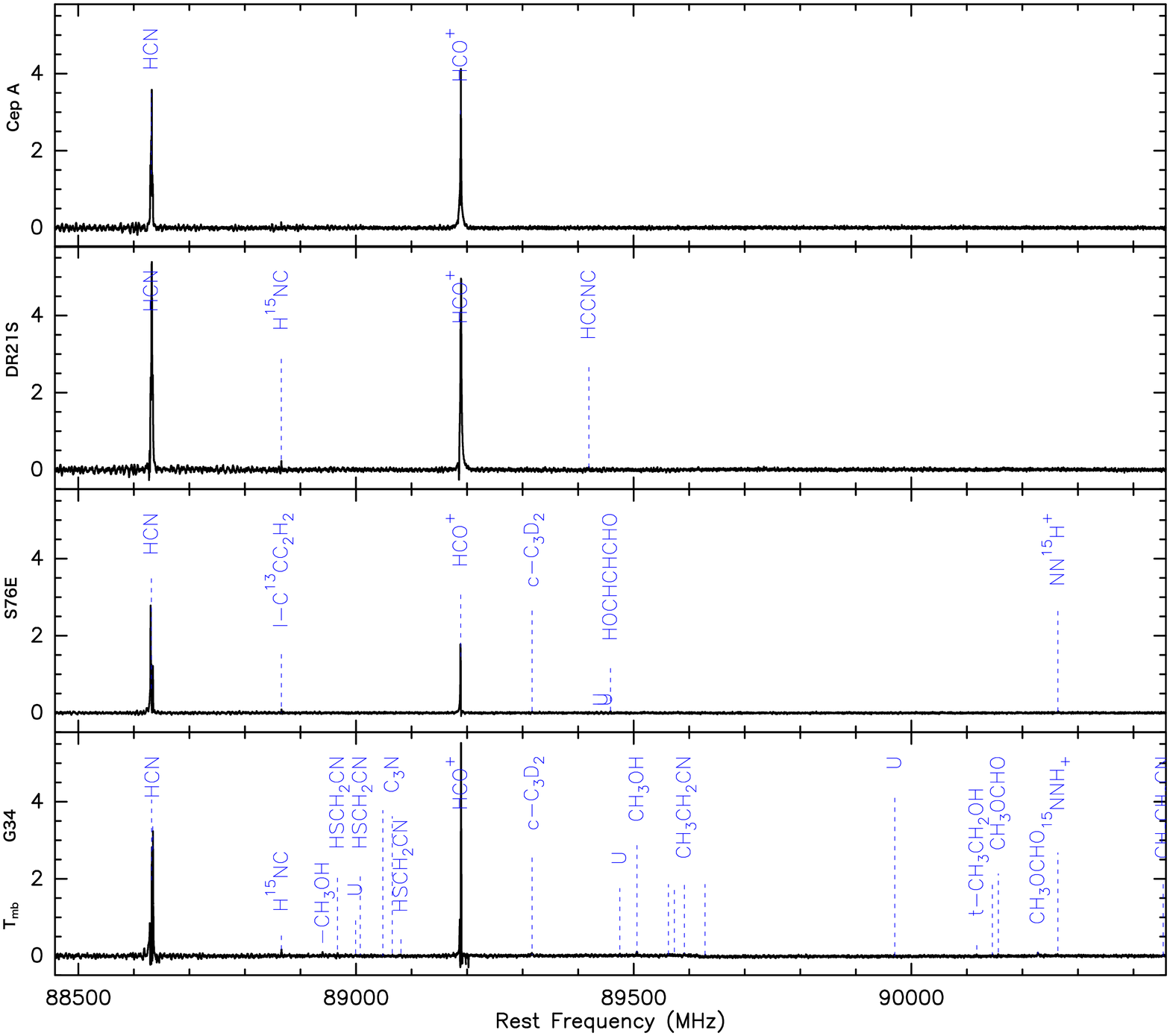}}
\scalebox{0.44}{ \includegraphics{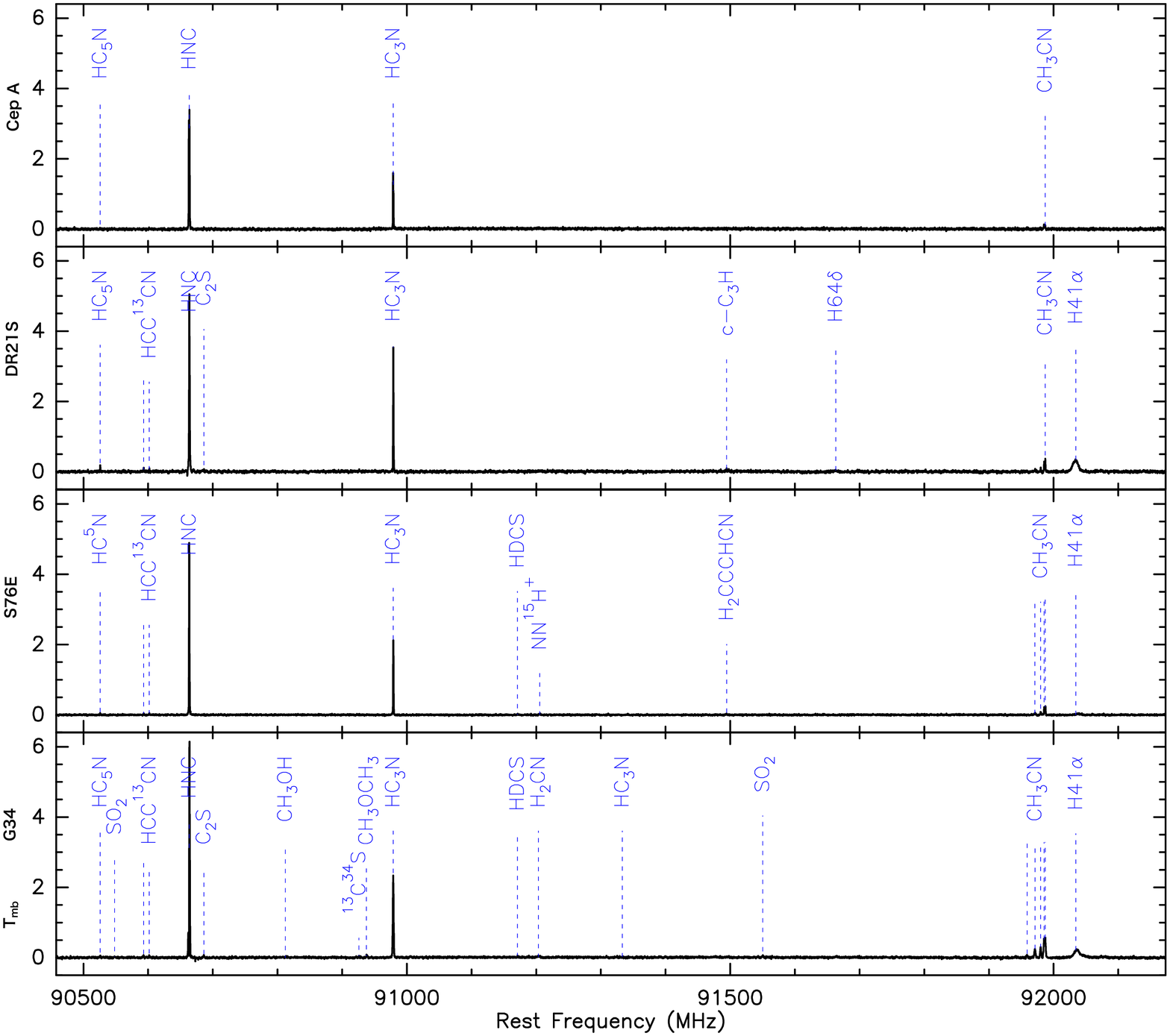}}
%\vspace*{-0.2 cm}
\caption{the map of 30$^{''} \times$ 30$^{''}$ for each source. The spectra velocity(frequency) resolution is 0.65 km/s (0.195MHz) and 0.4 km/s (0.195MHz) at 3mm and 2mm, respectively,  while the unit is K with main-beam temperature.
	\label{fig:1}}
%    \vskip-10pt
\end{figure*}

\addtocounter{figure}{-1}

\begin{figure*}
  \addtocounter{subfigure}{1}
	\centering
\scalebox{0.45}{ \includegraphics{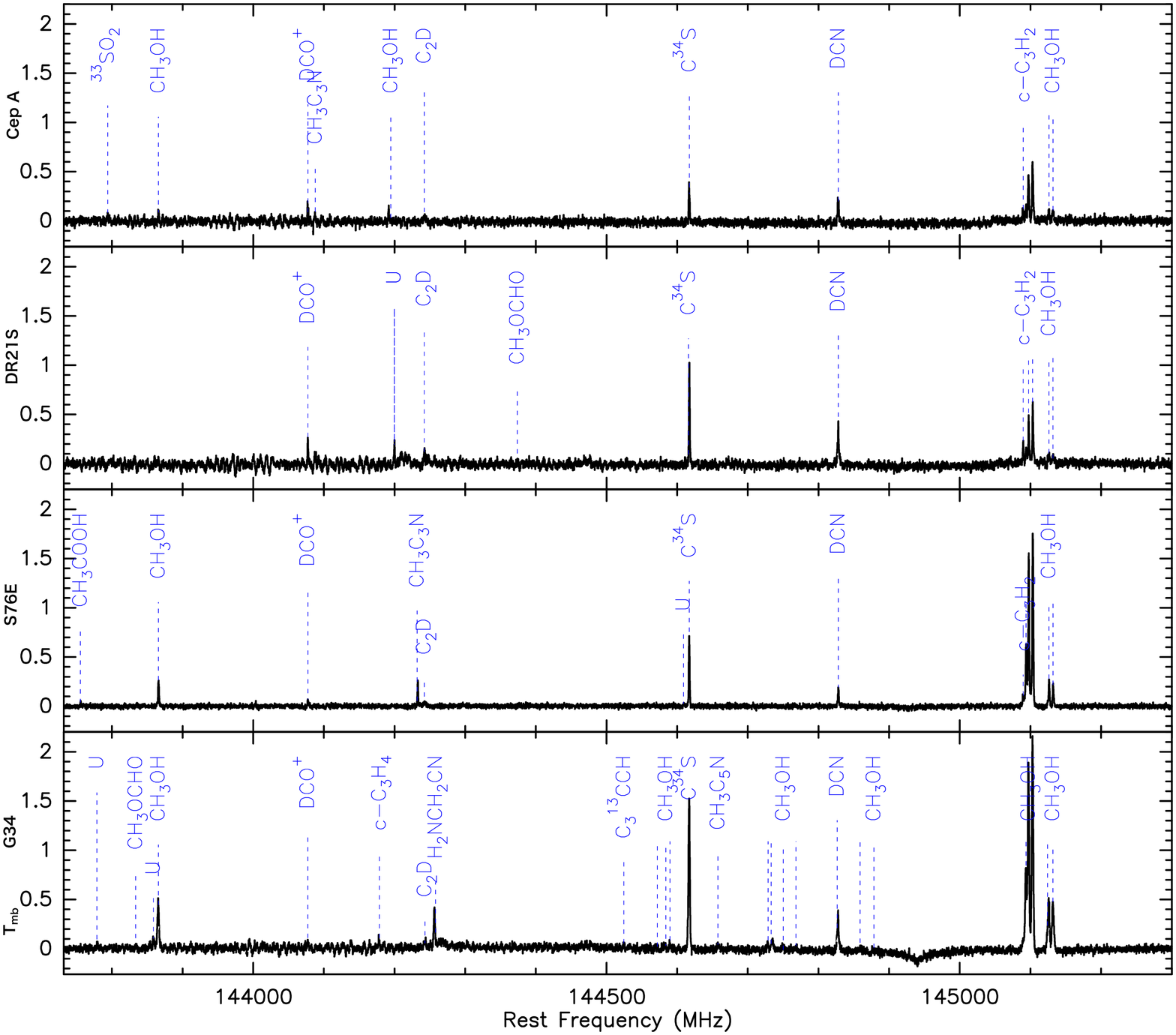}}
\scalebox{0.45}{ \includegraphics{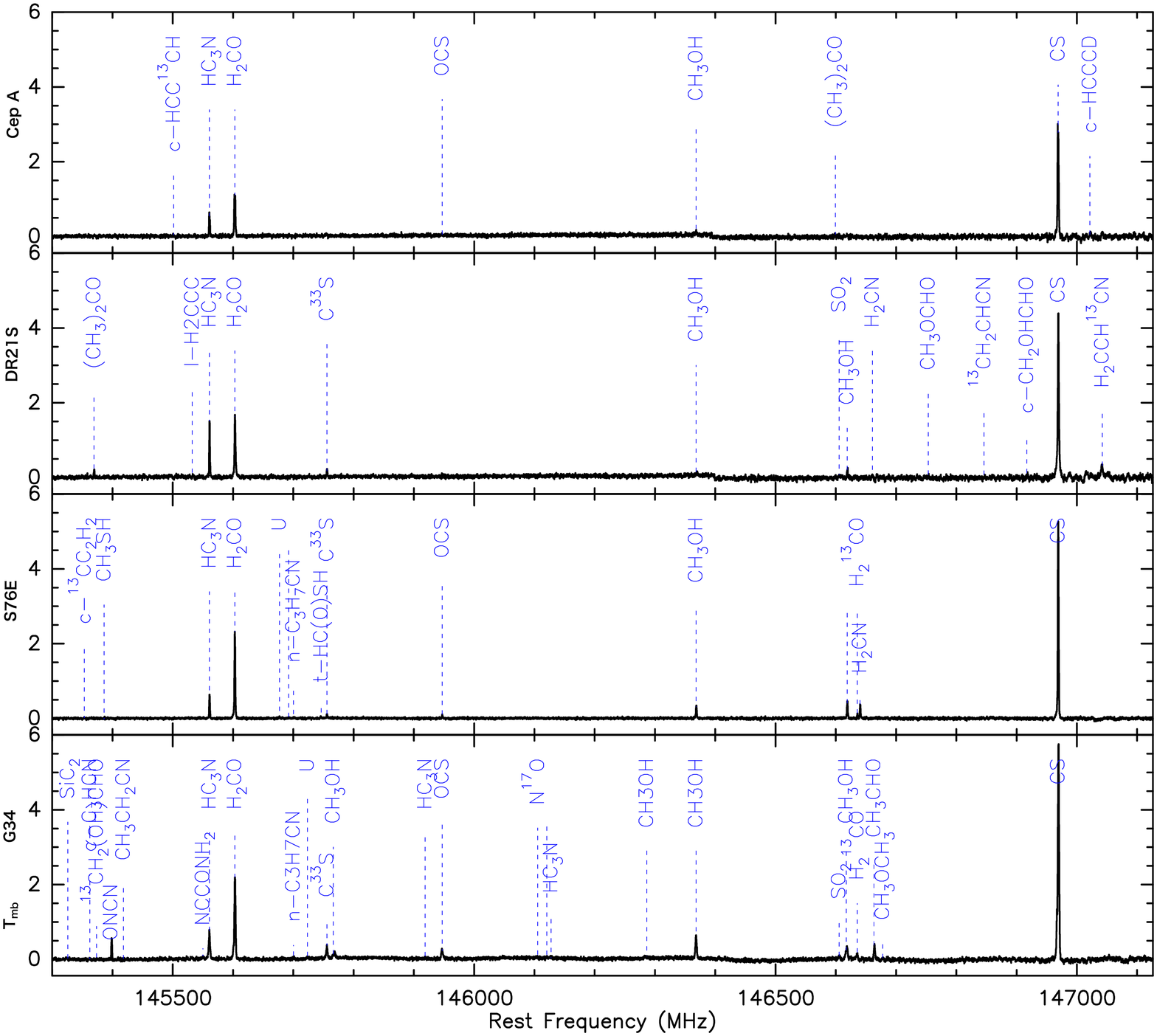}}
%\vspace*{-0.2 cm}
\caption{continued
	\label{fig:1}}
%    \vskip-10pt
\end{figure*}

\clearpage

\begin{figure*} 
	\centering 
\scalebox{1}{\includegraphics{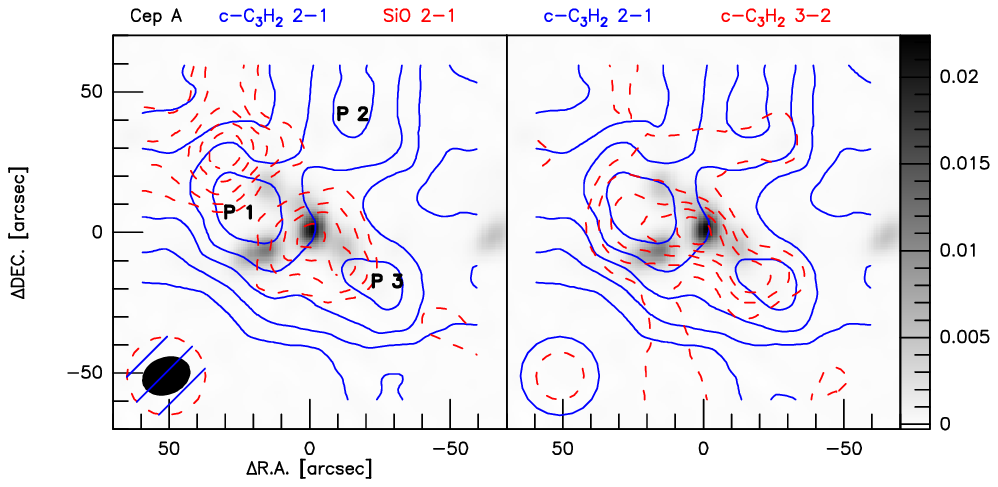}}	
\scalebox{0.25} {\includegraphics{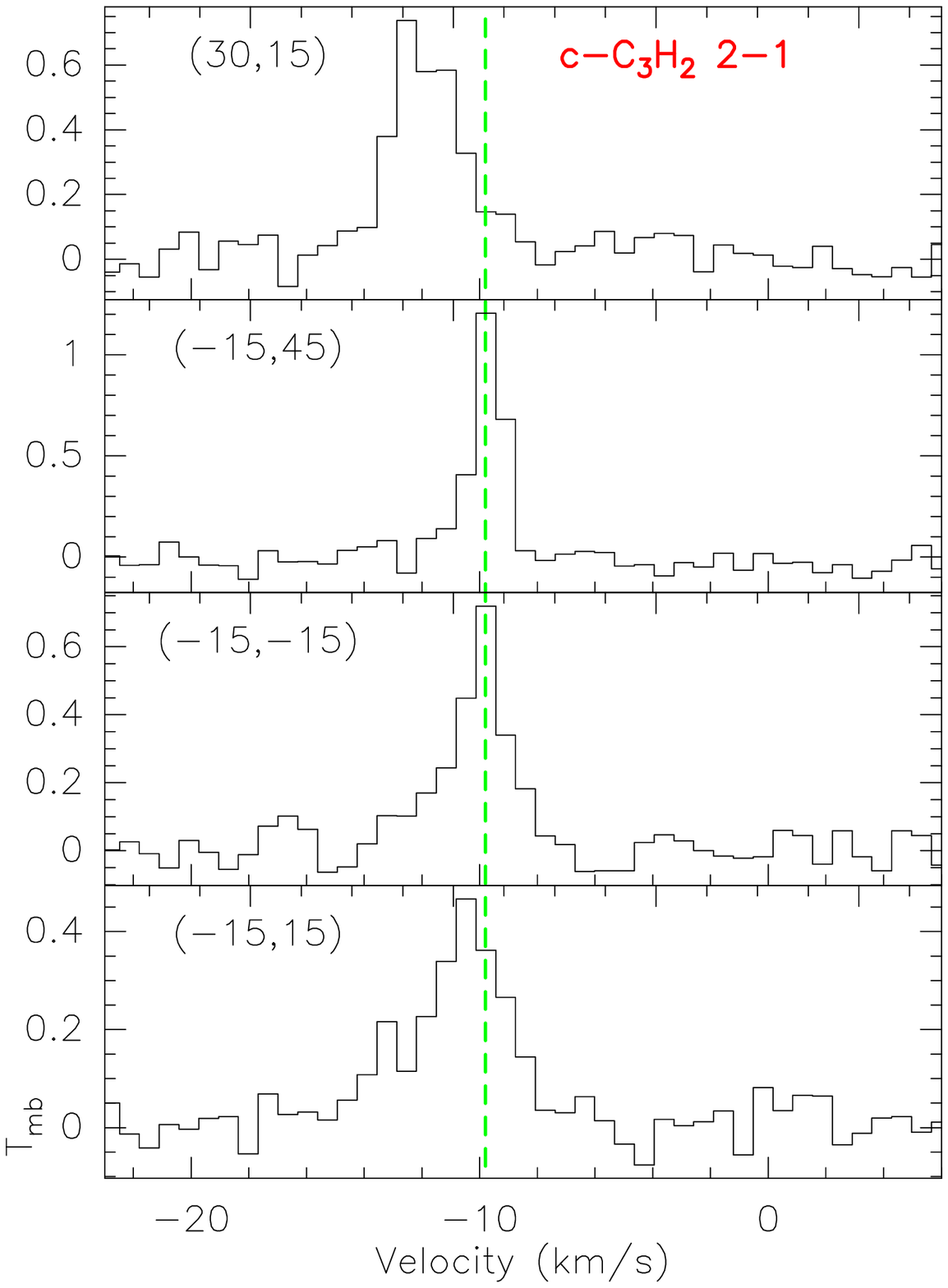}}
\scalebox{0.3}{\includegraphics{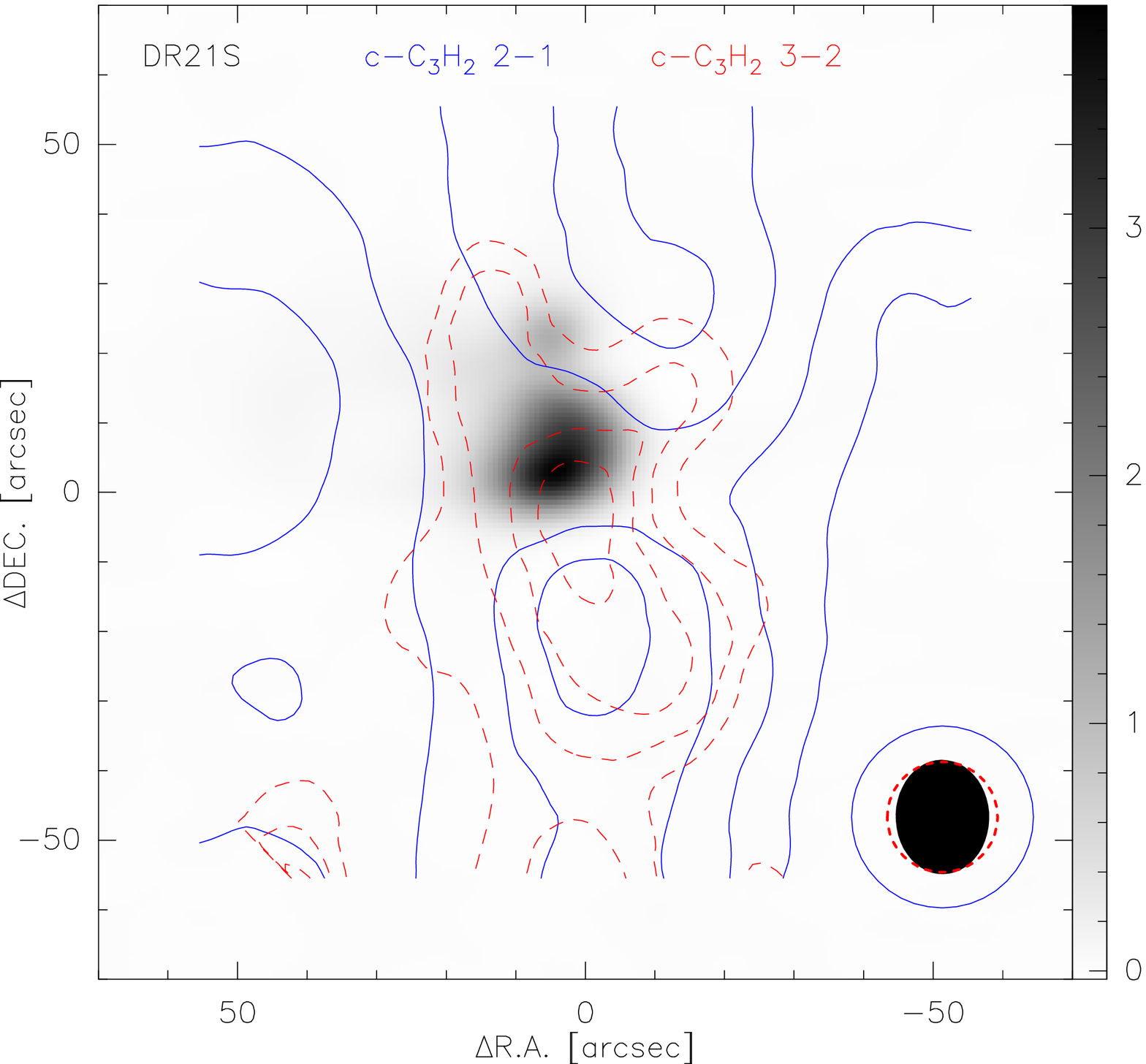} \includegraphics{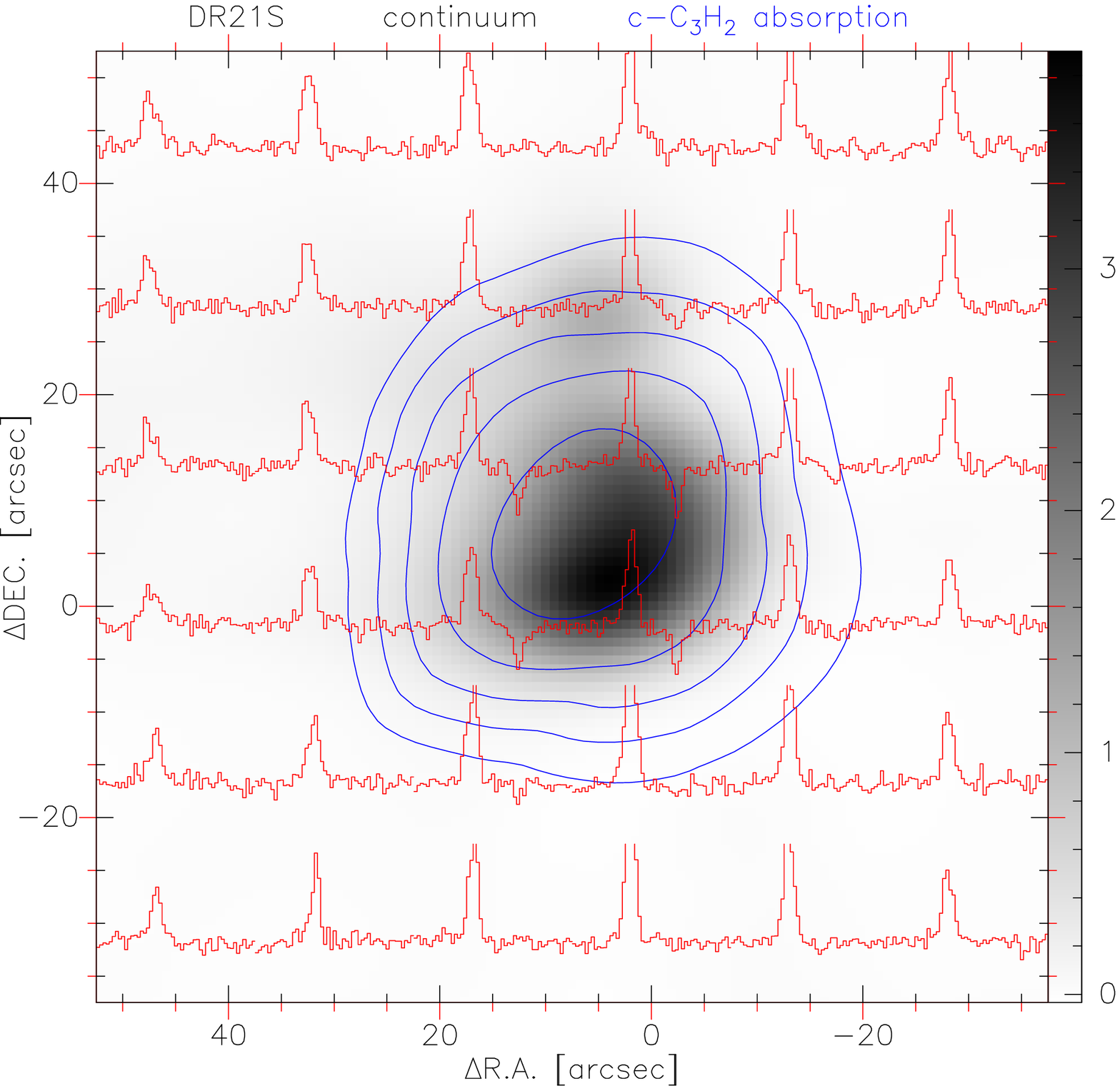}}	
\scalebox{0.3}{\includegraphics{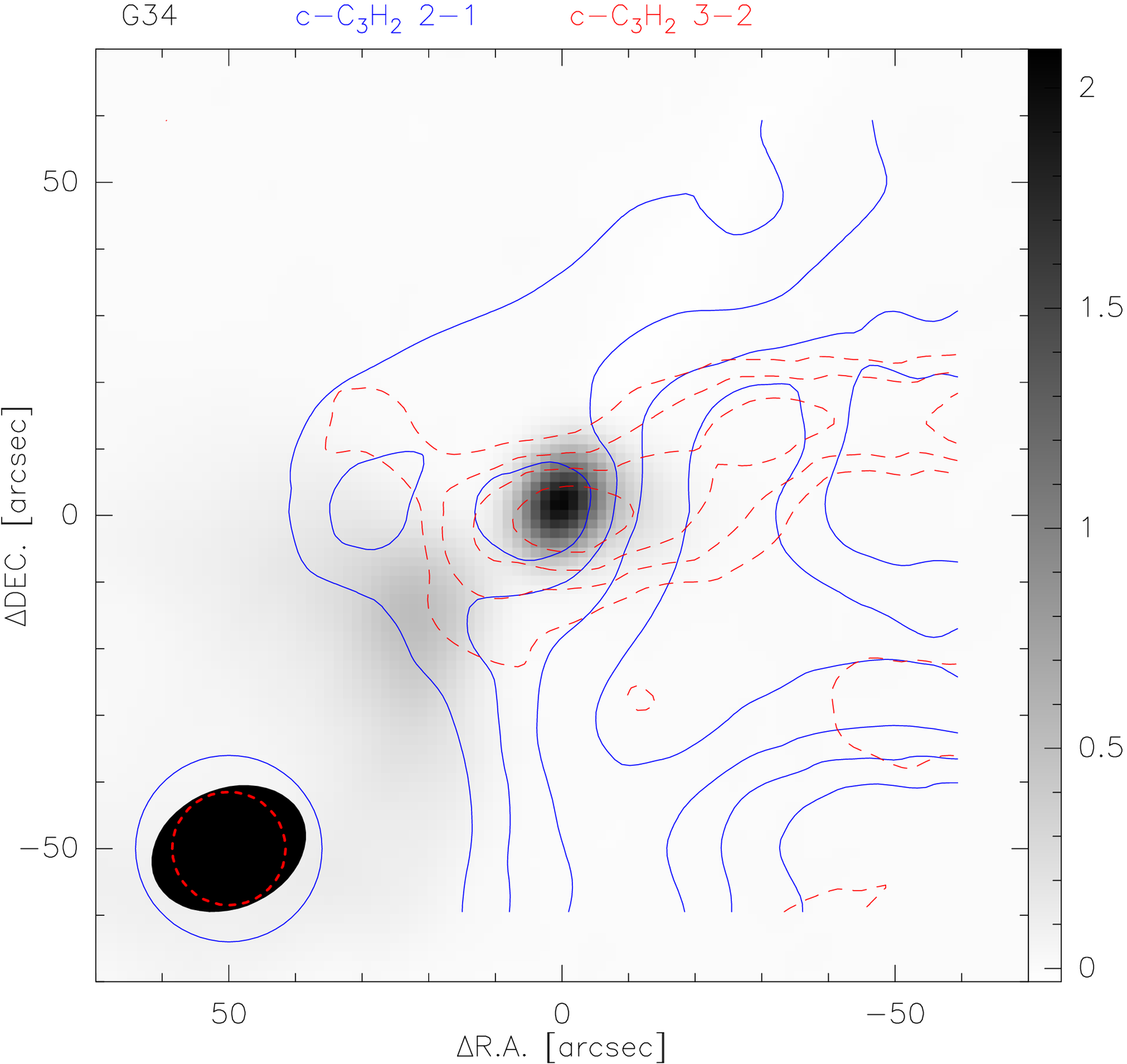}}
\scalebox{0.3}{ \includegraphics{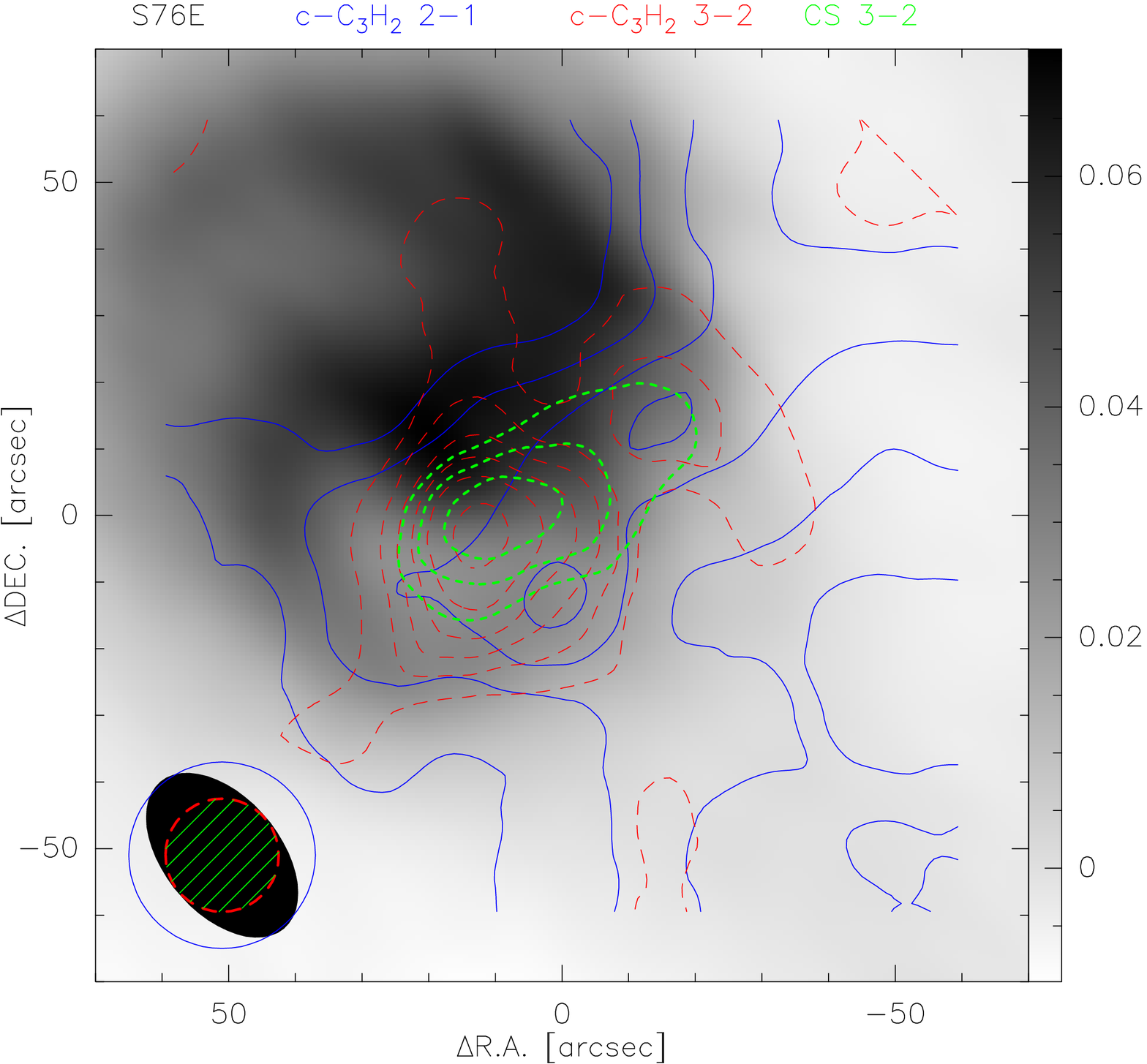}}
\vspace*{-0.2 cm} 
\caption{Images of $c$-C$ _{3}$H$ _{2}$ emissions in four sources. The gray-scale is radio continuum (VLA 8.4 GHz) and range indicated in the wedge is in Jy beam$^{-1}$. \newline
{\it Upper plane: }shows $c$-C$ _{3}$H$ _{2}$ emissions in Cep A. {\it Left:} Velocity-integrated intensity maps of SiO 2-1 (red dash line, -13.5 to -1.3 km s$^{-1}$) overlaid on  $c$-C$ _{3} $H$ _{2} $ 2-1 (-15 to -7 km s$^{-1}$). {\it Middle:} Velocity-integrated intensity maps of $c$-C$ _{3} $H$ _{2}$ 3-2 (red dash line, -15 to -7 km s$^{-1}$) overlaid on  $c$-C$ _{3} $H$ _{2} $ 2-1. {\it Right:} spectra of $c$-C$ _{3} $H$ _{2} $ 2-1 at four different positions. The contour levels are 30\% to 100\%, in step 15\% of the peak intensities, for SiO and $c$-C$ _{3} $H$ _{2} $ 2-1. $c$-C$ _{3} $H$ _{2} $ 3-2 are 45\% to 100\%, in step 15\% of the peak intensity. \newline
{\it Middle plane:} $c$-C$ _{3}$H$ _{2}$ emissions in DR21S. {\it Left:} velocity-integrated intensity maps of $c$-C$ _{3}$H$ _{2}$ 3-2 (red dash line, -6.5 to 2  km s$^{-1}$)  overlaid on $c$-C$ _{3}$H$ _{2}$ 2-1 (-6.5 to 2  km s$^{-1}$). {\it Right:} spectrum of $c$-C$ _{3}$H$ _{2}$ overlaid on velocity-integrated intensity map of $c$-C$ _{3}$H$ _{2}$ 2-1 absorption component (5.6 to 12  km s$^{-1}$). The contour level are 30\% to 100\%, in step 15\% of the peak intensity, for $c$-C$ _{3}$H$ _{2}$ 3-2 and $c$-C$ _{3}$H$ _{2}$ 2-1. \newline
{\it Bottom plane: Left:} velocity-integrated intensity maps of $c$-C$ _{3}$H$ _{2}$ 3-2 (red dash line) overlaid on $c$-C$ _{3}$H$ _{2}$ 2-1 in G34.  The contour level are 30\% to 100\%, in step 15\% of the peak intensity, for $c$-C$ _{3} $H$ _{2}$ 3-2 and $c$-C$ _{3}$H$ _{2}$ 2-1.
{\it Right:} Velocity-integrated intensity maps of $c$-C$ _{3}$H$ _{2}$ 3-2 (red dash line) and CS 3-2 (green dash line) overlaid on $c$-C$ _{3} $H$ _{2}$ 2-1 (blue line) in S76E. The contour level are 30\% to 100\%, in step 15\% of the peak intensity, for $c$-C$ _{3} $H$ _{2}$ 3-2 and $c$-C$ _{3}$H$ _{2}$ 2-1. CS are 80\%, 90\% and 100\%. 
The ellipse and circle in the bottom conner show the beam sizes (HPBW) of the observations. (A colour version of this figure is available in the online journal.)
	\label{fig:2}}
    \vskip-10pt
\end{figure*}

%%%%%%%%%%%%%%%%%%%%  shock & deuterium %%%%%%%%%%%%%%%%%%%%%%%

\begin{figure*}
	%\centering 
\scalebox{1.2}{\includegraphics{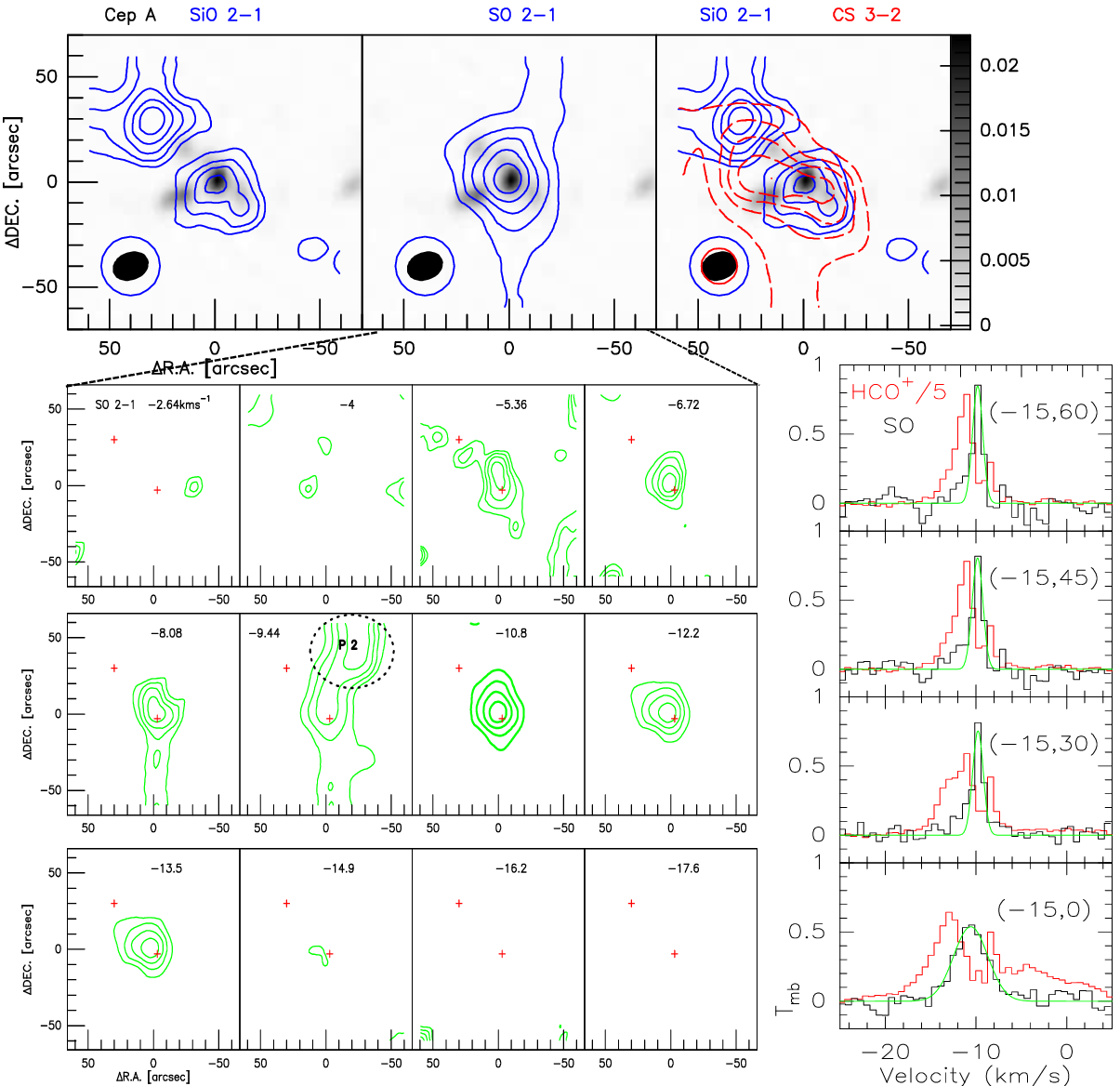}}
\scalebox{0.65}{\includegraphics{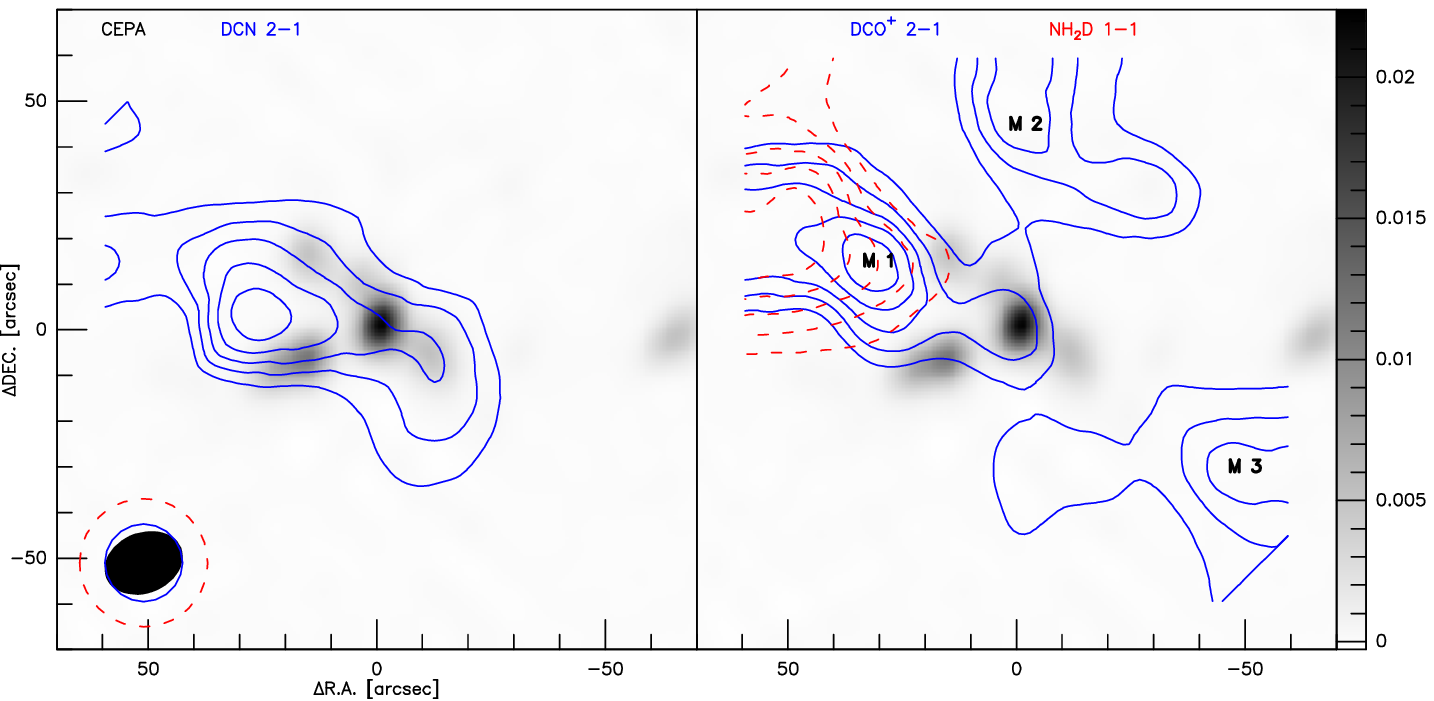} }
%\scalebox{0.25}{\includegraphics{fig3c}}
\vspace*{-0.2 cm} 
\caption{Shock tracers and deuterated molecules in Cep A. The gray-scale is radio continuum (VLA 8.4 GHz) and range indicated in the wedge is in Jy beam$^{-1}$. \newline
{\it Upper plane:  from left to right}: velocity-integrated intensity maps of SiO 2-1 (-13.5 to -1.3 km s$^{-1}$), SO 2-1 (-14 to -4 km s$^{-1}$), and CS 3-2 (red dash line, -17 to 0 km s$^{-1}$) overlaid on SiO 2-1 (blue line).  \newline
{\it Middle plane:  from left to right}: channel map of SO 2-1, the spectra of SO 2-1 (black line) overlaid on HCO$^{+}$ 1-0 (red line). \newline
{\it Bottom plane: from left to right}: velocity-integrated intensity maps of DCN 1-0 (-15 to -5 km s$^{-1}$),  NH$_{2}$D1-1 (red dash line,  -20 to -5 km s$^{-1}$) overlaid on DCO$^{+}$ 2-1 (blue line,  -15 to -5 km s$^{-1}$).  The contour levels are 30\% to 100\%, in step 15\% of the peak intensities, for these molecular lines. 
The ellipse and circle in the bottom-left conner show the beam sizes of the observations.
	\label{fig:3}}
    \vskip-10pt
\end{figure*}

\begin{figure*} 
	\centering 
\scalebox{1.5}{\includegraphics{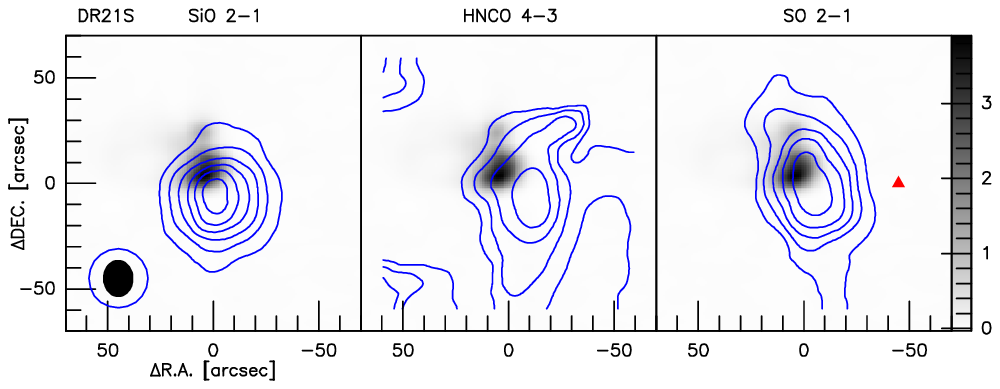}}
\scalebox{0.5}{\includegraphics{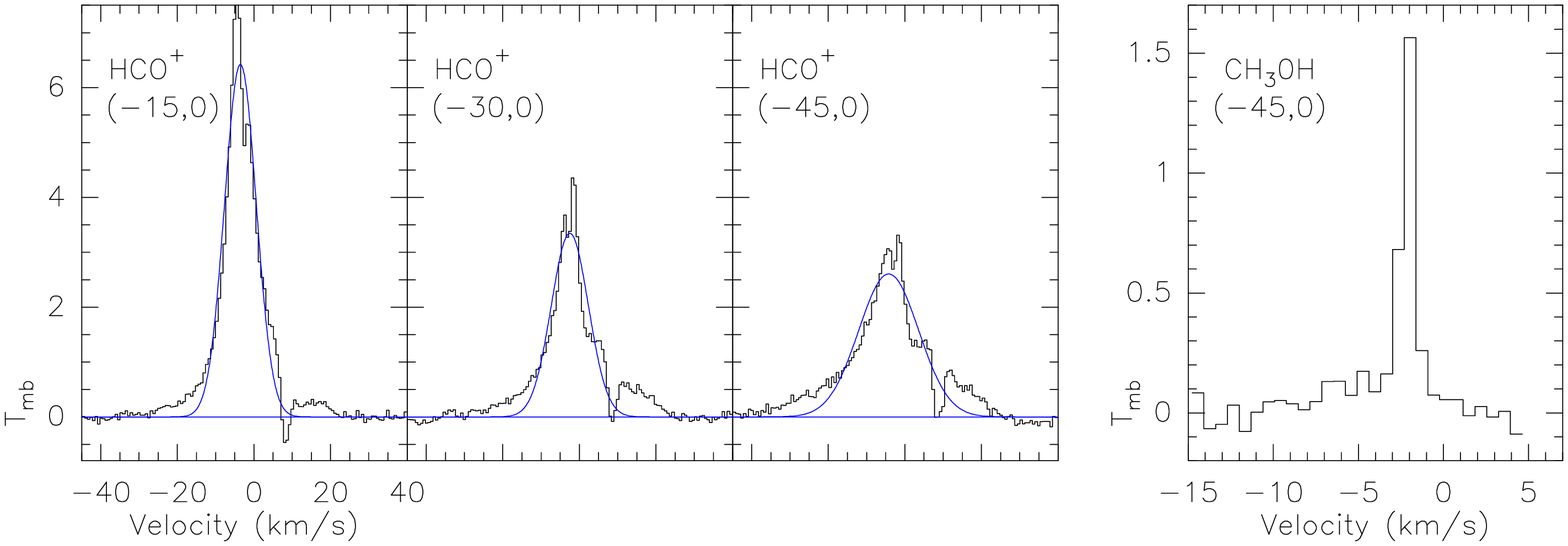}}
\scalebox{1}{\includegraphics{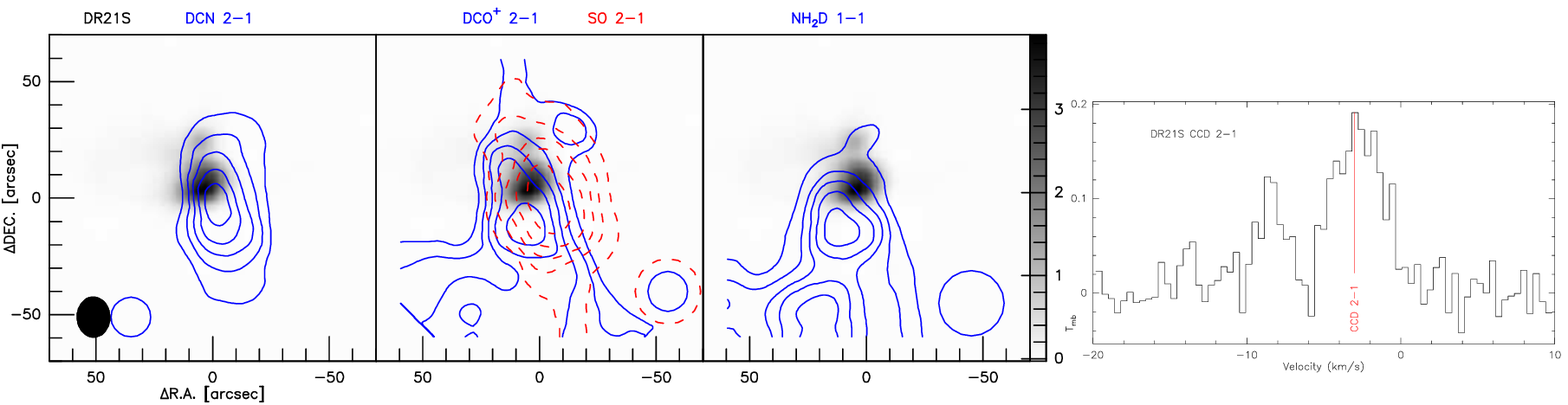}}
%\scalebox{0.2}{\includegraphics{fig4b}}
\vspace*{-0.2 cm} 
\caption{Shock tracers and deuterated molecules in DR21S. The gray-scale is radio continuum (VLA 8.4 GHz) and range indicated in the wedge is in Jy beam$^{-1}$. \newline
{\it Upper plane:  From left to right,} velocity-integrated intensity maps of SiO 2-1 (-20 to 20 km s$^{-1}$), HNCO 4-3 ( -8 to 3 km s$^{-1}$), SO 2-1( -8 to 3 km s$^{-1}$), red triangle indicates the position of CH$_{3}$OH 5$_{-1,5}$-4$_{0,4}$ E maser). The contour level are 30\% to 100\%, in step 15\% of the peak intensity, for SiO 2-1, HNCO 4-3, SO 2-1 and HCO$ ^{+}$ 1-0. \newline
 {\it Middle plane:} the spectra of HCO$ ^{+}$ 1-0 and CH$_{3}$OH 5$_{-1,5}$-4$_{0,4}$ E. \newline
 {\it Bottom plane: From left to right,} velocity-integrated intensity maps of DCN 2-1 (-9 to 5 km s$^{-1}$), SO 2-1 (red dash line) overlaid on DCO$^{+}$ 2-1 (-5.5 to 2 km s$^{-1}$), NH$_{2}$D 1-1 (-7.5 to 5 km s$^{-1}$) and spectra of C$_{2}$D 2-1. The contour level are 30\% to 100\%, in step 15\% of the peak intensity, for these molecular lines.
The ellipse and circle in the bottom-left conner show the beam sizes of the observations.
	\label{fig:4}}
    \vskip-10pt
\end{figure*}

\begin{figure*} 
	\centering
\scalebox{1.5}{\includegraphics{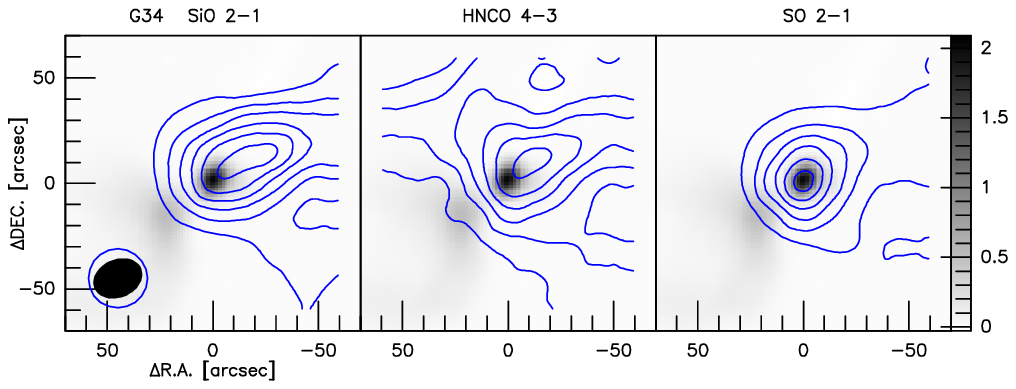}}
\scalebox{0.5}{\includegraphics{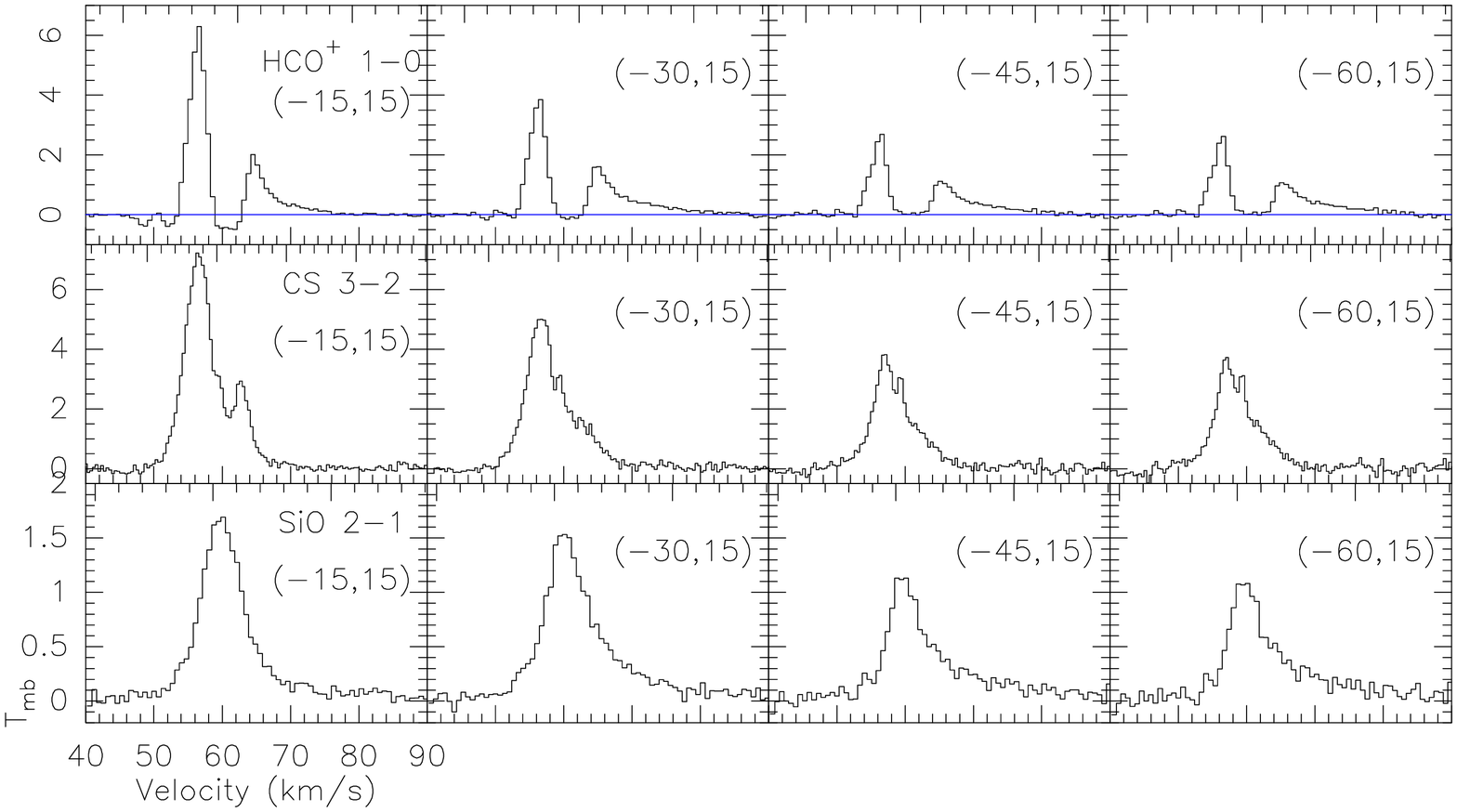}}
\scalebox{1}{\includegraphics{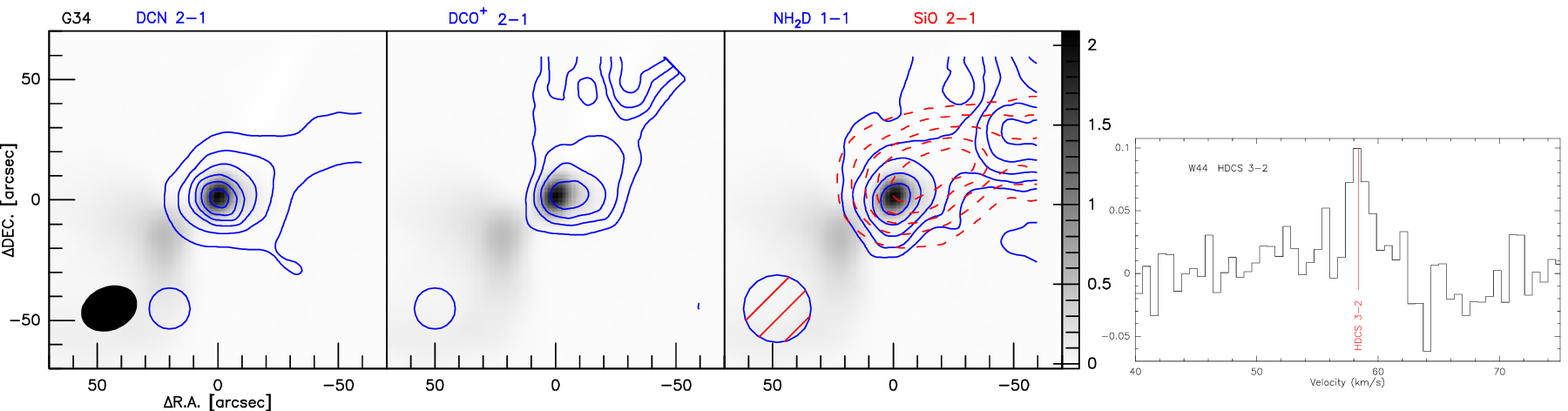}}	 	
\vspace*{-0.2 cm} 
\caption{Shock tracers and deuterated molecules in G34. The gray-scale is radio continuum (VLA 8.4 GHz) and range indicated in the wedge is in Jy beam$^{-1}$. \newline
{\it Upper plane:  from the left to right:} velocity-integrated intensity maps of SiO 2-1 (50 to 70 km s$^{-1}$), HNCO 4-3 (52 to 65 km s$^{-1}$), SO 2-1 (52 to 70 km s$^{-1}$). The contour level are 30\% to 100\%, in step 15\% of the peak intensity, for SiO 2-1, HNCO 4-3, SO 2-1 and HCO$^{+}$ 1-0. \newline
{\it Middle plane: }the first row is spectrum of HCO$^{+}$ 1-0,  the second row is CS 3-2, and the third row is SiO 2-1. \newline
{\it Bottom plane:  From left to right:} velocity-integrated intensity maps of DCN 2-1 (50 to 65 km s$^{-1}$), DCO$^{+}$ 2-1 (53 to 61 km s$^{-1}$), SiO 2-1 (red dash line) overlaid on NH$_{2}$D 1-1 (50 to 65 km s$^{-1}$), the spectra of HDCS 3-2. The contour level are 30\% to 100\%, in step 15\% of the peak intensity, for  DCN 2-1, DCO$^{+}$ 2-1, SiO 2-1 and NH$_{2}$D 1-1. The ellipse and circle in the bottom-left conner show the beam sizes of the observations. 
	\label{fig:5}}
    \vskip-10pt
\end{figure*}

\begin{figure*}
	\centering
\scalebox{1.4}{\includegraphics{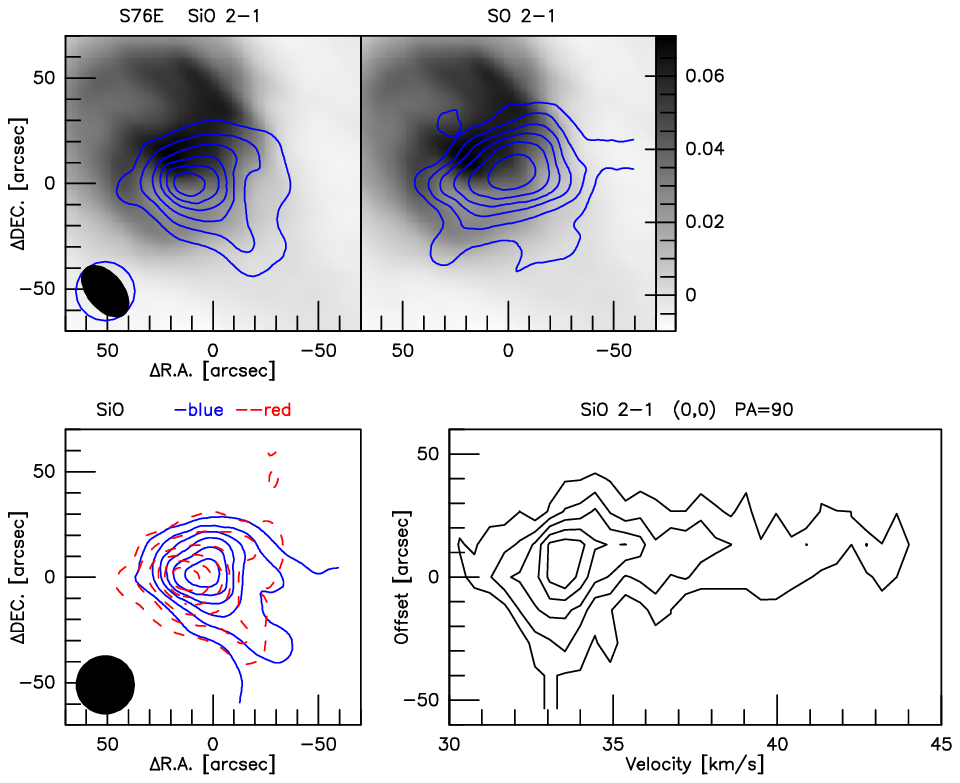}}
\scalebox{0.3}{\includegraphics{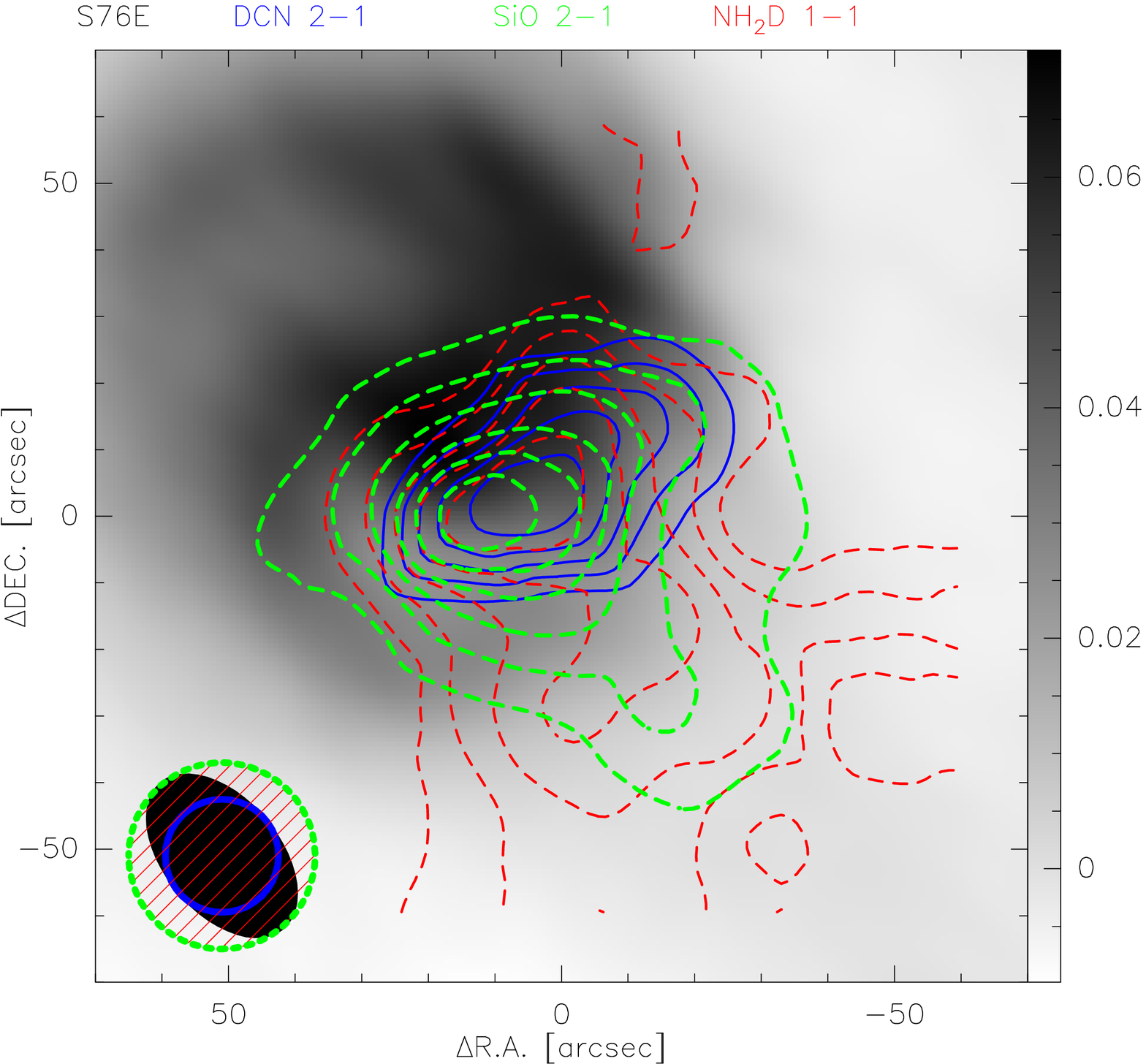} }\scalebox{0.25}{\includegraphics{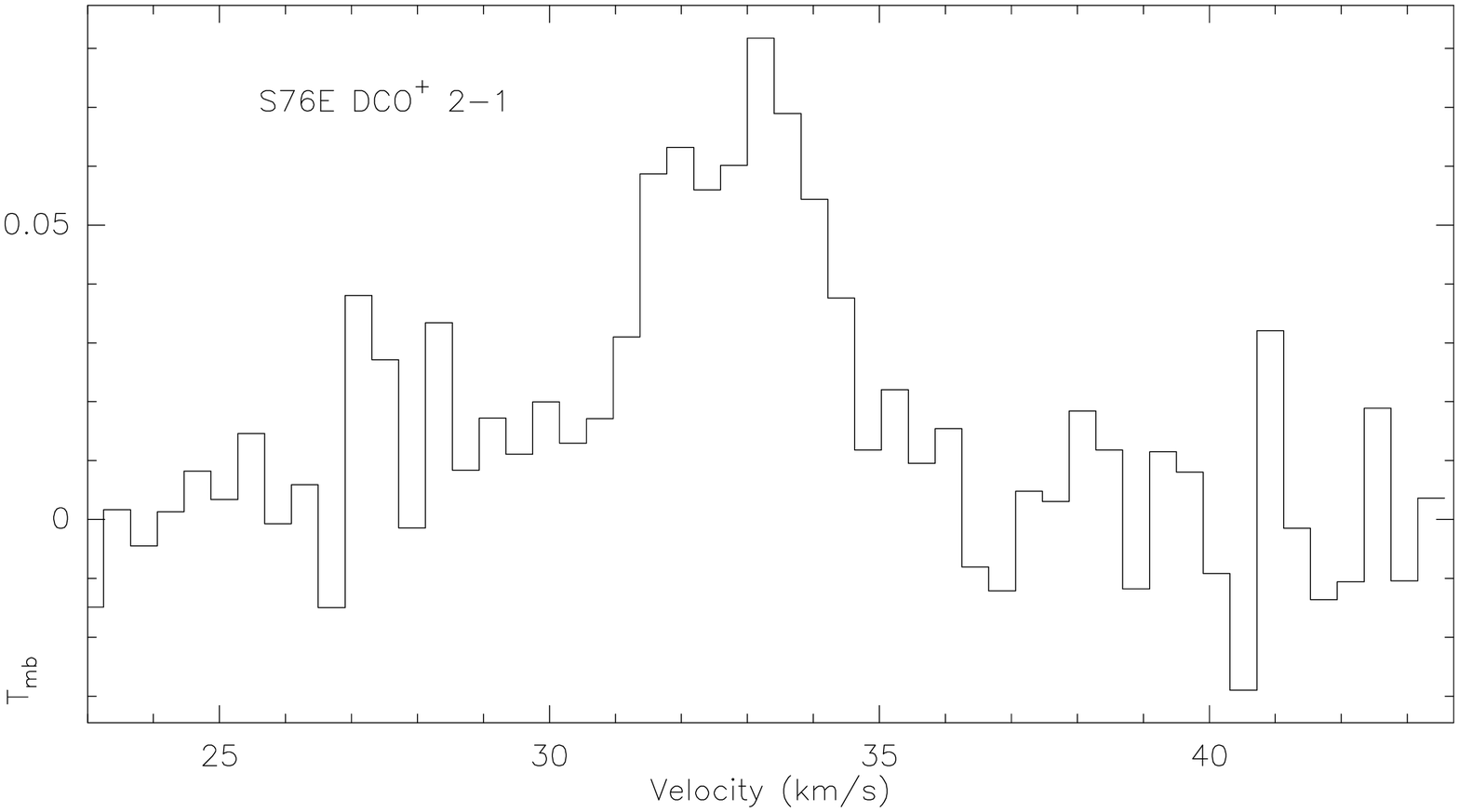} \includegraphics{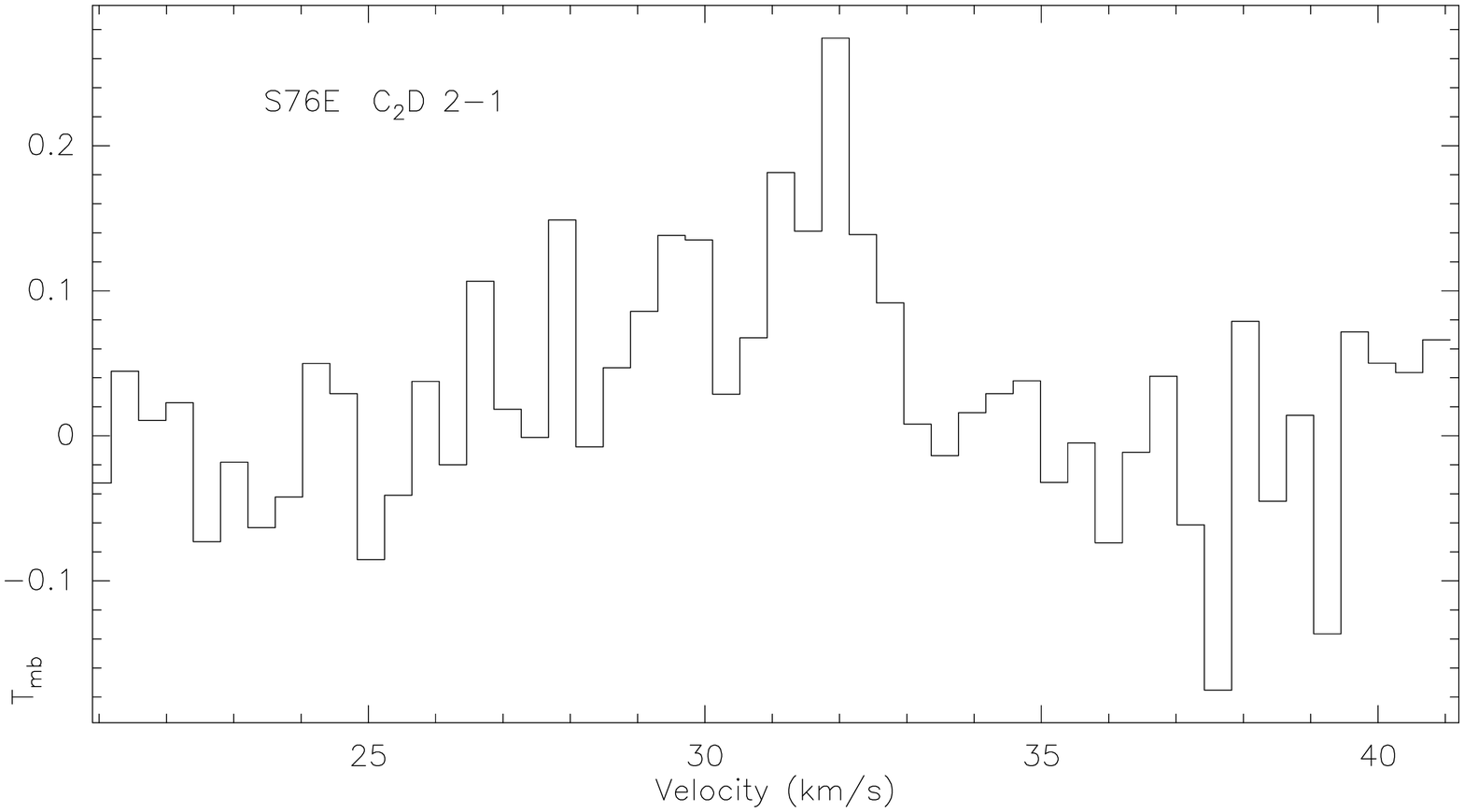}}
\vspace*{-0.2 cm}
\caption{Shock tracers and deuterated molecules in S76E. The gray-scale is radio continuum (VLA 8.4 GHz) and range indicated in the wedge is in Jy beam$^{-1}$. \newline
{\it Upper plane:  from left to right:} velocity-integrated intensity maps of SiO 2-1 (25 to 55 km s$^{-1}$), SO 2-1 (30 to 40 km s$^{-1}$). The contour level are 30\% to 100\%, in step 15\% of the peak intensity, for SiO 2-1 and SO 2-1. \newline
{\it Middle plane:  from left to right:} velocity-integrated intensity maps of redshift velocity component (25 to 33.5 km s$^{-1}$) of SiO 2-1 overlaid on its blueshift velocity component (33.5 to 55 km s$^{-1}$), position-velocity diagram of SiO 2-1 (PA=90$^{\circ}$). The contour level are 30\% to 100\%, in step 15\% of the peak intensity, for SiO 2-1. \newline
{\it Bottom plane:  Left:} velocity-integrated intensity maps of SiO 2-1 (green dash line) and  NH$_{2}$D 1-1 (red dash line, 28 to 38 km s$^{-1}$) overlaid on DCN 2-1 (blue line, 30 to 38 km s$^{-1}$). The contour level are 30\% to 100\%, in step 15\% of the peak intensity, for SiO 2-1, NH$_{2}$D 1-1 and DCN 2-1.  {\it Middle:} DCO$^{+}$ 2-1 line profile. {\it Right:} C$_{2}$D 2-1  line profile.
The ellipse and circle in the bottom-left conner show the beam sizes of the observations.
	\label{fig:6}}
    \vskip-10pt
\end{figure*}

%%%%%%%%%%%%%%%%%%%%%%%%%%%%%%%%%%%%%%%%%%%%%%%%%%%%%%%%%%%%

\begin{figure*} 
	\centering
\scalebox{0.3}{\includegraphics{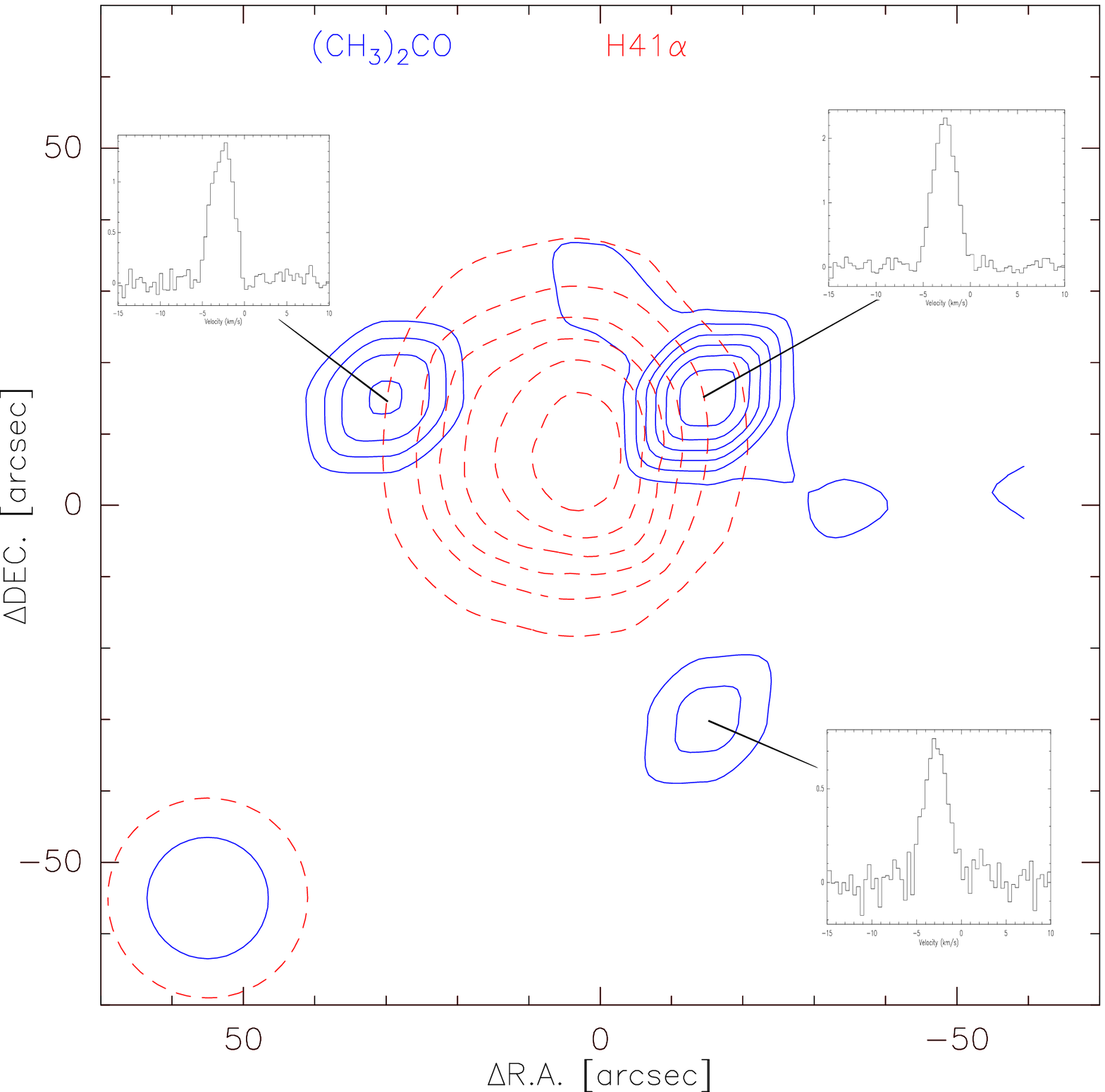}} \scalebox{0.3}{\includegraphics{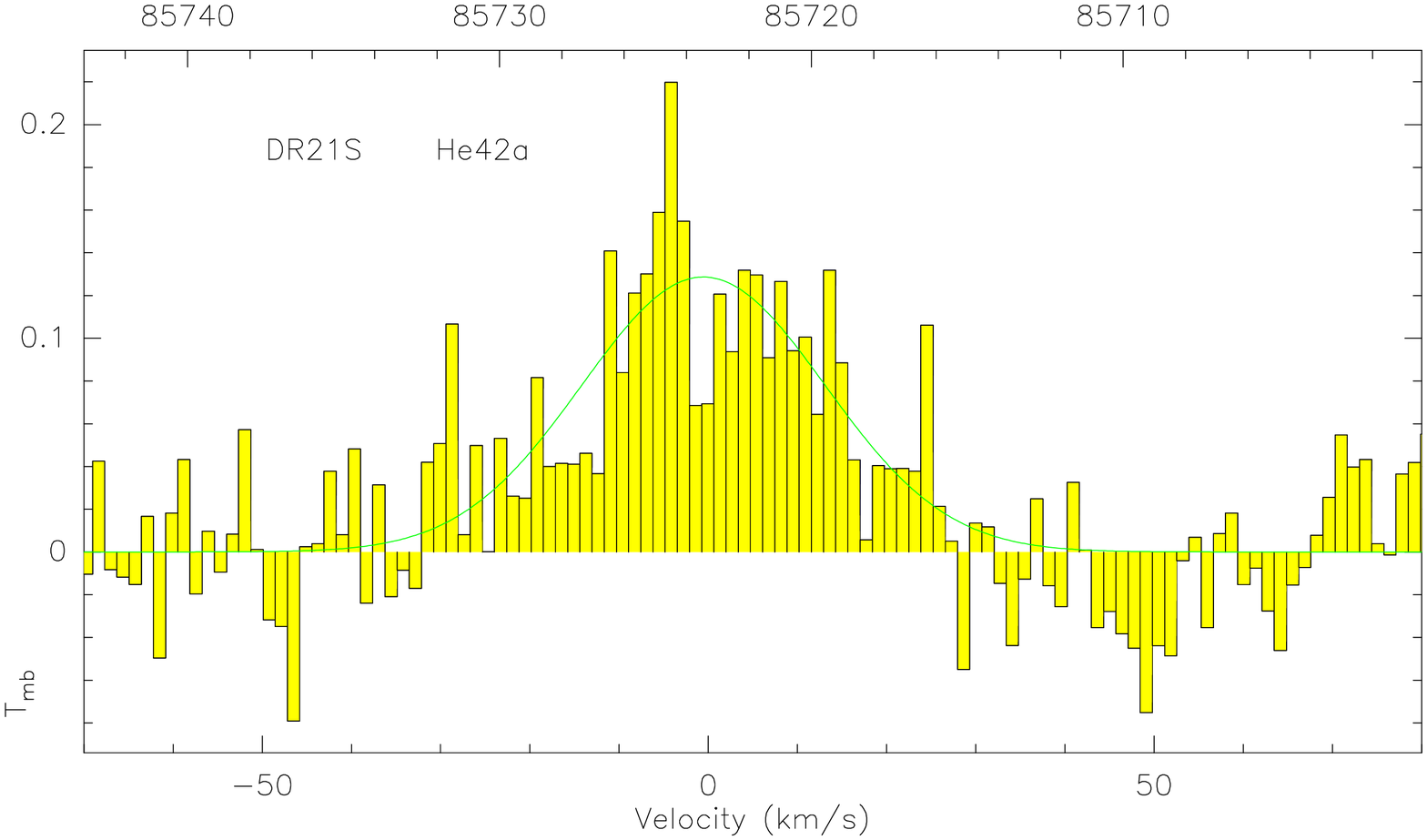}}
\vspace*{-0.2 cm} 
\caption{(CH$ _{3}$)$ _{2}$CO and He42$\alpha$ in DR21S. 
{\it Left:} velocity-integrated intensity maps of H41$\alpha$ (red dash line, -35 to 35 km s$^{-1}$) overlaid on (CH$ _{3}$)$ _{2}$CO (-6 to 1 km s$^{-1}$). The contour level are 15\% to 100\%, in step 15\% of the peak intensity, for CH$ _{3}$)$ _{2}$CO 23-22 and H41$\alpha$. The ellipse  in the bottom-left conner show the beam sizes  of the observations.
{\it Right:} He42$\alpha$ line profile at offset(0, 0).
	\label{fig:7}}
    \vskip-10pt
\end{figure*}

\begin{figure*} 
	\centering
\scalebox{0.25}{\includegraphics{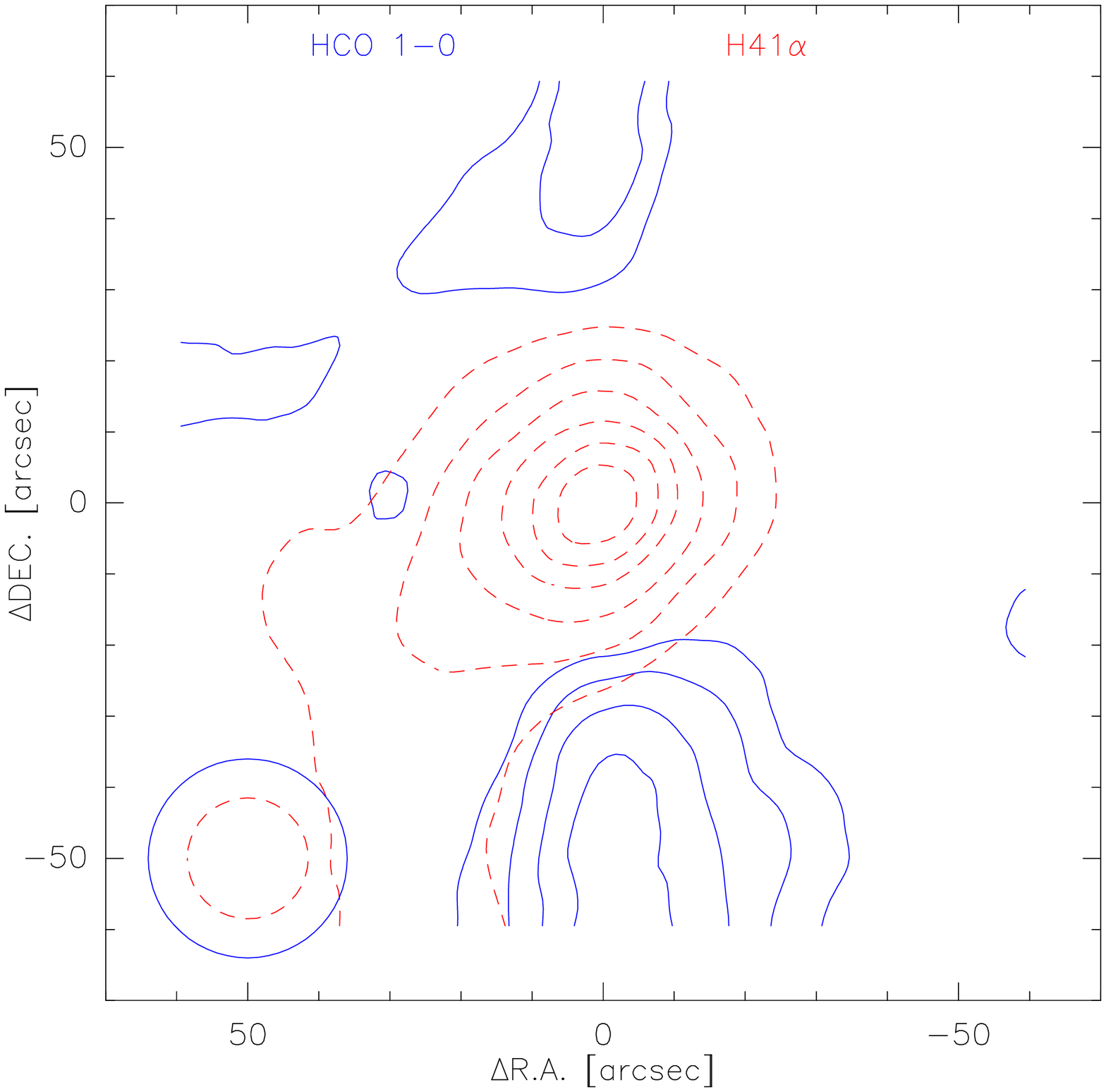} \includegraphics{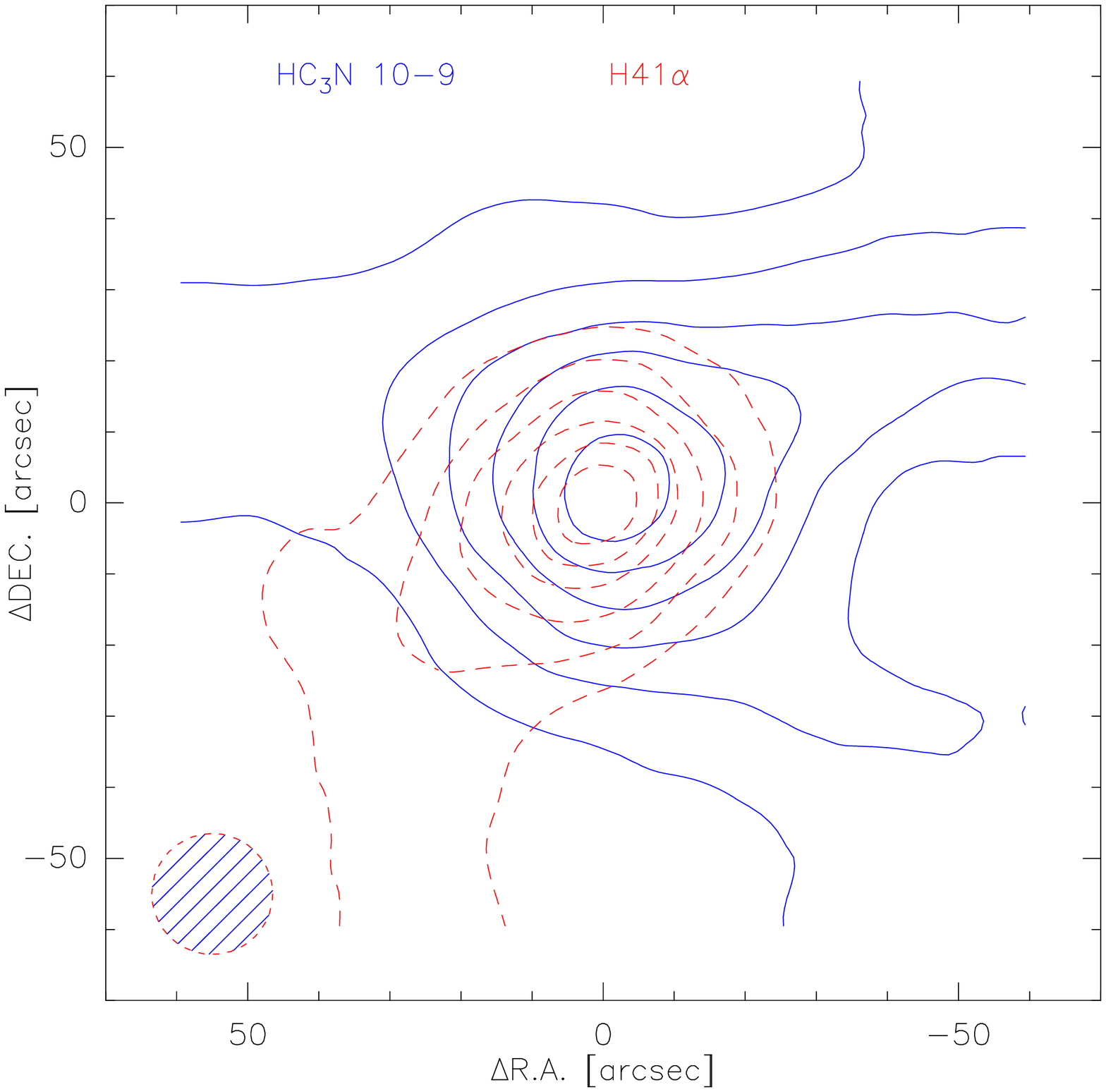}}
\vspace*{-0.2 cm} 
\caption{ HCO, HC$_{3}$N and H41$\alpha$ in G34.
{\it Left:} velocity-integrated intensity map of H41$\alpha$ (red dash line, -10 to 105 km s$^{-1}$) overlaid on HCO 1-0 (55 to 64 km s$^{-1}$). 
{\it Right:} velocity-integrated intensity map of H41$\alpha$  (red dash line) overlaid on HC$_{3}$N 10-9 (52 to 64 km s$^{-1}$).
The contour level are 30\% to 100\%, in step 15\% of the peak intensity, for HCO 1-0, HC$_{3}$N 10-9 and H41$\alpha$ .
The circle in the bottom-left conner show the beam sizes of the observations.
	\label{fig:8}}
    \vskip-10pt
\end{figure*}

\begin{figure*} 
	\centering 
\scalebox{0.4}{\includegraphics{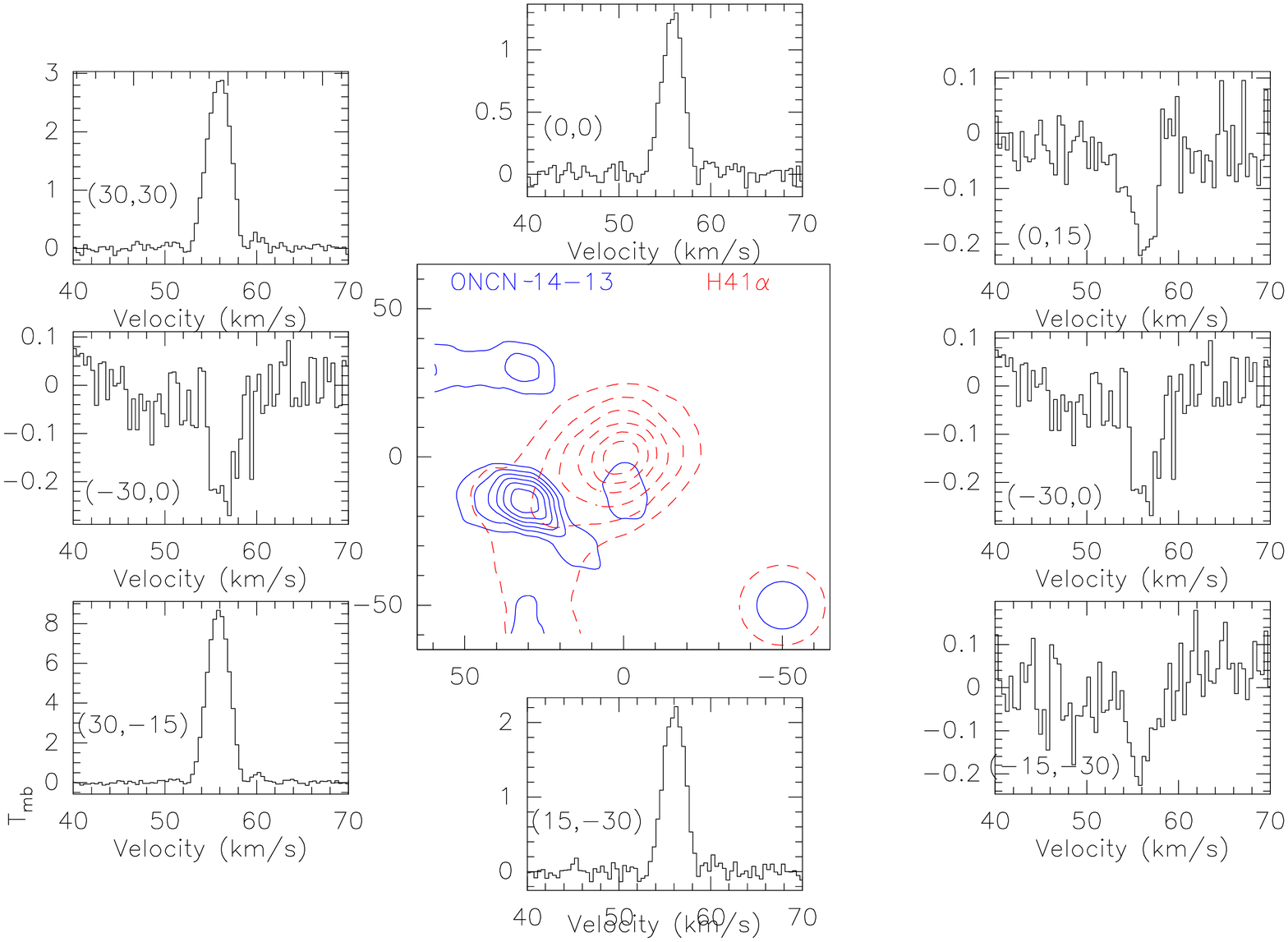} }
\vspace*{-0.2 cm} 
\caption{Velocity-integrated intensity map and spectra of ONCN 14-13 in G34.  Velocity-integrated intensity map of H41$\alpha$ (red dash line) overlaid on ONCN 14-13 (52.5 to 58.5 km s$^{-1}$). 
The contour level are 15\% to 100\%, in step 15\% of the peak intensity, for ONCN  14-13 and H41$\alpha$. 
The ellipse in the bottom-right conner show beam sizes of the observations.
 	\label{fig:9}}
    \vskip-10pt
\end{figure*}

%%%%%%%%%%%%%%%%%%%%%%%%%%%%%%%%%%%%%%%%%%%%%%%%%%%%%%%%%%%%%%%%%%%%%%%%%%%%%%%%%%%%%%%%%%%%%%%%%%
%%%%%%%%%%%%%%%%% APPENDICES %%%%%%%%%%%%%%%%%%%%%

\clearpage

\appendix

\section{The molecular line profiles} 
%\label{sec:fig}

\begin{figure*}
	\centering
\scalebox{0.28}{\includegraphics{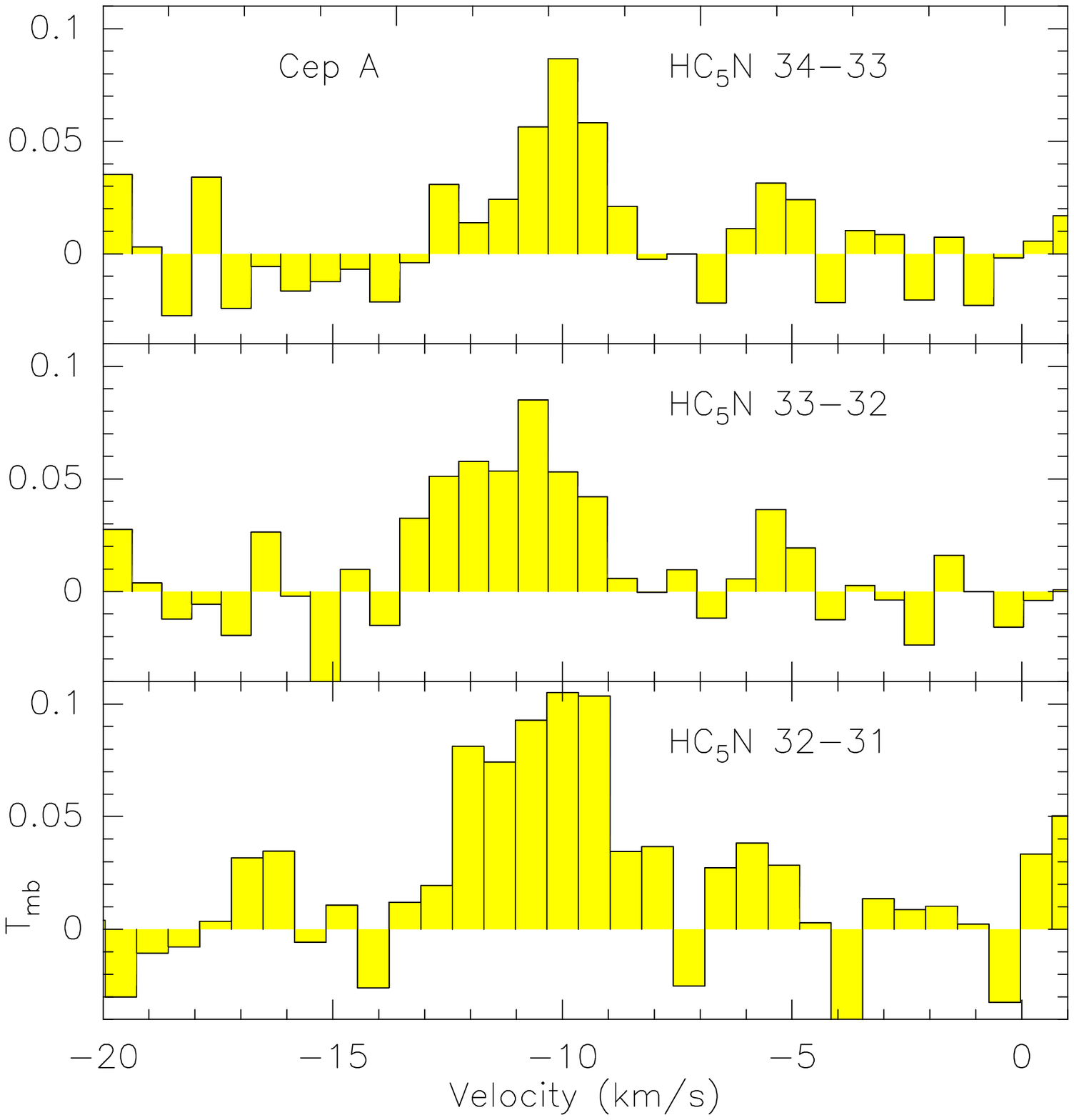} \includegraphics{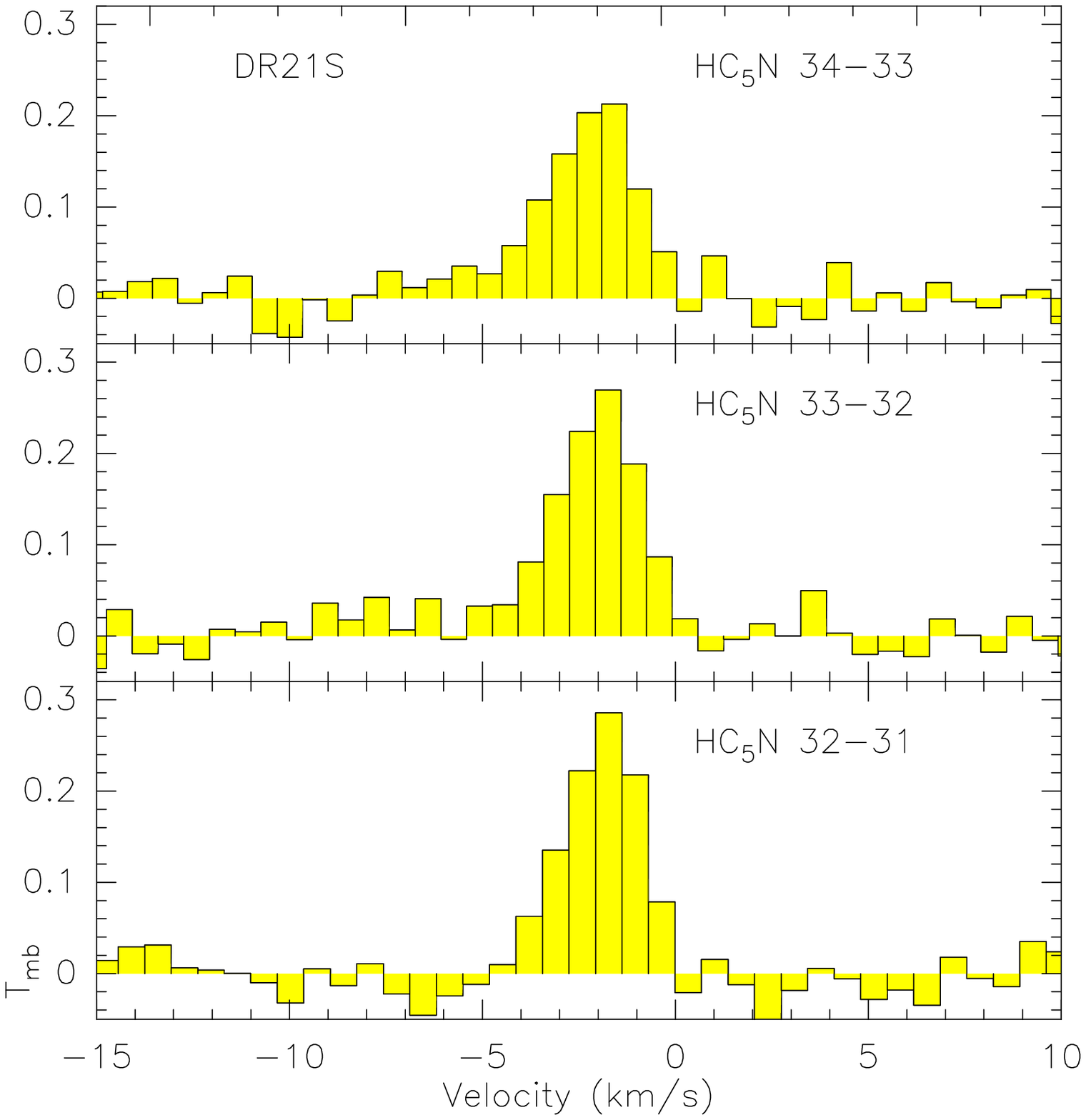} \includegraphics{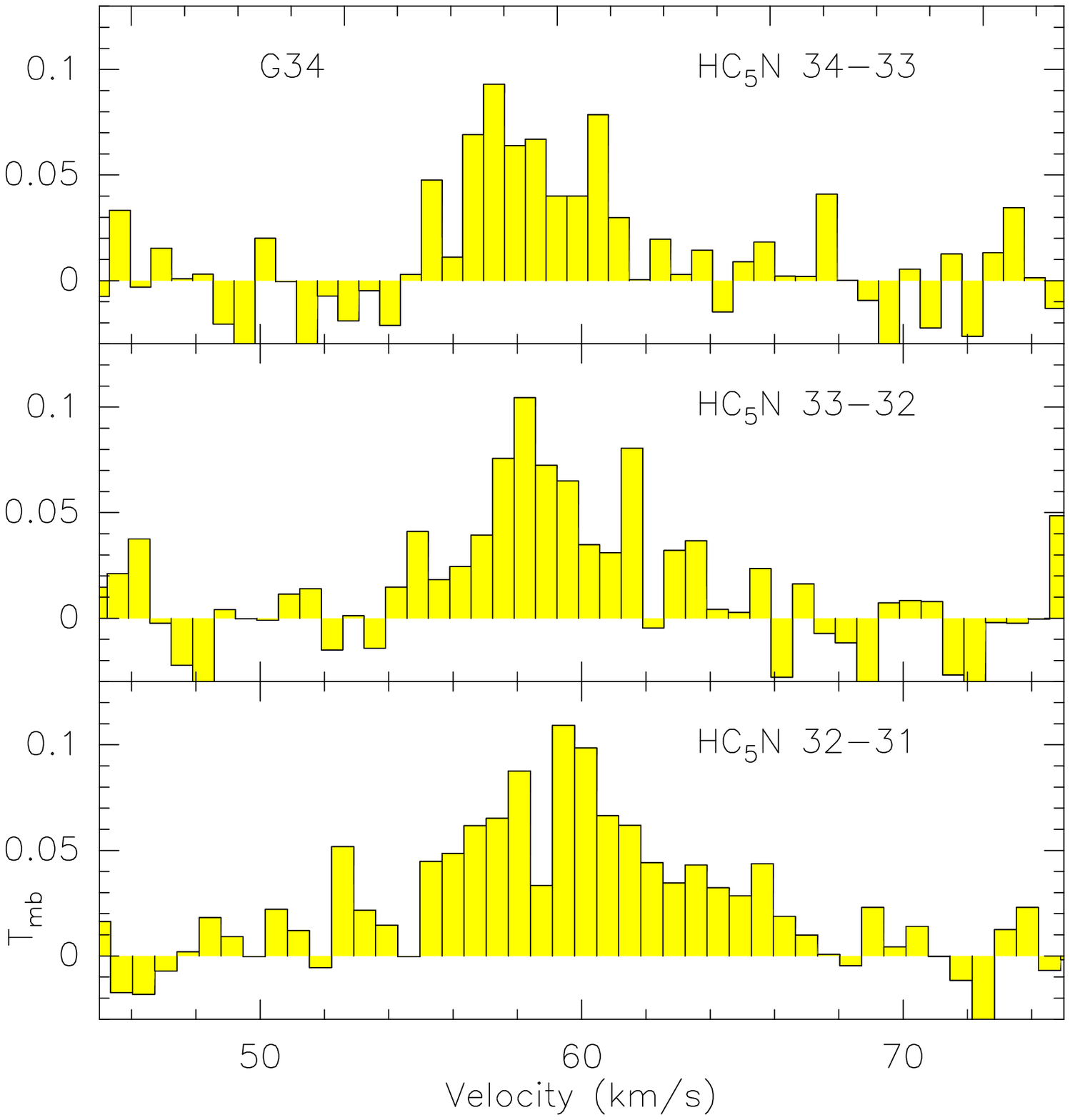}\includegraphics{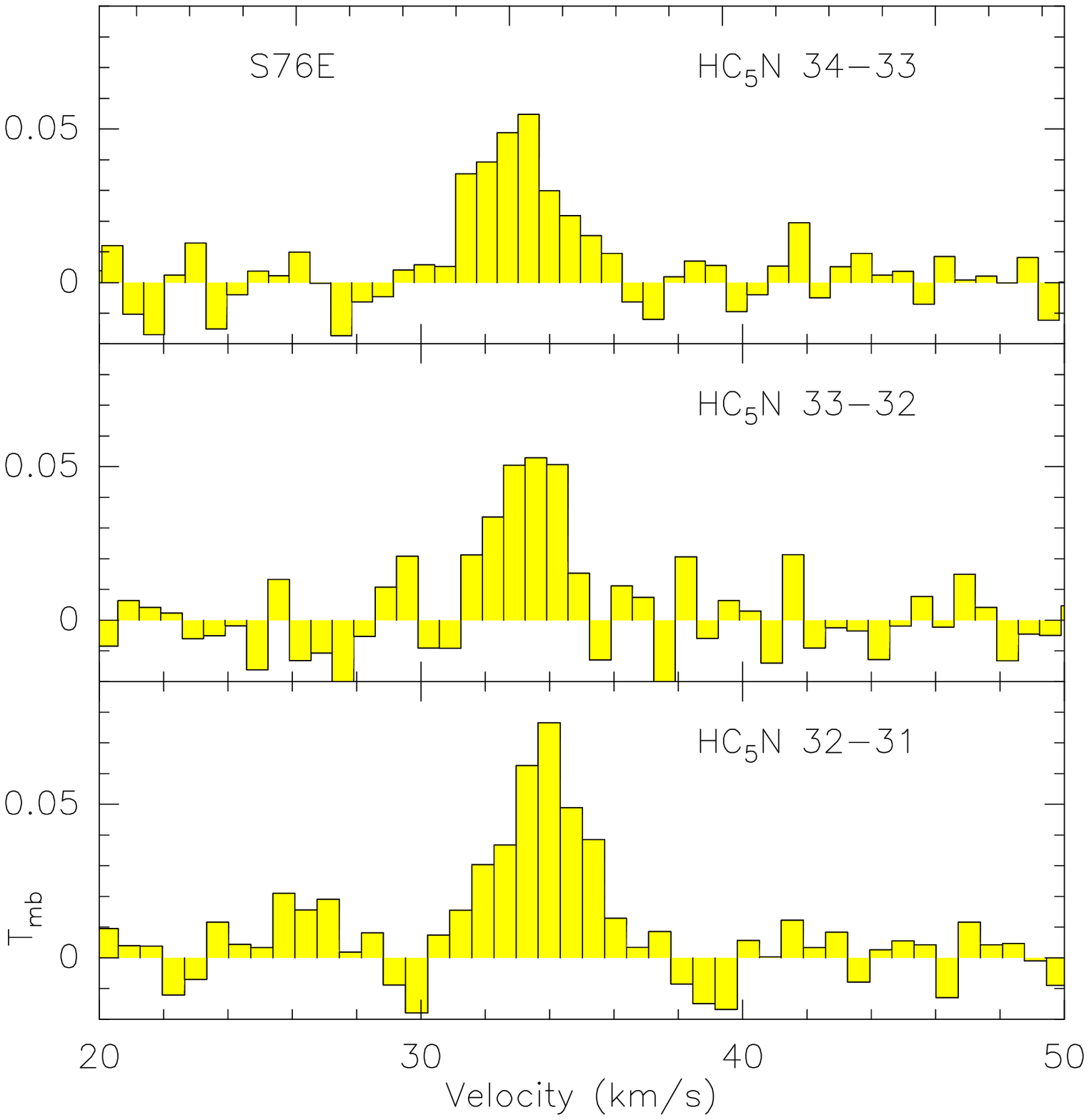}}
\scalebox{0.28}{\includegraphics{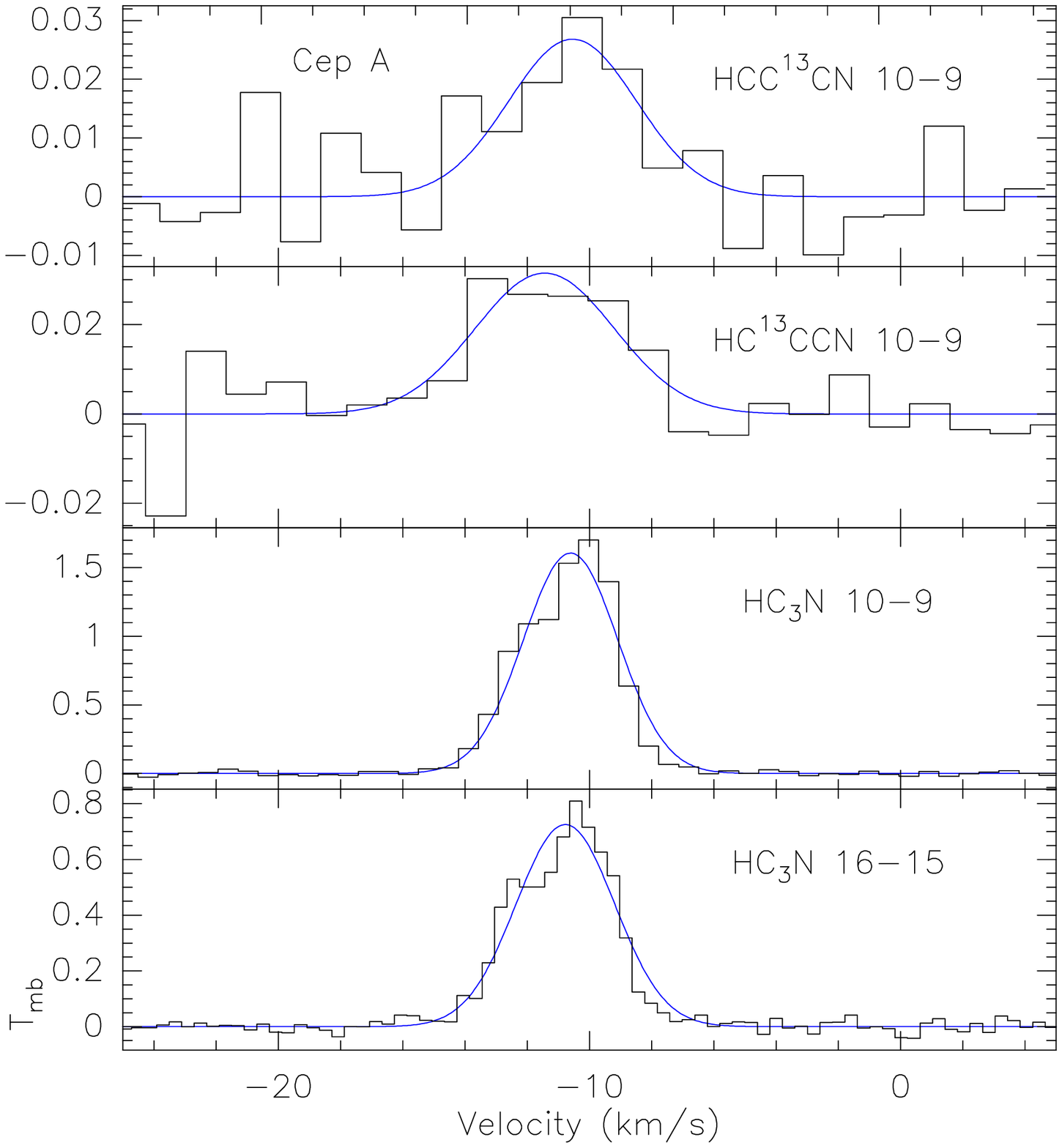} \includegraphics{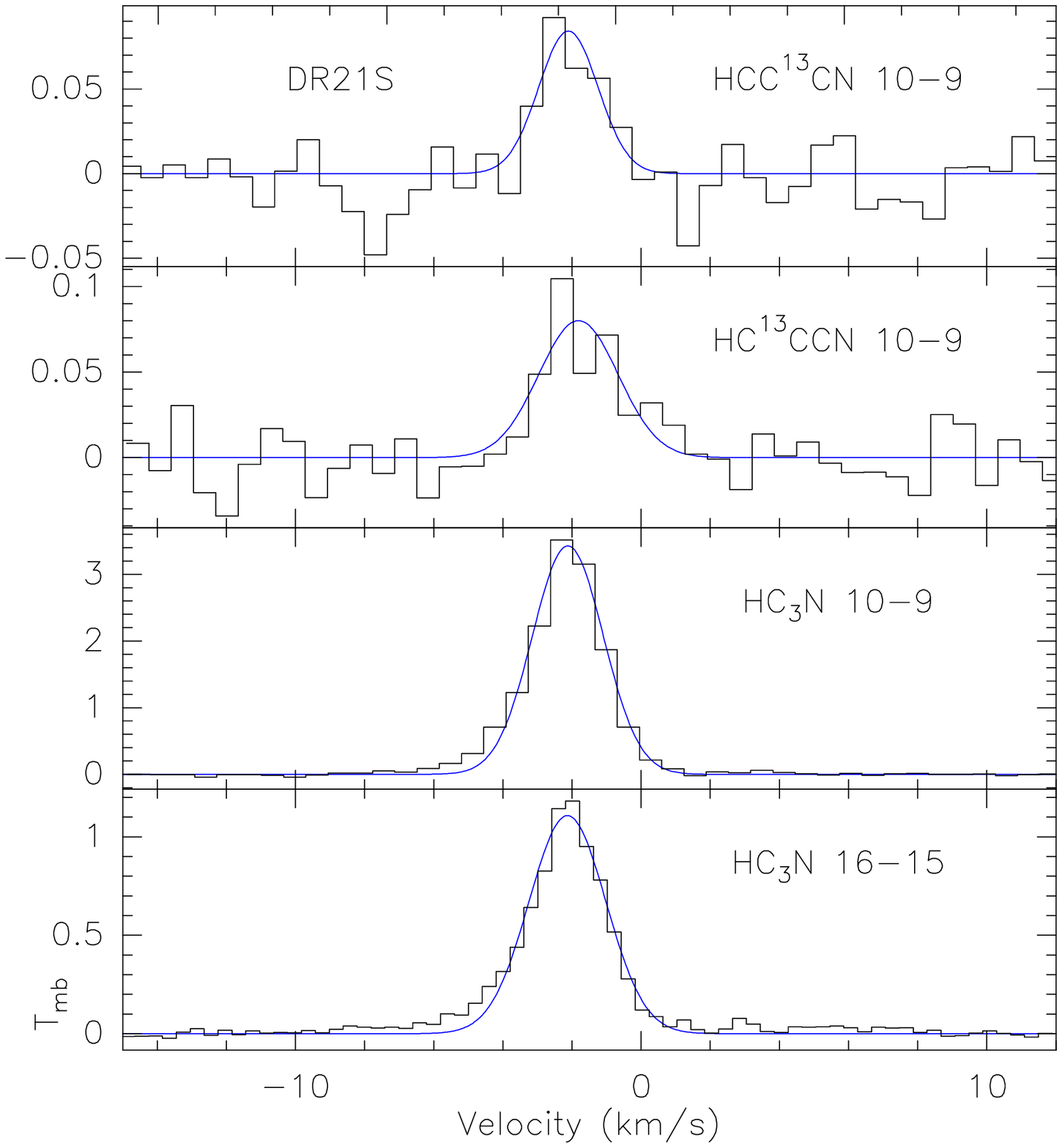} \includegraphics{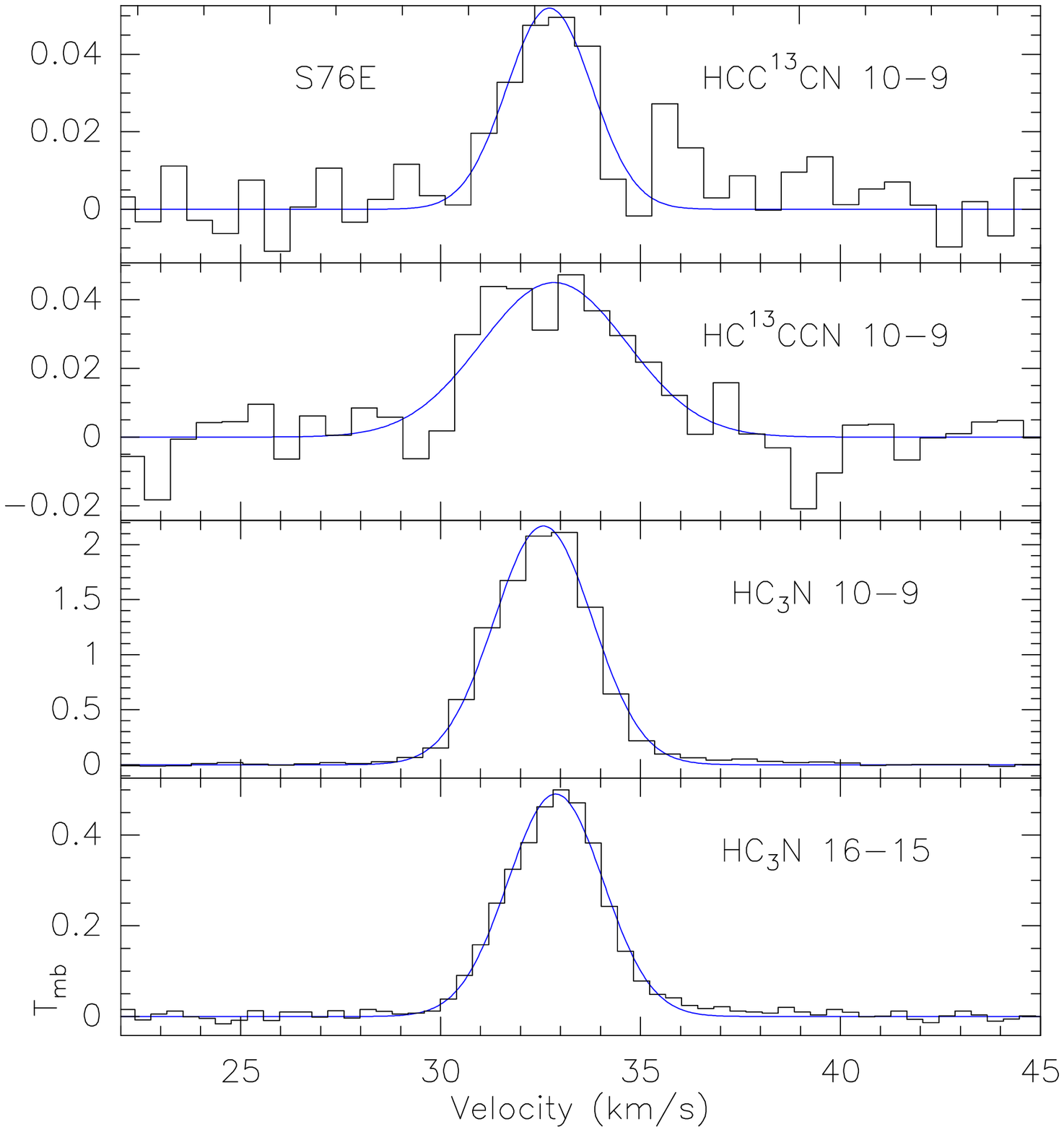} \includegraphics{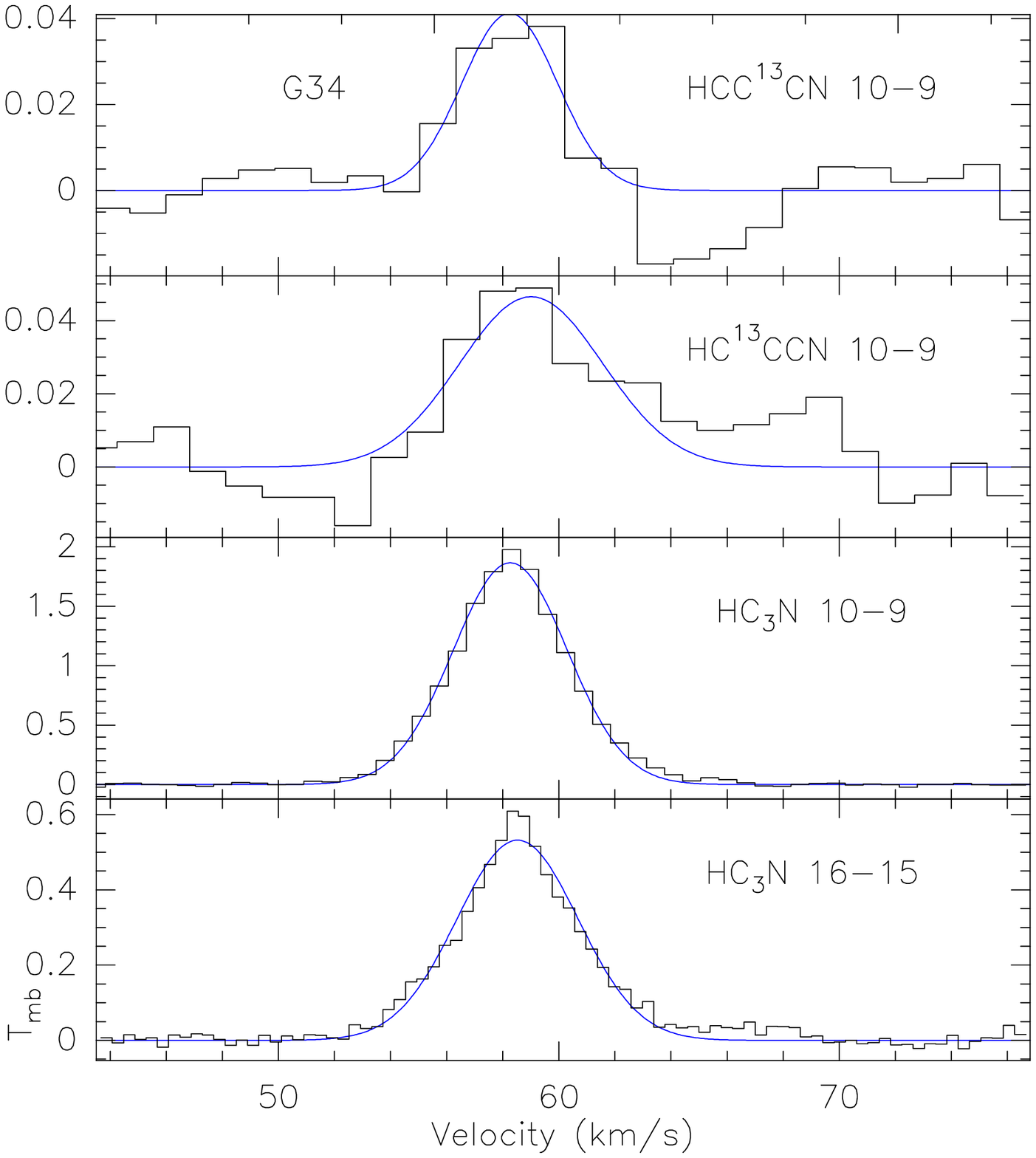}}
\scalebox{0.18}{\includegraphics{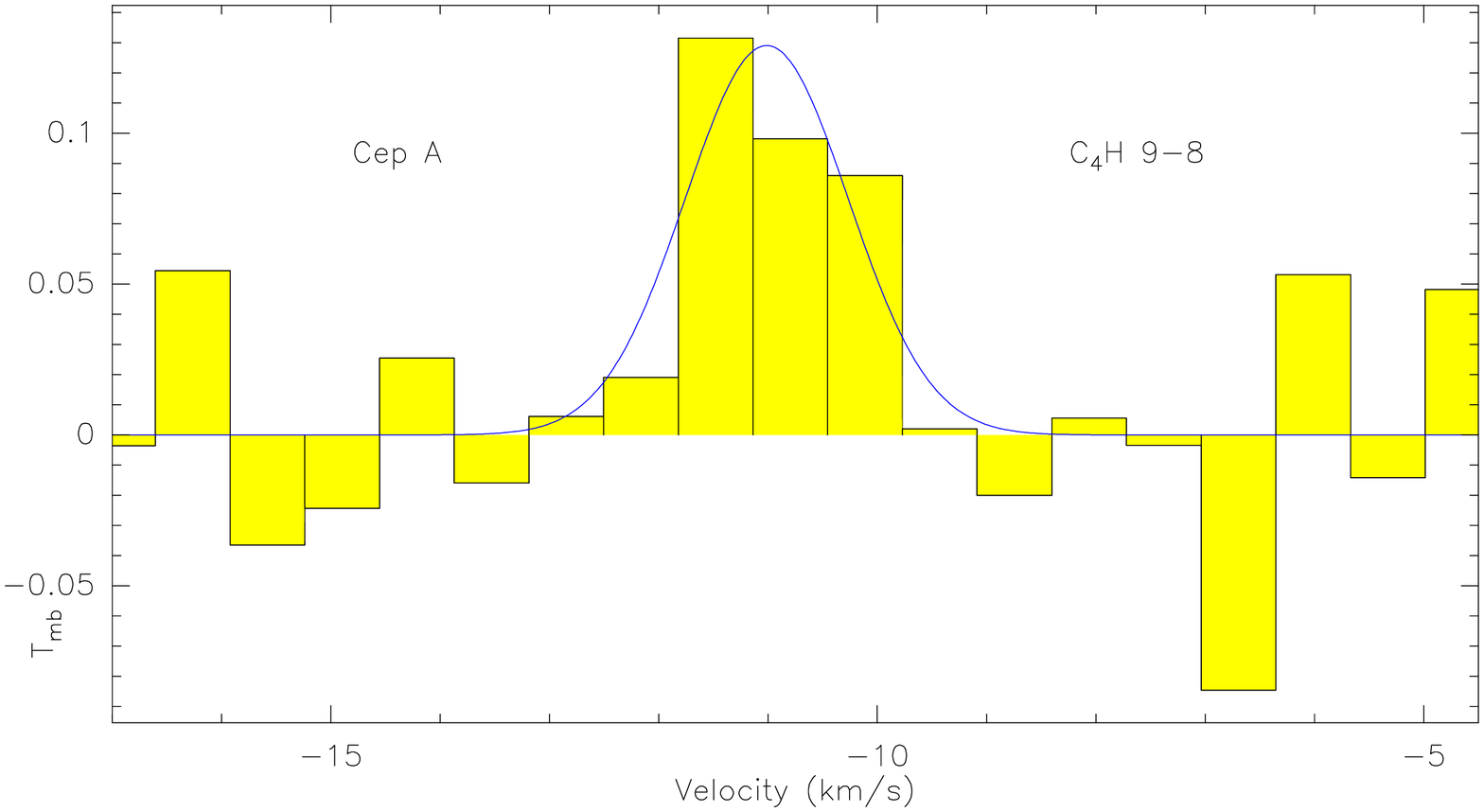} \includegraphics{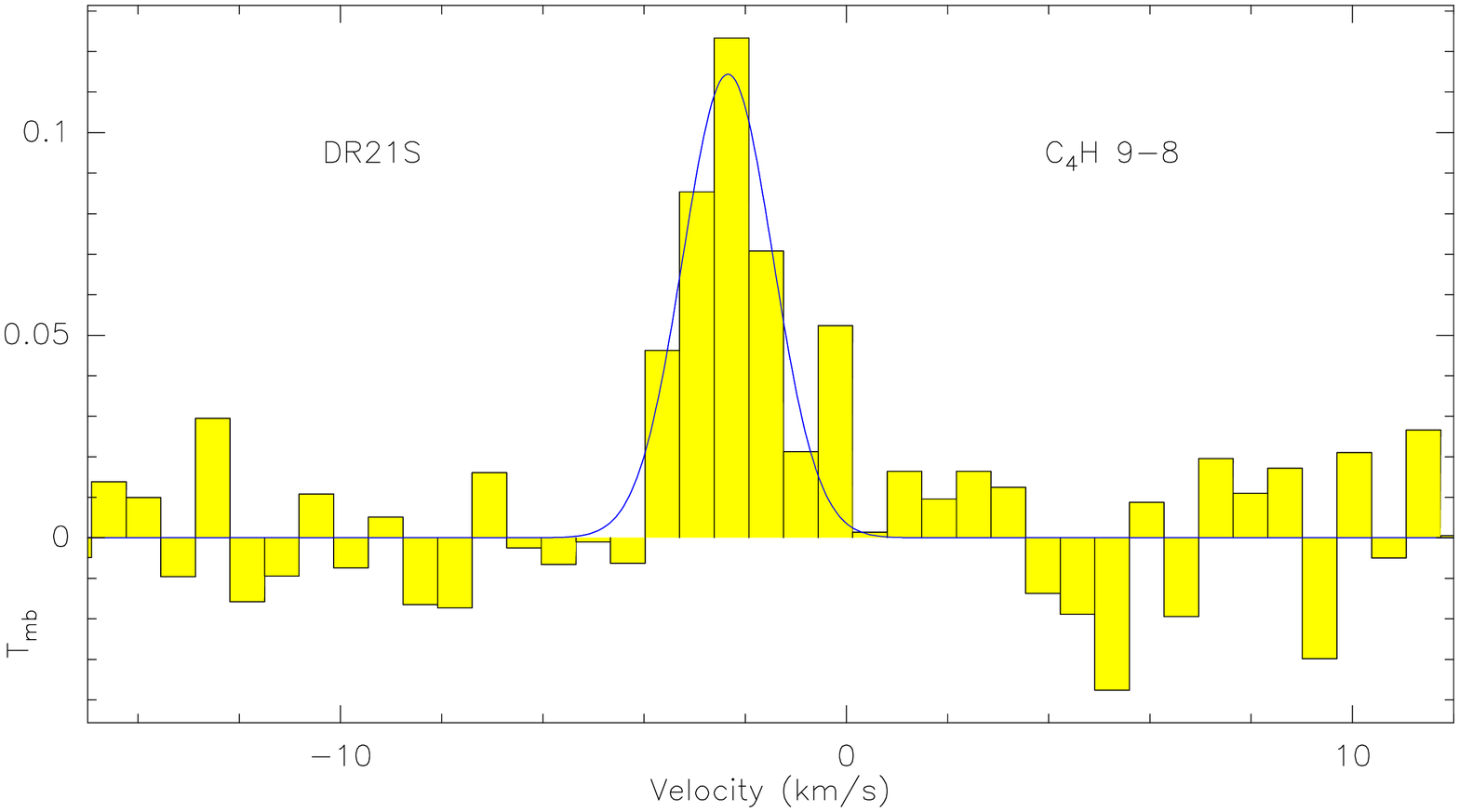} \includegraphics{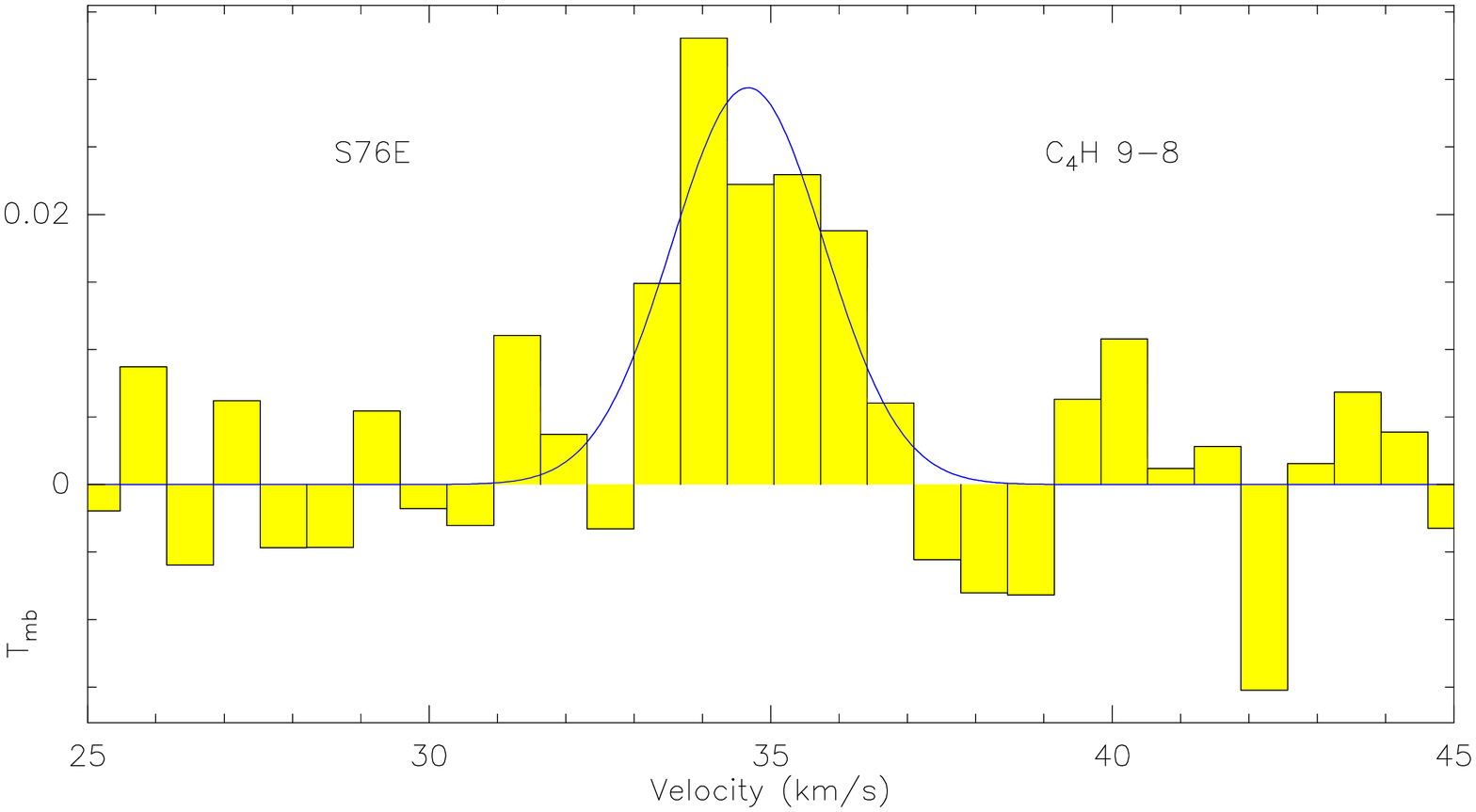} \includegraphics{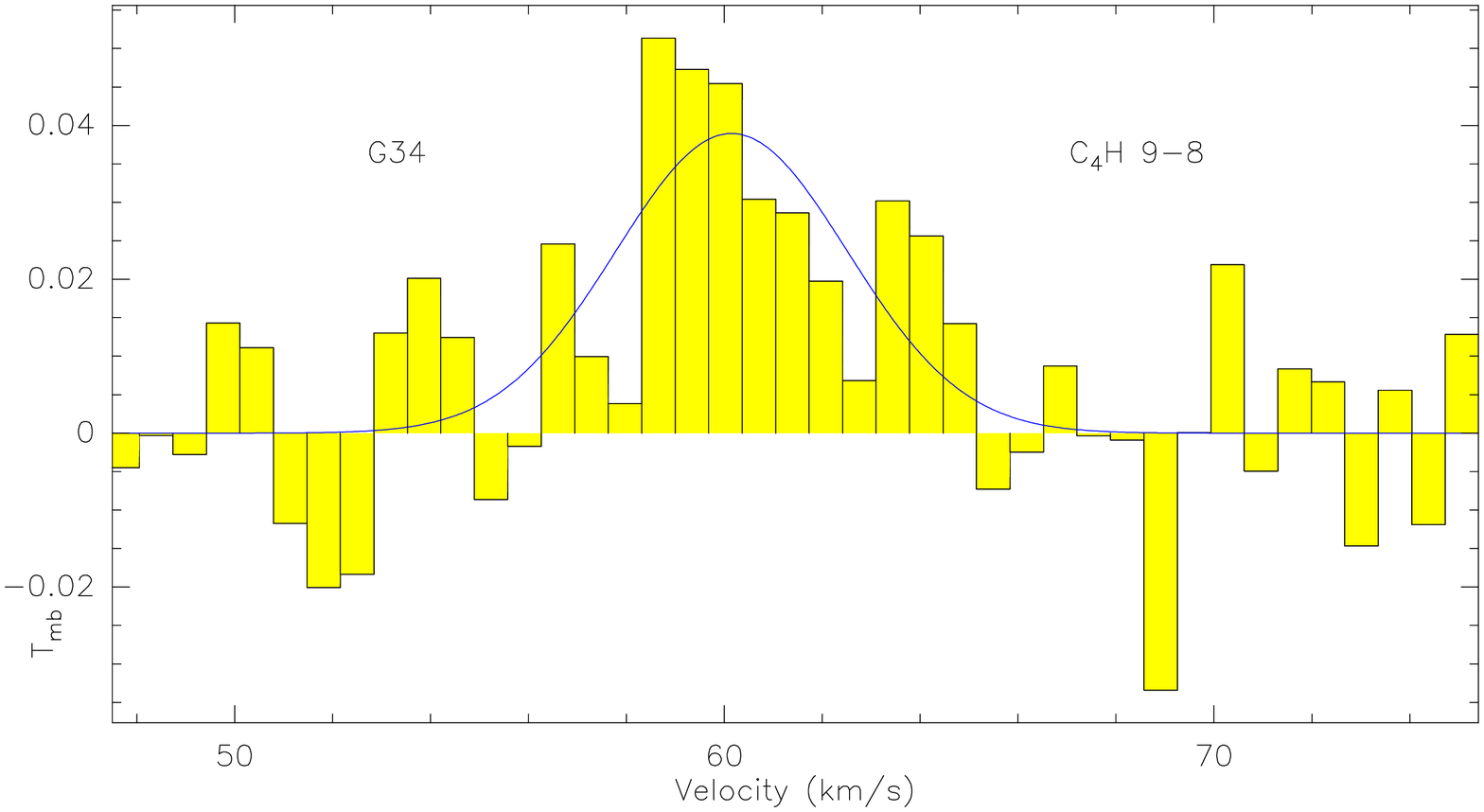}}
\scalebox{0.2}{\includegraphics{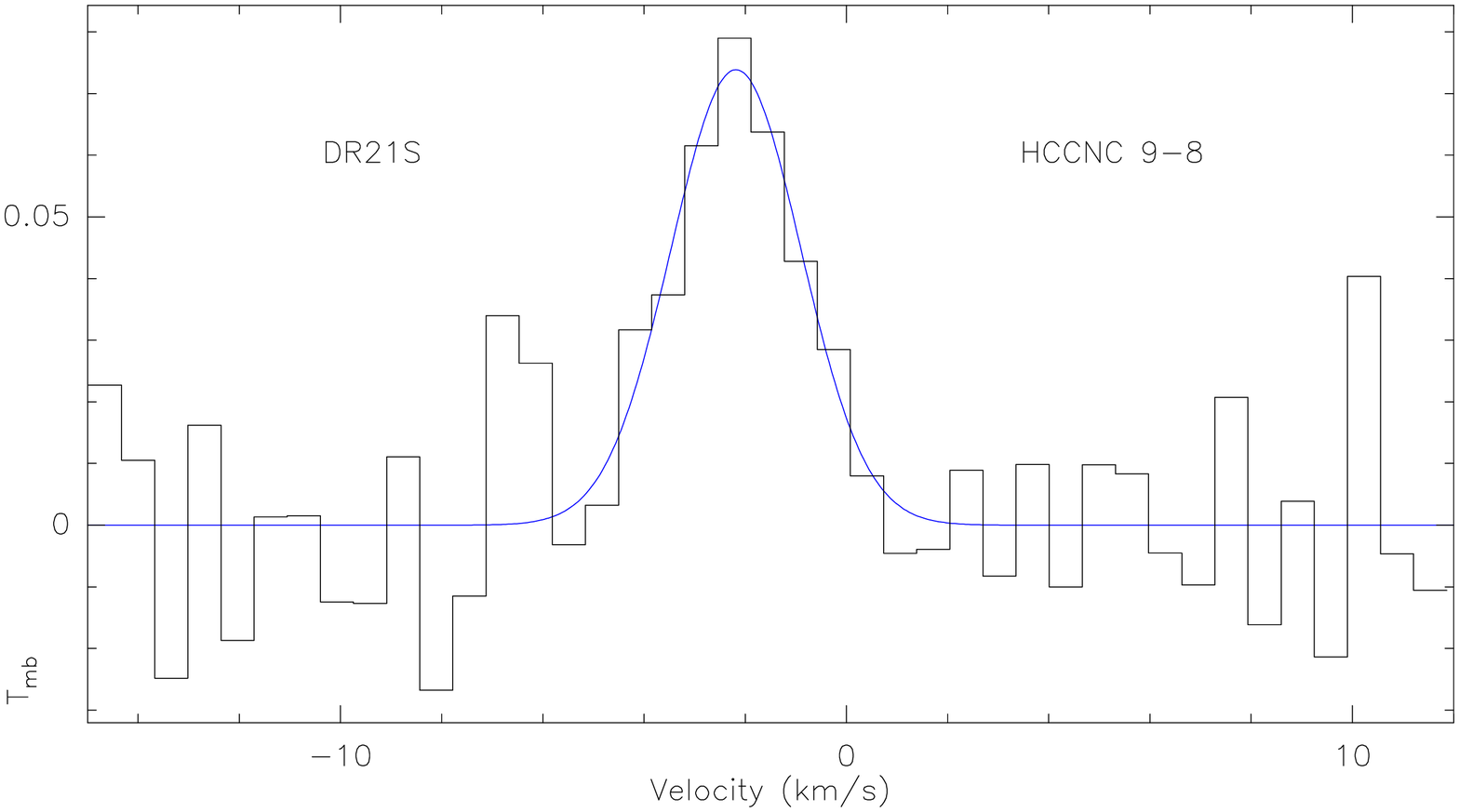} \includegraphics{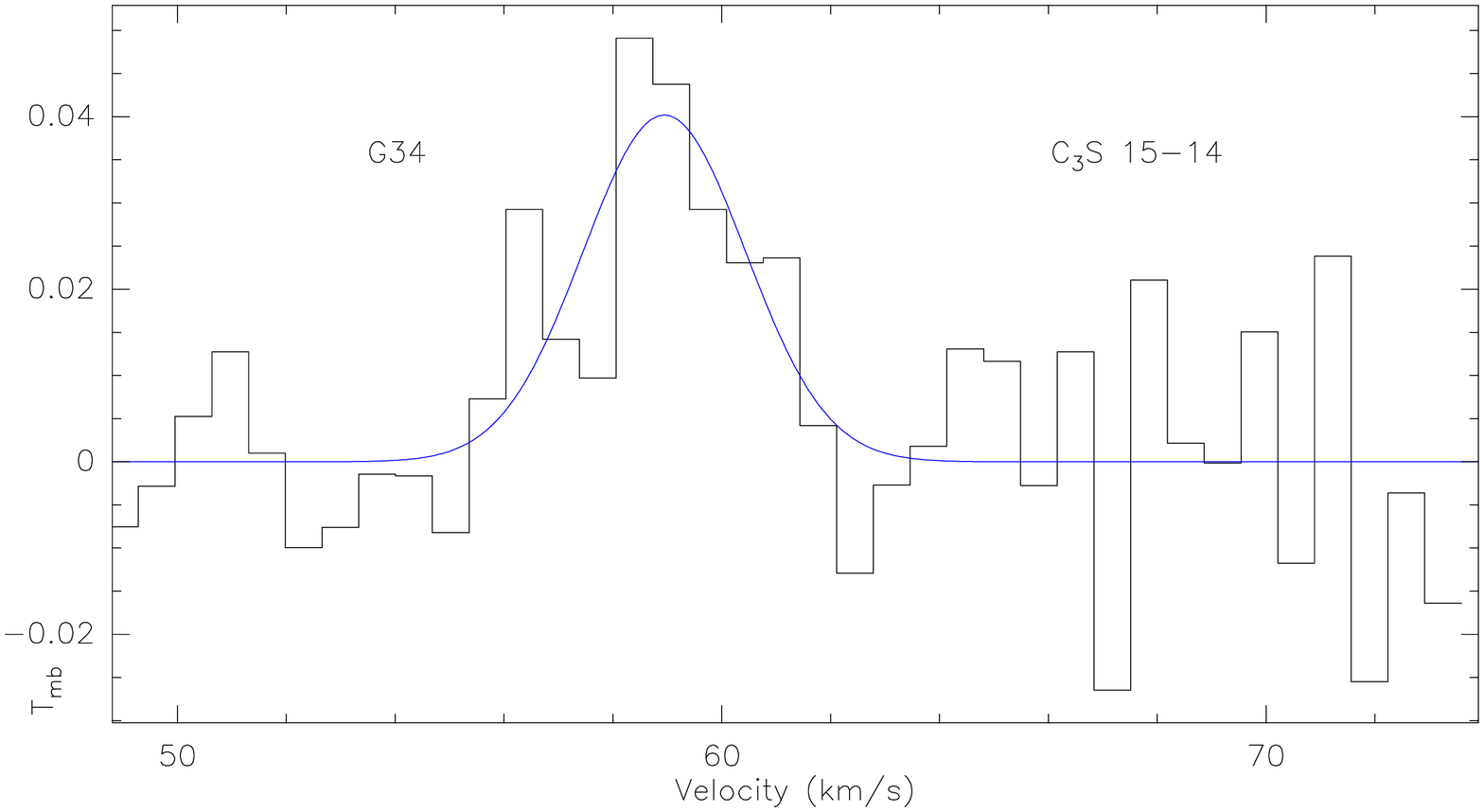} \includegraphics{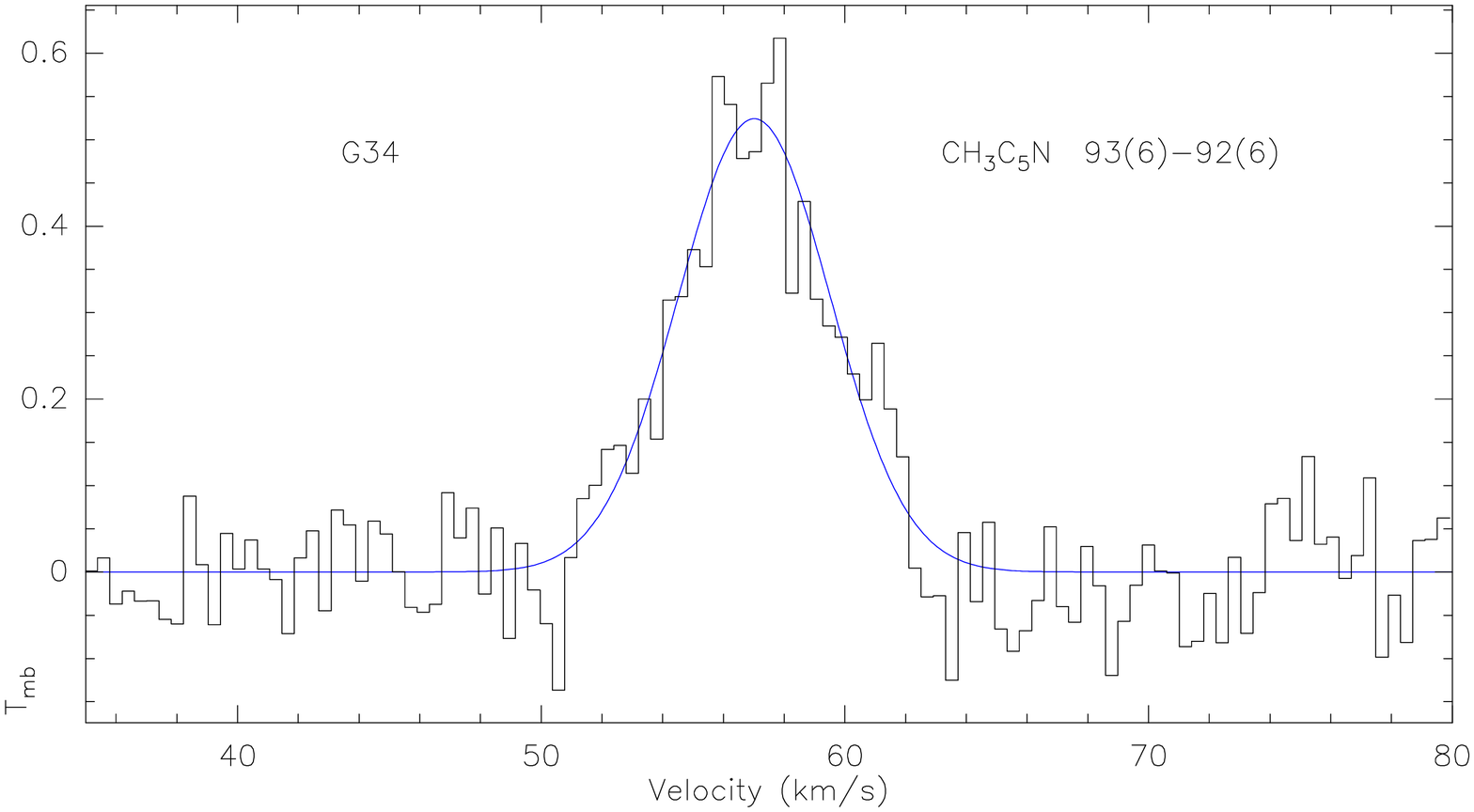}}
\vspace*{-0.2 cm}
\caption{The spectra of long carbon chain molecules.
	\label{fig:10}}
    \vskip-10pt
\end{figure*}

\begin{figure*}
	\centering
\scalebox{0.9}{\includegraphics{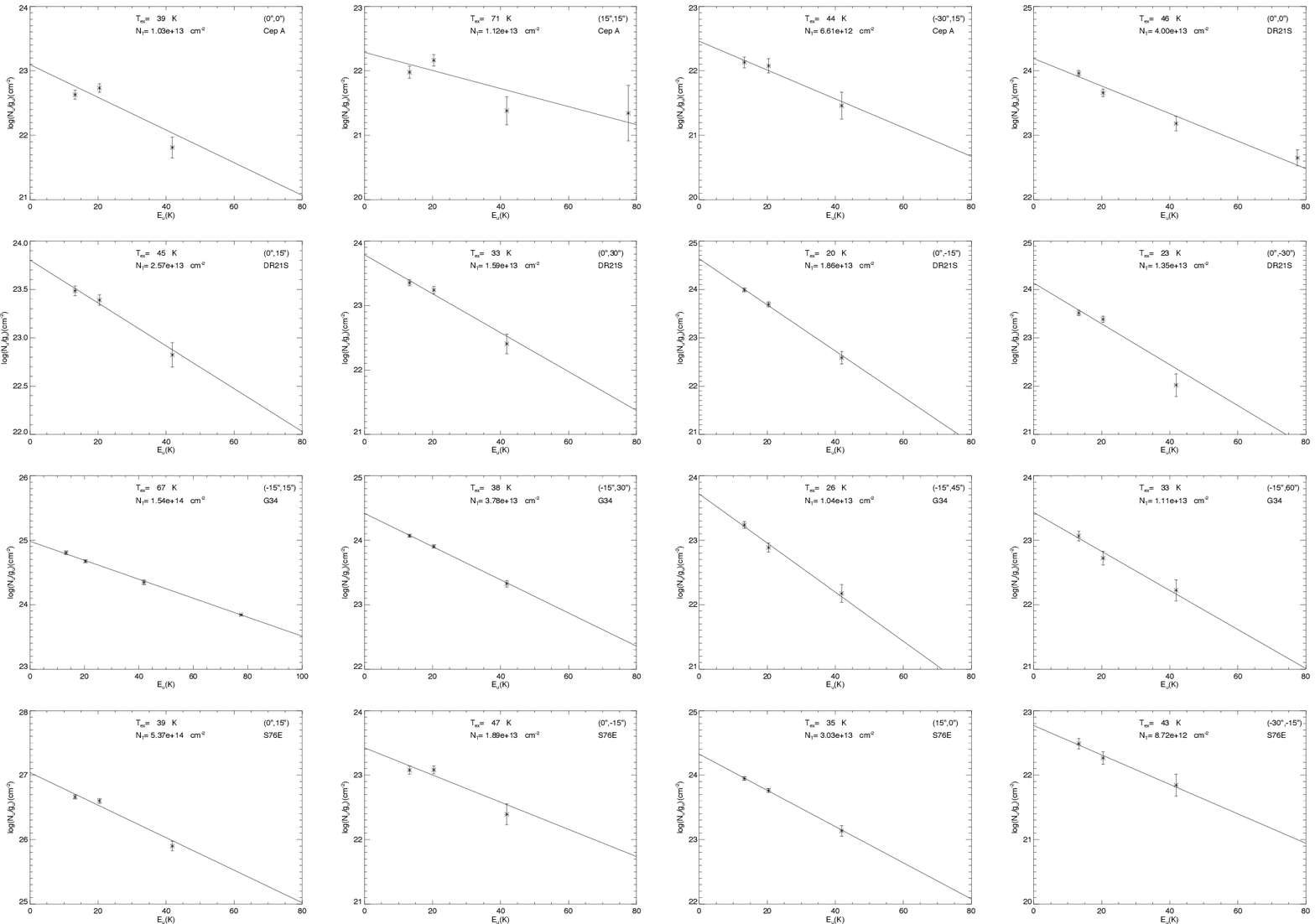}}
\vspace*{-0.2 cm} 
\caption{Rotation diagrams for the CN$_{3}$CN J=5$_{k}$-4$_{k}$ lines observed at Cep A, DR21S, G34 and S76E, respectively. Offset position and source name are present in upper right.
	\label{fig:11}}
    \vskip-10pt
\end{figure*}

\clearpage

\section{The parameters of molecular lines.} 
%\label{sec:table}

\begin{table*}
%\begin{minipage}{200mm}
 \centering
  \begin{minipage}{160mm}
     \begin{center}
	 \caption{The line information}
	  \label{Deuterated}
		\begin{tabular}{c c c c  c c}
 \hline
 \hline        
Species				&Transition	&Rest Frequency	& $E_{u}/k$	& $(\mu_{0})^{2}$ \\
					&			&MHz				&			&D	\\
	         \hline        
DCN				&2-1		&144828.11			&10.43		&13.42\\
H$^{13}$CN			&1-0		&86342.26			&4.14		&15.85\\
DCO$^{+}$			&2-1		&144077.29			&10.37		&7.89\\
H$^{13}$CO$^{+}$	&1-0		&86754.29			&4.16		&6.34\\
	         \hline	  
\end{tabular}
\end{center}
The parameters of molecular lines were assigned using CMDS spectroscopic databases.
\end{minipage}
\end{table*}

%%%%%%%%%%%%%%%%    CEPA         %%%%%%%%%%%%%%%%%%%%%%%%%%%%
\begin{table*}
%\begin{minipage}{200mm}
\caption{Cep A}
\label{tab:CEPA}
\centering
\begin{minipage}{160mm}
\begin{tabular}{lrrrrrr}
 \hline 
Molecule & Transition &Frequency & $\int T_{mb}dv$	& $v_{LSR}$		&FWHM & $T_{mb}$\\
		 &			& MHz		& (K km s$^{-1}$) & (km s$^{-1}$)	&(km s$^{-1}$)    &  K\\
\hline

CH$_{3}$OH	&	vt=0,1 5(-1,5)-4(0,4)E	&	84521.16	&	0.63	$\pm$	0.05	&	-10.80	$\pm$	0.13	&	3.66	$\pm$	0.32	&	0.16	\\
HC$^{18}$O$^{+}$	&	1-0						&	85162.22	&	0.39	$\pm$	0.04	&	-10.85	$\pm$	0.19	&	3.52	$\pm$	0.39	&	0.11	\\
HC$_{5}$N	&	v=0   32-31				&	85201.34	&	0.29	$\pm$	0.05	&	-11.62	$\pm$	0.24	&	3.15	$\pm$	0.66	&	0.05	\\
$c$-C$_{3}$H$_{2}$	&	2(1,2)-1(0,1)			&	85338.89	&	1.01	$\pm$	0.04	&	-10.49	$\pm$	0.07	&	3.38	$\pm$	0.17	&	0.28	\\
HCS$^{+}$			&	2-1						&	85347.89	&	0.34	$\pm$	0.04	&	-10.81	$\pm$	0.22	&	3.61	$\pm$	0.48	&	0.09	\\
CH$_{3}$CCH		&	5(3)-4(3)				&	85442.60	&	0.22	$\pm$	0.06	&	-10.73	$\pm$	0.67	&	4.71	$\pm$	1.91	&	0.04	\\
CH$_{3}$CCH		&	5(2)-4(2)				&	85450.76	&	0.49	$\pm$	0.05	&	-10.77	$\pm$	0.18	&	3.55	$\pm$	0.38	&	0.13	\\
CH$_{3}$CCH		&	5(1)-4(1)				&	85455.66	&	0.90	$\pm$	0.04	&	-10.70	$\pm$	0.07	&	2.81	$\pm$	0.15	&	0.30	\\
CH$_{3}$CCH		&	5(0)-4(0)				&	85457.29	&	1.18	$\pm$	0.04	&	-10.77	$\pm$	0.05	&	3.11	$\pm$	0.13	&	0.36	\\
C$_{4}$H	&	9-8 J=17/2-15/2			&	85672.58	&	0.13	$\pm$	0.04	&	-10.66	$\pm$	0.32	&	2.37	$\pm$	0.71	&	0.05	\\
NH$_{2}$D	&	1(1,1)0+-1(0,1)0- F=1-2	&	85926.27	&	1.02	$\pm$	0.09	&	-11.86	$\pm$	0.24	&	5.99	$\pm$	0.79	&	0.06	\\
HC$^{15}$N			&	v=0  1-0				&	86054.96	&	0.53	$\pm$	0.04	&	-10.69	$\pm$	0.12	&	3.24	$\pm$	0.27	&	0.15	\\
SO	&	v=0   N,J=2,2-1,1		&	86093.95	&	0.89	$\pm$	0.05	&	-10.45	$\pm$	0.09	&	3.69	$\pm$	0.24	&	0.23	\\
H$^{13}$CN	&	1-0 F=1-1				&	86338.73	&	0.67	$\pm$	0.04	&	-10.55	$\pm$	0.09	&	3.01	$\pm$	0.24	&	0.21	\\
H$^{13}$CN	&	1-0 F=2-1				&	86340.17	&	1.28	$\pm$	0.05	&	-10.60	$\pm$	0.06	&	3.70	$\pm$	0.17	&	0.33	\\
H$^{13}$CN	&	1-0 F=0-1				&	86342.25	&	0.26	$\pm$	0.04	&	-10.46	$\pm$	0.32	&	3.76	$\pm$	0.71	&	0.07	\\
H$^{13}$CO$^{+}$	&	1-0						&	86754.28	&	3.38	$\pm$	0.05	&	-10.71	$\pm$	0.02	&	3.41	$\pm$	0.05	&	0.93	\\
SiO	&	v=0-6  2-1 v=0			&	86846.96	&	0.50	$\pm$	0.07	&	-10.21	$\pm$	0.53	&	7.51	$\pm$	1.36	&	0.06	\\
HN$^{13}$C	&	1-0 F=1-1				&	87090.82	&	0.86	$\pm$	0.04	&	-10.40	$\pm$	0.06	&	2.92	$\pm$	0.14	&	0.28	\\
CCH	&	v=0  1-0 3/2-1/2 F=1-1	&	87284.10	&	0.74	$\pm$	0.04	&	-10.81	$\pm$	0.09	&	3.24	$\pm$	0.22	&	0.22	\\
CCH	&	v=0  1-0 3/2-1/2 F=2-1	&	87316.89	&	4.95	$\pm$	0.18	&	-10.88	$\pm$	0.06	&	3.35	$\pm$	0.14	&	1.39	\\
CCH		&	v=0  1-0 3/2-1/2 F=1-0	&	87328.58	&	2.72	$\pm$	0.05	&	-10.82	$\pm$	0.02	&	3.31	$\pm$	0.06	&	0.79	\\
CCH	&	v=0  1-0 1/2-1/2 F=1-1	&	87401.98	&	2.87	$\pm$	0.05	&	-10.82	$\pm$	0.02	&	3.38	$\pm$	0.06	&	0.81	\\
CCH	&	v=0 1-0 1/2-1/2 F=0-1	&	87407.16	&	1.34	$\pm$	0.05	&	-10.62	$\pm$	0.05	&	3.42	$\pm$	0.13	&	0.37	\\
CCH	&	v=0 1-0 1/2-1/2 F=1-0	&	87446.47	&	0.65	$\pm$	0.04	&	-10.89	$\pm$	0.09	&	3.27	$\pm$	0.20	&	0.19	\\
HC$_{5}$N	&	v=0    33-32			&	87863.63	&	0.14	$\pm$	0.04	&	-12.37	$\pm$	0.21	&	1.54	$\pm$	0.68	&	0.09	\\
HNCO	&	4(0,4)-3(0,3)			&	87925.23	&	0.41	$\pm$	0.05	&	-10.69	$\pm$	0.25	&	4.07	$\pm$	0.70	&	0.10	\\
HCN		&	1-0 F=1-1				&	88630.41	&	12.67	$\pm$	0.43	&	-12.53	$\pm$	0.14	&	8.66	$\pm$	0.31	&	1.38	\\
HCN		&	1-0 F=2-1				&	88631.84	&	4.52	$\pm$	0.23	&	-12.17	$\pm$	0.02	&	2.23	$\pm$	0.07	&	1.90	\\
HCN		&	1-0 F=0-1				&	88633.93	&	3.74	$\pm$	0.20	&	-10.71	$\pm$	0.11	&	4.74	$\pm$	0.25	&	0.74	\\
HCO$^{+}$	&	v=0  1-0				&	89188.52	&	13.02	$\pm$	0.43	&	-12.13	$\pm$	0.05	&	4.07	$\pm$	0.20	&	3.01	\\
HC$_{5}$N	&	v=0 34-33				&	90525.88	&	0.21	$\pm$	0.04	&	-10.53	$\pm$	0.45	&	3.35	$\pm$	0.87	&	0.04	\\
HNC			&	v=0 1-0					&	90663.56	&	11.39	$\pm$	0.04	&	-10.91	$\pm$	0.00	&	4.40	$\pm$	0.01	&	2.43	\\
CCS	&	7-6	&	90686.38	&	0.18	$\pm$	0.04	&	-10.32	$\pm$	0.36	&	3.57	$\pm$	0.77	&	0.05	\\
HCCCN		&	v=0 10-9				&	90979.02	&	3.44	$\pm$	0.04	&	-10.43	$\pm$	0.01	&	2.95	$\pm$	0.03	&	1.09	\\
CH$_{3}$CN	&	v=0  5(1)-4(1)			&	91985.31	&	0.22	$\pm$	0.05	&	-10.47	$\pm$	0.28	&	3.29	$\pm$	0.90	&	0.07	\\
CH$_{3}$CN	&	v=0  5(0)-4(0)			&	91987.08	&	0.21	$\pm$	0.04	&	-10.24	$\pm$	0.25	&	2.91	$\pm$	0.59	&	0.07	\\
$^{33}$SO$_{2}$		&	4(2,2)-4(1,3) F=5.5-5.5	&	143793.88	&	0.27	$\pm$	0.04	&	-11.25	$\pm$	0.26	&	3.39	$\pm$	0.57	&	0.08	\\
$^{33}$SO$_{2}$		&	4(2,2)-4(1,3)			&	143795.86	&	0.22	$\pm$	0.04	&	-10.65	$\pm$	0.34	&	3.15	$\pm$	0.74	&	0.06	\\
CH$_{3}$OH		&	3(1,3)-2(1,2) A++		&	143865.79	&	0.25	$\pm$	0.04	&	-11.11	$\pm$	0.26	&	3.33	$\pm$	0.47	&	0.07	\\
DCO$^{+}$	&	2-1						&	144077.28	&	0.46	$\pm$	0.04	&	-10.60	$\pm$	0.17	&	3.37	$\pm$	0.32	&	0.13	\\
CH$_{3}$C$_{3}$N	&	35(19)-34(19)			&	144087.77	&	0.33	$\pm$	0.04	&	 -9.32	$\pm$	0.31	&	4.09	$\pm$	0.53	&	0.08	\\
CH$_{3}$OH		&	vt=0 24(-6,18)-25(-5,20)	&	144194.75	&	0.33	$\pm$	0.04	&	 -4.79	$\pm$	0.21	&	3.62	$\pm$	0.47	&	0.09	\\
CCD	&	2-1 J=5/2-3/2F=7/2-5/2	&	144241.96	&	0.27	$\pm$	0.04	&	-11.83	$\pm$	0.36	&	4.30	$\pm$	0.73	&	0.04	\\
CCD		&	2-1 J=5/2-3/2 F=3/2-1/2	&	144243.05	&	0.13	$\pm$	0.03	&	-14.87	$\pm$	0.27	&	2.14	$\pm$	0.56	&	0.05	\\
C$^{34}$S			&	2-1						&	144617.10	&	0.73	$\pm$	0.03	&	-10.55	$\pm$	0.06	&	2.73	$\pm$	0.14	&	0.25	\\
DCN		&	2-1 F1=2-1				&	144828.00	&	0.61	$\pm$	0.05	&	-10.71	$\pm$	0.18	&	4.66	$\pm$	0.45	&	0.13	\\
$c$-C$_{3}$H$_{2}$	&	3(1,2)-2(2,1)			&	145089.62	&	0.23	$\pm$	0.03	&	-10.33	$\pm$	0.18	&	2.64	$\pm$	0.39	&	0.09	\\
CH$_{3}$OH	&	3(0,3)-2(0,2) E			&	145093.76	&	0.54	$\pm$	0.05	&	-11.55	$\pm$	0.31	&	5.78	$\pm$	0.82	&	0.09	\\
CH$_{3}$OH		&	3(-1,3)-2(-1,2) E		&	145097.44	&	1.48	$\pm$	0.05	&	-10.71	$\pm$	0.05	&	3.67	$\pm$	0.16	&	0.38	\\
CH$_{3}$OH	&	3(0,3)-2(0,2) A++		&	145103.19	&	1.86	$\pm$	0.04	&	-10.69	$\pm$	0.04	&	3.75	$\pm$	0.11	&	0.47	\\
CH$_{3}$OH	&	3(-2,2)-2(-2,1) E		&	145126.39	&	0.23	$\pm$	0.17	&	-10.14	$\pm$	1.53	&	3.81	$\pm$	2.89	&	0.06	\\
CH$_{3}$OH		&	3(1,2)-2(1,1) E			&	145131.87	&	0.25	$\pm$	0.04	&	-11.08	$\pm$	0.37	&	4.57	$\pm$	0.79	&	0.05	\\
HCCCN	&	16-15	&	145560.95	&	1.04	$\pm$	0.04	&	-10.56	$\pm$	0.05	&	2.97	$\pm$	0.12	&	0.33	\\
H$_{2}$CO	&	2(0,2)-1(0,1)	&	145602.95	&	4.19	$\pm$	0.04	&	-10.76	$\pm$	0.02	&	4.38	$\pm$	0.05	&	0.90	\\
OCS	&	12-11	&	145946.81	&	0.11	$\pm$	0.04	&	-10.09	$\pm$	0.33	&	1.90	$\pm$	1.05	&	0.05	\\
CH$_{3}$OH		&	3(1,2)-2(1,1) A--		&	146368.34	&	0.46	$\pm$	0.08	&	-10.55	$\pm$	0.48	&	6.36	$\pm$	1.82	&	0.07	\\
(CH$_{3}$)2CO	&	v=0 43(16,27)-42(19,24)EE	&	146599.28	&	0.17	$\pm$	0.04	&	 -9.28	$\pm$	0.26	&	2.23	$\pm$	0.73	&	0.07	\\
CS	&	3-2						&	146969.02	&	9.80	$\pm$	0.09	&	-10.65	$\pm$	0.02	&	4.57	$\pm$	0.04	&	2.01	\\
$c$-HCCCD	&	19(6,13)-18(8,10)		&	147021.69	&	0.47	$\pm$	0.09	&	-12.81	$\pm$	0.38	&	4.19	$\pm$	1.10	&	0.11	\\

    \hline
\end{tabular}
\end{minipage}
\end{table*}

%%%%%%%%%%%%%%%%%%%%%%%%%%%%%%%%%%%%%%%%%%%%%%%%%%%%%%%%%%%%%%%%%%%%%%%%%%%%%%%%%%%%%%%%%%%%%%%%%%%

%%%%%%%%%%%%%%%%%%%%%%%%%%%%%%%%%%%%%%%%%%%%%%%%%%%%%%%%%%%%%%%%%%%%%%%%%%%%%%%%%%%%%%%%%%%%%%%%%%%

\begin{table*}
%\begin{minipage}{200mm}
\caption{DR21S}
\label{tab:DR21S}
\centering
\begin{minipage}{200mm}
\begin{tabular}{lrrrrrr}
 \hline 
Molecule & Transition &Frequency & $\int T_{mb}dv$	& $v_{LSR}$		&FWHM & $T_{mb}$\\
		 &			& MHz		& (K km s$^{-1}$) & (km s$^{-1}$)	&(km s$^{-1}$)    &  K\\
\hline

CH$_{3}$OH		&	5(-1,5)-4(0,4)				&	84521.20	&	0.62	$\pm$	0.05	&	-2.03	$\pm$	0.10	&	2.74	$\pm$	0.32	&	0.22	\\
H$\gamma$			&	H60$\gamma$					&	84914.39	&	3.80	$\pm$	0.16	&	-0.16	$\pm$	0.68	&	30.80	$\pm$	1.68	&	0.05	\\
HC$^{18}$O$^{+}$		&	 1-0						&	85162.22	&	0.15	$\pm$	0.02	&	-1.69	$\pm$	0.16	&	1.60	$\pm$	0.46	&	0.10	\\
HC$_{5}$N			&	32-31						&	85201.34	&	0.35	$\pm$	0.04	&	-2.07	$\pm$	0.12	&	2.32	$\pm$	0.28	&	0.14	\\
C$^{13}$CH	&	1-0 3/2-1/2F=2,2.5-1,1.5	&	85229.27	&	0.12	$\pm$	0.02	&	-2.00	$\pm$	0.14	&	1.30	$\pm$	0.48	&	0.09	\\
C$^{13}$CH			&	1-0 3/2-1/2F=2,1.5-1,0.5	&	85232.76	&	0.95	$\pm$	0.04	&	-2.84	$\pm$	0.39	&	1.97	$\pm$	0.83	&	0.04	\\
C$^{13}$CH			&	1-0 3/2-1/2F=1,1.5-0,0.5	&	85256.96	&	0.14	$\pm$	0.05	&	-2.10	$\pm$	0.69	&	3.36	$\pm$	1.99	&	0.03	\\
c-C$_{3}$H$_{2}$	&	2(1,2)-1(0,1)				&	85338.90	&	2.36	$\pm$	0.05	&	-2.13	$\pm$	0.03	&	3.44	$\pm$	0.08	&	0.64	\\
HCS$^{+}$	&	   2-1						&	85347.86	&	0.34	$\pm$	0.04	&	-1.99	$\pm$	0.19	&	3.19	$\pm$	0.46	&	0.10	\\
CH$_{3}$CCH			&	5(3)-4(3)					&	85442.60	&	0.13	$\pm$	0.02	&	-1.91	$\pm$	0.26	&	2.02	$\pm$	0.52	&	0.06	\\
CH$_{3}$CCH		&	 5(2)-4(2)					&	85450.76	&	0.34	$\pm$	0.04	&	-1.77	$\pm$	0.11	&	2.27	$\pm$	0.28	&	0.14	\\
CH$_{3}$CCH		&	 5(1)-4(1)					&	85455.66	&	0.81	$\pm$	0.04	&	-1.87	$\pm$	0.05	&	2.30	$\pm$	0.13	&	0.33	\\
CH$_{3}$CCH				&	5(0)-4(0)					&	85457.29	&	0.90	$\pm$	0.04	&	-1.84	$\pm$	0.05	&	2.37	$\pm$	0.13	&	0.36	\\
C4H			&	 9-8 J=19/2-17/2			&	85634.00	&	0.25	$\pm$	0.04	&	-2.34	$\pm$	0.18	&	2.48	$\pm$	0.53	&	9.58	\\
$c$-C$_{3}$H$_{2}$	&	4(3,2)-4(2,3)				&	85656.41	&	0.25	$\pm$	0.05	&	-3.57	$\pm$	0.63	&	5.59	$\pm$	1.40	&	0.04	\\
He$\alpha$			&	He42$\alpha$				&	85723.31	&	5.55	$\pm$	0.40	&	-0.74	$\pm$	0.12	&	32.51	$\pm$	0.64	&	0.02	\\
H$\alpha$			&	H42$\alpha$					&	85688.39	&	26.20	$\pm$	0.20	&	0.78	$\pm$	1.16	&	37.49	$\pm$	2.87	&	0.34	\\
SO				&	N,J=2,2-1,1					&	86093.98	&	0.63	$\pm$	0.05	&	-2.30	$\pm$	0.12	&	3.29	$\pm$	0.34	&	0.18	\\
CCS					&	N,J=7,6-6,5					&	86181.41	&	0.16	$\pm$	0.04	&	-2.07	$\pm$	0.31	&	2.69	$\pm$	0.79	&	0.06	\\
H$^{13}$CN	&	 1-0 F=1-1					&	86338.73	&	1.07	$\pm$	0.05	&	-2.03	$\pm$	0.06	&	2.75	$\pm$	0.17	&	0.36	\\
H$^{13}$CN	&	1-0 F=2-1					&	86340.17	&	1.90	$\pm$	0.05	&	-2.02	$\pm$	0.03	&	2.83	$\pm$	0.10	&	0.63	\\
H$^{13}$CN	&	 1-0 F=0-1					&	86342.25	&	0.50	$\pm$	0.07	&	-1.66	$\pm$	0.17	&	3.06	$\pm$	0.66	&	0.16	\\
H$\varepsilon$			&	H70$\varepsilon$			&	86488.42	&	1.08	$\pm$	0.14	&	-1.99	$\pm$	1.80	&	27.62	$\pm$	4.42	&	0.03	\\
HCO		&	1(0,1)-0(0,0) 3/2-1/2 F=2-1	&	86670.82	&	0.57	$\pm$	0.04	&	-2.17	$\pm$	0.11	&	3.04	$\pm$	0.25	&	0.18	\\
HCO		&	1(0,1)-0(0,0) 3/2-1/2 F=1-0	&	86708.35	&	0.40	$\pm$	0.05	&	-2.29	$\pm$	0.18	&	3.18	$\pm$	0.46	&	0.12	\\
H$^{13}$CO$^{+}$		&	1-0							&	86754.28	&	2.57	$\pm$	0.04	&	-2.00	$\pm$	0.02	&	2.72	$\pm$	0.05	&	0.89	\\
HCO		&	1(0,1)-0(0,0) 1/2-1/2 F=1-1	&	86777.43	&	0.35	$\pm$	0.04	&	-2.52	$\pm$	0.20	&	3.28	$\pm$	0.51	&	0.10	\\
SiO			&	 2-1 v=0					&	86846.99	&	2.79	$\pm$	0.08	&	-1.51	$\pm$	0.22	&	5.06	$\pm$	0.59	&	0.17	\\
HN$^{13}$C	&	1-0 F=2-1					&	87090.85	&	1.30	$\pm$	0.04	&	-1.60	$\pm$	0.03	&	2.39	$\pm$	0.09	&	0.51	\\
CCH				&	1-0 3/2-1/2 F=1-1			&	87284.15	&	1.01	$\pm$	0.04	&	-2.03	$\pm$	0.05	&	2.77	$\pm$	0.15	&	0.34	\\
CCH				&	1-0 3/2-1/2 F=2-1			&	87316.92	&	7.67	$\pm$	0.05	&	-2.14	$\pm$	0.01	&	3.11	$\pm$	0.02	&	2.32	\\
CCH		&	1-0 3/2-1/2 F=1-0			&	87328.62	&	4.16	$\pm$	0.05	&	-2.08	$\pm$	0.01	&	3.01	$\pm$	0.04	&	1.30	\\
CCH		&	1-0 1/2-1/2 F=1-1			&	87402.00	&	4.12	$\pm$	0.05	&	-2.16	$\pm$	0.01	&	2.98	$\pm$	0.04	&	1.30	\\
CCH					&	1-0 1/2-1/2 F=0-1			&	87407.16	&	1.94	$\pm$	0.04	&	-2.17	$\pm$	0.03	&	2.94	$\pm$	0.08	&	0.62	\\
CCH					&	1-0 1/2-1/2 F=1-0			&	87446.47	&	1.01	$\pm$	0.04	&	-2.10	$\pm$	0.05	&	2.75	$\pm$	0.13	&	0.35	\\
H$\beta$			&	H65$\beta$					&	87615.00	&	2.29	$\pm$	0.16	&	-0.14	$\pm$	1.39	&	36.62	$\pm$	3.22	&	0.02	\\
HC$_{5}$N			&	33-32						&	87863.63	&	0.27	$\pm$	0.04	&	-1.99	$\pm$	0.15	&	2.13	$\pm$	0.33	&	0.12	\\
HNCO				&	4(0,4)-3(0,3)				&	87925.23	&	0.43	$\pm$	0.05	&	-2.22	$\pm$	0.14	&	2.54	$\pm$	0.37	&	0.16	\\
HCN	&	 1-0 F=1-1					&	88630.41	&	7.24	$\pm$	0.18	&	-1.22	$\pm$	0.04	&	3.95	$\pm$	0.11	&	1.72	\\
HCN					&	1-0 F=2-1					&	88631.84	&	19.00	$\pm$	0.41	&	-2.43	$\pm$	0.03	&	4.66	$\pm$	0.09	&	3.84	\\
HCN					&	1-0 F=0-1					&	88633.93	&	8.43	$\pm$	0.61	&	-2.26	$\pm$	0.66	&	5.00	$\pm$	0.66	&	1.59	\\
H$^{15}$NC				&	1-0							&	88865.69	&	0.15	$\pm$	0.04	&	-1.97	$\pm$	0.22	&	1.65	$\pm$	0.40	&	0.09	\\
HCO$^{+}$	&	v=0,1-0						&	89188.52	&	28.74	$\pm$	0.41	&	-3.08	$\pm$	0.06	&	9.44	$\pm$	0.18	&	2.87	\\
HCCNC	&	9-8							&	89419.26	&	0.23	$\pm$	0.05	&	-2.10	$\pm$	0.30	&	3.00	$\pm$	0.72	&	0.07	\\
t-HCOOH	&	4(3,1)-3(3,0)	&	89950.33	&	0.29	$\pm$	0.05	&	-1.31	$\pm$	0.52	&	4.99	$\pm$	1.04	&	0.05	\\
t-HCOOH	&	4(2,2)-3(2,1)	&	90164.63	&	0.28	$\pm$	0.07	&	-4.49	$\pm$	0.78	&	8.02	$\pm$	2.90	&	0.03	\\
HC$_{5}$N			&	34-33						&	90525.89	&	0.25	$\pm$	0.02	&	-2.20	$\pm$	0.17	&	2.86	$\pm$	0.38	&	0.08	\\
HC$^{13}$CCN	&	10--9						&	90593.05	&	0.27	$\pm$	0.02	&	-1.75	$\pm$	0.18	&	3.09	$\pm$	0.46	&	0.08	\\
HCC$^{13}$CN	&	10-9						&	90601.79	&	0.19	$\pm$	0.02	&	-2.08	$\pm$	0.16	&	2.10	$\pm$	0.35	&	0.09	\\
HNC		&	v=0 1-0						&	90663.56	&	13.72	$\pm$	0.06	&	-2.29	$\pm$	0.00	&	3.82	$\pm$	0.02	&	3.39	\\
CCS				&	N,J=7,7-6,6					&	90686.38	&	0.13	$\pm$	0.04	&	-2.88	$\pm$	0.61	&	3.80	$\pm$	1.58	&	0.03	\\
HCCCN	&	v=0							&	90979.02	&	4.82	$\pm$	0.04	&	-0.93	$\pm$	0.00	&	2.44	$\pm$	0.02	&	1.86	\\
c-C$_{3}$H		&	2(1,2)-1(1,1) 5/2,3-3/2,2	&	91494.34	&	0.22	$\pm$	0.04	&	-2.50	$\pm$	0.22	&	2.76	$\pm$	0.55	&	0.08	\\
c-C$_{3}$H			&	2(1,2)-1(1,1) 5/2,2-3/2,1	&	91497.60	&	0.19	$\pm$	0.04	&	-2.24	$\pm$	0.27	&	2.86	$\pm$	0.86	&	0.06	\\
H$\beta$			&	H64$\beta$					&	91663.13	&	1.87	$\pm$	0.13	&	-0.87	$\pm$	0.99	&	28.14	$\pm$	2.11	&	0.05	\\
CH$_{3}$CN			&	5(3)-4(3) F=4-3				&	91971.46	&	0.45	$\pm$	0.06	&	-1.67	$\pm$	0.16	&	6.46	$\pm$	1.15	&	0.06	\\
CH$_{3}$CN			&	5(2)-4(2) F=6-5				&	91979.99	&	0.14	$\pm$	0.02	&	-2.71	$\pm$	0.16	&	1.71	$\pm$	0.32	&	0.08	\\
CH$_{3}$CN			&	5(1)-4(1)					&	91985.31	&	0.43	$\pm$	0.04	&	-2.32	$\pm$	0.11	&	2.69	$\pm$	0.29	&	0.16	\\
CH$_{3}$CN			&	5(0)-4(0)					&	91987.08	&	0.63	$\pm$	0.04	&	-2.13	$\pm$	0.09	&	3.08	$\pm$	0.28	&	0.19	\\

 \hline
\end{tabular}
\end{minipage}
\end{table*}

\begin{table*}
%\begin{minipage}{200mm}
\contcaption{}
%\label{tab:DR21S}
\centering
\begin{minipage}{200mm}
\begin{tabular}{lrrrrrr}
 \hline 
Molecule & Transition &Frequency & $\int T_{mb}dv$	& $v_{LSR}$		&FWHM & $T_{mb}$\\
		 &			& MHz		& (K km s$^{-1}$) & (km s$^{-1}$)	&(km s$^{-1}$)    &  K\\
\hline

H$\alpha$			&	H41$\alpha$						&	92034.43	&	23.86	$\pm$	0.14	&	-0.95	$\pm$	0.09	&	31.55	$\pm$	0.21	&	0.32	\\
DCO$^{+}$	&	2-1							&	144077.28	&	0.43	$\pm$	0.04	&	-1.86	$\pm$	0.10	&	2.24	$\pm$	0.22	&	0.18	\\
unidentified	&		&	144199.7	&	0.49	$\pm$	0.05	&	-2.26	$\pm$	0.11	&	2.05	$\pm$	0.27	&	0.23	\\
CCD	&	N=2-1,J=5/2-3/2,F=7/2-5/2	&	144241.93	&	0.32	$\pm$	0.04	&	-3.12	$\pm$	0.20	&	3.09	$\pm$	0.42	&	0.10	\\
CH$_{3}$OCHO	&	v=1 18(14, 5)-19(13, 6) A	&	144373.86	&	0.23	$\pm$	0.04	&	-2.94	$\pm$	0.68	&	3.49	$\pm$	0.90	&	0.03	\\
C$^{34}$S	&	3-2							&	144616.13	&	1.50	$\pm$	0.04	&	-4.37	$\pm$	0.03	&	2.59	$\pm$	0.08	&	0.55	\\
DCN			&	J= 2 - 1, F=2-1				&	144828.00	&	0.97	$\pm$	0.05	&	-1.76	$\pm$	0.13	&	5.22	$\pm$	0.45	&	0.17	\\
$c$-C$_{3}$H$_{2}$		&	3(1,2)-2(2,1)				&	145089.62	&	0.27	$\pm$	0.09	&	-2.35	$\pm$	0.37	&	2.06	$\pm$	0.88	&	0.12	\\
CH$_{3}$OH	&	3(-1,3)-2(-1,2) E			&	145097.44	&	0.80	$\pm$	0.10	&	-2.29	$\pm$	0.17	&	2.78	$\pm$	0.45	&	0.27	\\
CH$_{3}$OH	&	3(0,3)-2(0,2) A++			&	145103.19	&	1.08	$\pm$	0.09	&	-2.19	$\pm$	0.11	&	2.91	$\pm$	0.32	&	0.35	\\
CH$_{3}$OH	&	3(-2,2)-2(-2,1) E			&	145126.39	&	0.92	$\pm$	0.06	&	-1.65	$\pm$	0.25	&	6.26	$\pm$	0.47	&	0.14	\\
CH$_{3}$OH	&	3(1,2)-2(1,1) E				&	145131.87	&	0.46	$\pm$	0.05	&	-2.84	$\pm$	0.45	&	6.92	$\pm$	1.09	&	0.06	\\
(CH$_{3}$)2CO		&	23(11,13)-22(12,10) AE		&	145369.23	&	0.65	$\pm$	0.04	&	-2.71	$\pm$	0.09	&	3.01	$\pm$	0.19	&	0.06	\\
$l$-H$_{2}$CCC	&	7(2,5)-6(2,4)				&	145532.32	&	0.31	$\pm$	0.06	&	-4.54	$\pm$	0.41	&	3.99	$\pm$	0.85	&	0.07	\\
HCCCN	&	16-15						&	145560.95	&	1.95	$\pm$	0.04	&	-2.21	$\pm$	0.02	&	2.61	$\pm$	0.06	&	0.70	\\
H$_{2}$CO	&	(0,2)-1(0,1)				&	145602.95	&	4.32	$\pm$	0.05	&	-2.42	$\pm$	0.02	&	3.72	$\pm$	0.05	&	1.09	\\
C$^{33}$S	&	3-2 F=3/2-1/2				&	145755.62	&	0.28	$\pm$	0.04	&	-2.66	$\pm$	0.20	&	3.05	$\pm$	0.65	&	0.09	\\
CH$_{3}$OH			&	3( 1, 2)- 2( 1, 1) - -		&	146368.34	&	2.60	$\pm$	0.22	&	-3.76	$\pm$	0.31	&	7.51	$\pm$	0.80	&	0.33	\\
SO$^{2}$	&	4(2,2)-4(1,3)				&	146605.51	&	2.48	$\pm$	0.20	&	-4.84	$\pm$	0.32	&	7.72	$\pm$	0.73	&	0.30	\\
CH$_{3}$OH	&	9(0,9)-8(1,8) A++			&	146618.83	&	0.33	$\pm$	0.04	&	-2.64	$\pm$	0.21	&	3.14	$\pm$	0.49	&	0.10	\\
H$_{2}$CN	&	2(0,2)-1(0,1),F=5/2-5/2		&	146660.74	&	0.43	$\pm$	0.08	&	-1.83	$\pm$	0.43	&	4.46	$\pm$	0.76	&	0.09	\\
CH$_{3}$OCHO		&	26(6,21)-25( 7,18) A		&	146753.39	&	0.23	$\pm$	0.05	&	-3.66	$\pm$	0.31	&	2.86	$\pm$	0.79	&	0.08	\\
$^{13}$CH$_{2}$CHCN		&	13(1,13)-12(0,12)			&	146846.03	&	0.22	$\pm$	0.05	&	-1.41	$\pm$	0.55	&	3.84	$\pm$	0.77	&	0.06	\\
$c$-CH$_{2}$OHCHO	&	v=0  40(5,35)-41(4,38)		&	146916.98	&	0.48	$\pm$	0.08	&	-3.96	$\pm$	0.59	&	5.90	$\pm$	1.08	&	0.08	\\
CS					&	 3-2						&	146969.02	&	12.52	$\pm$	0.13	&	-2.44	$\pm$	0.01	&	4.00	$\pm$	0.05	&	2.94	\\
H$_{2}$CCH$^{13}$CN	&	18(0,18) - 17(1,17)	&	147042.13	&	2.10	$\pm$	0.14	&	-1.00	$\pm$	0.23	&	7.07	$\pm$	0.58	&	0.28	\\

 \hline
\end{tabular}
\end{minipage}
\end{table*}

%%%%%%%%%%%%%%%%%%%%%%%%%%%%%%%%%%%%%%%%%%%%%%%%%%%%%%%%%%%%%%%%%%%%%%%%%%%%%%%%%%%%%%%%%%%%%%%%%%%

\begin{table*}
%\begin{minipage}{200mm}
\caption{G34}
\label{tab:G34}
\centering
\begin{minipage}{200mm}
\begin{tabular}{lrrrrrr}
 \hline 
Molecule & Transition &Frequency & $\int T_{mb}dv$	& $v_{LSR}$		&FWHM & $T_{mb}$\\
		 &			& MHz		& (K km s$^{-1}$) & (km s$^{-1}$)	&(km s$^{-1}$)    &  K\\
\hline

C$_{6}$H	&	23/2 J=61/2-59/2 f							&	84574.59	&	0.13	$\pm$	0.04	&	60.97	$\pm$	0.33	&	2.71	$\pm$	0.91	&	0.05	\\
t-CH$_{3}$CH$_{2}$OH	&	4(2,3)-4(1,4)								&	84595.78	&	0.27	$\pm$	0.06	&	60.04	$\pm$	0.65	&	6.26	$\pm$	1.64	&	0.04	\\
$^{30}$SiO				&	2-1 v=0										&	84745.99	&	0.43	$\pm$	0.07	&	60.90	$\pm$	0.96	&	11.14	$\pm$	1.87	&	0.04	\\
NH$_{2}$CHO				&	4(2,3)-3(2,2) n,t							&	84807.79	&	0.41	$\pm$	0.05	&	58.08	$\pm$	0.37	&	6.01	$\pm$	0.83	&	0.07	\\
O$^{13}$CS				&	7--6										&	84865.15	&	0.22	$\pm$	0.04	&	58.04	$\pm$	0.50	&	4.69	$\pm$	1.01	&	0.05	\\
H$\alpha$				&	H60$\gamma$								&	84914.39	&	0.71	$\pm$	0.09	&	55.64	$\pm$	1.33	&	20.87	$\pm$	2.49	&	0.03	\\
$^{13}$CH$_{3}$OH		&	8(0,8)-7(1,7) A++							&	84970.23	&	0.63	$\pm$	0.06	&	59.76	$\pm$	0.37	&	7.42	$\pm$	0.80	&	0.08	\\
NH$_{2}$CHO				&	4(2,2)-3(2,1)								&	85093.26	&	0.13	$\pm$	0.04	&	59.38	$\pm$	0.42	&	3.29	$\pm$	1.14	&	0.04	\\
OCS						&	7-6											&	85139.10	&	1.02	$\pm$	0.05	&	59.28	$\pm$	0.12	&	5.39	$\pm$	0.29	&	0.18	\\
HC$^{18}$O$^{+}$		&	1-0											&	85162.22	&	0.65	$\pm$	0.04	&	59.02	$\pm$	0.12	&	4.10	$\pm$	0.32	&	0.15	\\
C$^{13}$CH	&	1-0 3/2-1/2F=1,0.5-0,0.5					&	85247.65	&	0.13	$\pm$	0.04	&	59.61	$\pm$	0.43	&	2.65	$\pm$	0.74	&	0.04	\\
C$^{13}$CH				&	1-0 3/2-1/2 F=1,1.5-0,0.5					&	85256.96	&	0.18	$\pm$	0.05	&	58.90	$\pm$	0.96	&	5.89	$\pm$	2.11	&	0.03	\\
t-CH$_{3}$CH$_{2}$OH		&	6(0,6)-5(1,5)										&	85265.50	&	0.32	$\pm$	0.05	&	59.25	$\pm$	0.40	&	5.52	$\pm$	1.07	&	0.06	\\
$c$-C$_{3}$H$_{2}$		&	2(1,2)-1(0,1)								&	85338.90	&	1.74	$\pm$	0.05	&	59.30	$\pm$	0.07	&	5.41	$\pm$	0.20	&	0.30	\\
HCS$^{+}$				&	2-1											&	85347.86	&	0.84	$\pm$	0.05	&	58.99	$\pm$	0.13	&	5.15	$\pm$	0.33	&	0.15	\\
CH$_{3}$CCH				&	5(3)-4(3)									&	85442.60	&	0.23	$\pm$	0.19	&	58.88	$\pm$	1.58	&	3.65	$\pm$	3.15	&	0.06	\\
CH$_{3}$CCH				&	5(2)-4(2)									&	85450.76	&	0.62	$\pm$	0.05	&	58.85	$\pm$	0.16	&	4.44	$\pm$	0.43	&	0.13	\\
CH$_{3}$CCH				&	5(1)-4(1)									&	85455.66	&	1.40	$\pm$	0.06	&	58.96	$\pm$	0.08	&	4.35	$\pm$	0.20	&	0.30	\\
CH$_{3}$CCH				&	5(0)-4(0)									&	85457.29	&	1.64	$\pm$	0.06	&	58.97	$\pm$	0.07	&	4.24	$\pm$	0.17	&	0.36	\\
CH$_{3}$OH				&	6(-2,5)-7(-1,7) E							&	85568.07	&	0.16	$\pm$	0.04	&	59.06	$\pm$	0.37	&	3.68	$\pm$	1.00	&	0.04	\\
C$_{4}$H	&	9(10,10)-8(9,9)	&	85634.02	&	0.26	$\pm$	0.05	&	60.08	$\pm$	0.54	&	6.19	$\pm$	1.48	&	0.04	\\
CH$_{3}$OCHO			&	4(2,3)-3(1.2) A								&	85655.80	&	0.16	$\pm$	0.04	&	58.71	$\pm$	0.71	&	6.01	$\pm$	1.91	&	0.03	\\
H$\alpha$				&	H42$\alpha$									&	85688.39	&	3.38	$\pm$	0.09	&	54.48	$\pm$	0.39	&	27.11	$\pm$	0.98	&	0.12	\\
$^{29}$SiO				&	2-1 v=0										&	85759.18	&	0.19	$\pm$	0.04	&	58.71	$\pm$	0.33	&	3.48	$\pm$	0.72	&	0.05	\\
CH$_{3}$OCHO			&	7(6,2)-6(6,1) A+E							&	85926.50	&	0.47	$\pm$	0.06	&	59.52	$\pm$	0.45	&	7.85	$\pm$	1.36	&	0.06	\\
CH$_{3}$OCHO			&	7(5,2)-6(5,1) E								&	86021.00	&	0.26	$\pm$	0.04	&	59.27	$\pm$	0.44	&	5.68	$\pm$	1.05	&	0.04	\\
CH$_{3}$OCHO			&	7(5,3)-6(5,2) E								&	86027.67	&	0.06	$\pm$	0.02	&	60.30	$\pm$	0.38	&	1.82	$\pm$	0.64	&	0.04	\\
CH$_{3}$OCHO			&	7(5,2)-6(5,1) A								&	86030.21	&	0.77	$\pm$	0.11	&	60.56	$\pm$	1.08	&	16.09	$\pm$	3.14	&	0.05	\\
HC$^{15}$N				&	1-0											&	86054.96	&	1.08	$\pm$	0.04	&	58.96	$\pm$	0.07	&	4.00	$\pm$	0.19	&	0.25	\\
SO	&	N,J=2,2-1,1									&	86093.98	&	1.54	$\pm$	0.05	&	59.39	$\pm$	0.07	&	5.23	$\pm$	0.20	&	0.28	\\
CCS	&	N,J=7,6-6,5									&	86181.41	&	0.21	$\pm$	0.04	&	59.52	$\pm$	0.39	&	4.58	$\pm$	1.19	&	0.05	\\
CH$_{3}$OCHO			&	7(4,4)-6(4,3) A								&	86210.07	&	0.41	$\pm$	0.07	&	59.24	$\pm$	0.60	&	6.98	$\pm$	1.25	&	0.06	\\
CH$_{3}$OCH$_{3}$		&	7(4,3)-6(4,2) E								&	86223.54	&	0.66	$\pm$	0.05	&	58.04	$\pm$	0.29	&	7.37	$\pm$	0.69	&	0.08	\\
CH$_{3}$OCHO			&	7(4,3)-6(4,2) A								&	86250.57	&	0.26	$\pm$	0.04	&	59.33	$\pm$	0.50	&	5.42	$\pm$	0.91	&	0.04	\\
CH$_{3}$OCHO			&	7(3,5)-6(3,4) A								&	86265.82	&	0.36	$\pm$	0.06	&	60.28	$\pm$	0.38	&	5.36	$\pm$	1.16	&	0.06	\\
CH$_{3}$OCHO			&	7(3,5)-6(3,4) E								&	86268.65	&	0.54	$\pm$	0.06	&	58.58	$\pm$	0.38	&	7.55	$\pm$	1.06	&	0.07	\\
H$^{13}$CN				&	1-0 F=1-1									&	86338.73	&	2.46	$\pm$	0.08	&	59.20	$\pm$	0.06	&	4.18	$\pm$	0.14	&	0.55	\\
H$^{13}$CN				&	1-0 F=2-1									&	86340.17	&	4.45	$\pm$	0.14	&	58.95	$\pm$	0.67	&	4.40	$\pm$	0.67	&	0.95	\\
H$^{13}$CN				&	1-0 F=0-1									&	86342.25	&	1.51	$\pm$	0.11	&	59.69	$\pm$	0.22	&	6.49	$\pm$	0.54	&	0.22	\\
t-HCOOH	&	4(1,4)-3(1,3)								&	86546.19	&	0.08	$\pm$	0.02	&	58.68	$\pm$	0.26	&	1.56	$\pm$	0.60	&	0.05	\\
CH$_{3}$OH	&	7(2,6)-6(3,3) A--							&	86615.60	&	0.36	$\pm$	0.06	&	58.66	$\pm$	0.87	&	10.42	$\pm$	1.81	&	0.03	\\
SO$_{2}$	&	8(3,5)-9(2,8)								&	86639.09	&	0.42	$\pm$	0.04	&	58.03	$\pm$	0.24	&	4.80	$\pm$	0.59	&	0.08	\\
HCO	&	(0,1)-0(0,0) 3/2-1/2 F=2-1					&	86670.82	&	0.25	$\pm$	0.04	&	60.01	$\pm$	0.32	&	3.98	$\pm$	0.77	&	0.06	\\
CCCS					&	15-14										&	86708.37	&	0.16	$\pm$	0.04	&	59.08	$\pm$	0.42	&	3.74	$\pm$	1.05	&	0.04	\\
H$^{13}$CO$^{+}$		&	1-0											&	86754.28	&	4.57	$\pm$	0.04	&	58.73	$\pm$	0.01	&	4.49	$\pm$	0.04	&	0.98	\\
CH$_{3}$CH$_{2}$CN		&	10(1,10)-9(1,9)								&	86819.84	&	0.45	$\pm$	0.05	&	57.84	$\pm$	0.32	&	5.59	$\pm$	0.92	&	0.08	\\
SiO						&	2-1 v=0										&	86846.99	&	2.96	$\pm$	0.07	&	59.55	$\pm$	0.07	&	7.21	$\pm$	0.23	&	0.38	\\
t-HCOOD					&	4(2,3)-3(2,2)								&	86895.46	&	0.14	$\pm$	0.04	&	59.03	$\pm$	0.71	&	4.60	$\pm$	1.30	&	0.03	\\
CH$_{3}$OH	&	7(2,5)-6(3,4) A++							&	86902.94	&	0.40	$\pm$	0.05	&	57.96	$\pm$	0.66	&	9.45	$\pm$	1.25	&	0.04	\\
HC$^{17}$O$^{+}$		&	1-0 F=7/2-5/2								&	87057.25	&	0.30	$\pm$	0.05	&	56.89	$\pm$	0.52	&	6.91	$\pm$	1.27	&	0.04	\\
HN$^{13}$C				&	1-0 F=2-1									&	87090.85	&	2.15	$\pm$	0.04	&	59.08	$\pm$	0.03	&	4.03	$\pm$	0.09	&	0.50	\\
C$_{4}$H	&	23/2 J=19/2-17/2 1v7e						&	87142.30	&	0.50	$\pm$	0.06	&	56.51	$\pm$	0.54	&	8.91	$\pm$	1.65	&	0.05	\\
unidentified	&		&	87161.31	&	1.06	$\pm$	0.02	&	59.19	$\pm$	0.21	&	1.81	$\pm$	0.52	&	0.05	\\
C$_{2}$H	&	1-0 3/2-1/2 F=1-1							&	87284.15	&	1.04	$\pm$	0.05	&	59.08	$\pm$	0.09	&	4.50	$\pm$	0.25	&	0.22	\\
C$_{2}$H	&	1-0 3/2-1/2 F=2-1							&	87316.92	&	7.03	$\pm$	0.05	&	58.71	$\pm$	0.01	&	4.36	$\pm$	0.03	&	1.51	\\
C$_{2}$H	&	1-0 3/2-1/2 F=1-0							&	87328.62	&	4.01	$\pm$	0.05	&	58.91	$\pm$	0.02	&	4.37	$\pm$	0.06	&	0.86	\\
C$_{2}$H	&	1-0 1/2-1/2 F=1-1							&	87402.00	&	4.01	$\pm$	0.04	&	58.80	$\pm$	0.02	&	4.34	$\pm$	0.06	&	0.87	\\
C$_{2}$H	&	1-0 1/2-1/2 F=0-1							&	87407.16	&	1.87	$\pm$	0.04	&	58.89	$\pm$	0.04	&	4.21	$\pm$	0.11	&	0.42	\\
C$_{2}$H	&	1-0 1/2-1/2 F=1-0							&	87446.51	&	1.17	$\pm$	0.04	&	58.94	$\pm$	0.08	&	4.76	$\pm$	0.20	&	0.23	\\
HNCO					&	4(1,4)-3(1,3)								&	87597.33	&	0.47	$\pm$	0.05	&	57.60	$\pm$	0.34	&	6.21	$\pm$	0.94	&	0.07	\\
t-CH$_{3}$CH$_{2}$OH	&	5(2,4)-5(1,5)								&	87716.02	&	0.22	$\pm$	0.04	&	57.30	$\pm$	0.55	&	5.41	$\pm$	1.04	&	0.04	\\
NH$_{2}$CHO				&	4(1,3)-3(1,2)								&	87848.87	&	0.40	$\pm$	0.05	&	58.11	$\pm$	0.58	&	8.27	$\pm$	1.12	&	0.05	\\

 \hline
\end{tabular}
\end{minipage}
\end{table*}

\begin{table*}
%\begin{minipage}{200mm}
\contcaption{}
%\label{tab:G34}
\centering
\begin{minipage}{200mm}
\begin{tabular}{lrrrrrr}
 \hline 
Molecule & Transition &Frequency & $\int T_{mb}dv$	& $v_{LSR}$		&FWHM & $T_{mb}$\\
		 &			& MHz		& (K km s$^{-1}$) & (km s$^{-1}$)	&(km s$^{-1}$)    &  K\\
\hline

HC$_{5}$N	&	33-32										&	87863.63	&	0.22	$\pm$	0.04	&	58.64	$\pm$	0.34	&	4.07	$\pm$	1.20	&	0.05	\\
H$_{2}$CCCHCN			&	17( 4,14)-16( 4,13)							&	87894.36	&	0.22	$\pm$	0.04	&	55.45	$\pm$	0.56	&	5.23	$\pm$	1.03	&	0.04	\\
HNCO	&	4(2,2)-3(2,1)								&	87898.62	&	0.25	$\pm$	0.05	&	47.34	$\pm$	0.65	&	6.19	$\pm$	1.15	&	0.04	\\
unidentified	&		&	87911.49	&	0.15	$\pm$	0.05	&	57.73	$\pm$	0.43	&	2.87	$\pm$	1.26	&	0.05	\\
HNCO					&	4(0,4)-3(0,3)								&	87925.23	&	1.62	$\pm$	0.04	&	58.84	$\pm$	0.06	&	4.91	$\pm$	0.15	&	0.31	\\
CH$_{3}$OCH$_{3}$		&	2(2,0)-2(1,1) EE							&	88226.72	&	0.15	$\pm$	0.05	&	58.86	$\pm$	0.65	&	3.93	$\pm$	1.36	&	0.04	\\
CH$_{3}$CH$_{2}$CHO		&	30( 6,24)- 30( 6,25)						&	88628.80	&	2.23	$\pm$	0.16	&	59.44	$\pm$	0.16	&	4.53	$\pm$	0.44	&	0.46	\\
HCN		&	1-0 F=1-1									&	88630.41	&	0.80	$\pm$	0.08	&	58.49	$\pm$	0.07	&	1.43	$\pm$	0.17	&	0.53	\\
HCN	&	1-0 F=2-11									&	88631.84	&	6.02	$\pm$	0.11	&	56.92	$\pm$	0.03	&	3.39	$\pm$	0.06	&	1.67	\\
HCN	&	1-0 F=0-1									&	88633.93	&	5.18	$\pm$	0.11	&	57.15	$\pm$	0.03	&	3.06	$\pm$	0.08	&	1.59	\\
H$^{15}$NC				&	1-0											&	88865.69	&	0.21	$\pm$	0.05	&	58.87	$\pm$	0.35	&	2.81	$\pm$	0.95	&	0.07	\\
CH$_{3}$OH				&	15(3,12)-14(4,11) A--						&	88939.99	&	0.46	$\pm$	0.05	&	58.59	$\pm$	0.36	&	5.74	$\pm$	0.82	&	0.08	\\
HSCH$_{2}$CN	&	15(6,10)-14(6,9)	&	88966.54	&	0.30	$\pm$	0.06	&	57.93	$\pm$	0.54	&	5.37	$\pm$	0.94	&	0.05	\\
HSCH$_{2}$CN	&	15(4,11)-14(4,10)	&	88990.96	&	0.16	$\pm$	0.04	&	57.87	$\pm$	0.68	&	5.14	$\pm$	1.39	&	0.03	\\
unidentified	&		&	88999.14	&	0.18	$\pm$	0.04	&	58.77	$\pm$	0.94	&	6.49	$\pm$	2.15	&	0.26	\\
unidentified	&		&	89007.42	&	0.07	$\pm$	0.02	&	57.29	$\pm$	0.35	&	1.87	$\pm$	0.99	&	0.04	\\
HSCH$_{2}$CN	&	15(10,6)-14(10,5)	&	89026.90	&	0.18	$\pm$	0.40	&	56.48	$\pm$	0.58	&	4.66	$\pm$	0.88	&	0.04	\\
C$_{3}$N	&	9(10,9)-8(9,9)	&	89048.27	&	0.13	$\pm$	0.03	&	55.92	$\pm$	0.80	&	5.24	$\pm$	1.36	&	0.03	\\
C$_{3}$N	&	9(9,8)-8(8,8)	&	89065.02	&	0.13	$\pm$	0.03	&	56.01	$\pm$	0.42	&	3.61	$\pm$	1.25	&	0.03	\\
HSCH$_{2}$CN	&	15(12,3)-14(12,2)	&	89080.78	&	0.21	$\pm$	0.04	&	60.25	$\pm$	0.70	&	6.24	$\pm$	1.39	&	0.03	\\
c-C$_{3}$D$_{2}$	&	5(4,2)-5(3,3)	&	89316.91	&	0.62	$\pm$	0.07	&	59.85	$\pm$	0.88	&	12.00	$\pm$	1.75	&	0.04	\\
CH$_{3}$CH$_{2}$CN		&	10(5)-9(5)									&	89568.10	&	0.45	$\pm$	0.06	&	56.97	$\pm$	0.37	&	5.64	$\pm$	0.87	&	0.08	\\
CH$_{3}$CH$_{2}$CN		&	10(8)-9(8)									&	89573.05	&	0.26	$\pm$	0.04	&	57.73	$\pm$	0.36	&	4.04	$\pm$	0.73	&	0.06	\\
CH$_{3}$CH$_{2}$CN		&	10(4,6)-9(4,5)								&	89591.01	&	0.79	$\pm$	0.06	&	59.28	$\pm$	0.27	&	6.58	$\pm$	0.69	&	0.11	\\
CH$_{3}$CH$_{2}$CN		&	10(3,8)-9(3,7)								&	89628.44	&	0.45	$\pm$	0.05	&	57.15	$\pm$	0.31	&	5.39	$\pm$	0.67	&	0.08	\\
CH$_{3}$CH$_{2}$CN		&	10(3,7)-9(3,6)								&	89684.71	&	0.18	$\pm$	0.04	&	57.22	$\pm$	0.50	&	3.79	$\pm$	1.01	&	0.05	\\
(CH$_{3}$)$_{2}$CO		&	9(7,3)-8(8,0) AE							&	89684.71	&	0.12	$\pm$	0.04	&	67.65	$\pm$	0.83	&	4.18	$\pm$	1.59	&	0.03	\\
unidentified	&		&	89970.13	&	0.18	$\pm$	0.04	&	60.42	$\pm$	0.42	&	4.02	$\pm$	0.93	&	0.04	\\
t-CH$_{3}$CH$_{2}$OH	&	4(1,4)-3(0,3)								&	90117.59	&	0.19	$\pm$	0.02	&	58.53	$\pm$	0.31	&	3.82	$\pm$	0.66	&	0.05	\\
CH$_{3}$OCHO			&	7(2,5)-6(2,4) E								&	90145.63	&	0.28	$\pm$	0.04	&	58.16	$\pm$	0.47	&	6.46	$\pm$	0.92	&	0.04	\\
CH$_{3}$OCHO			&	7(2,5)-6(2,4) A								&	90156.51	&	0.25	$\pm$	0.04	&	58.60	$\pm$	0.37	&	4.92	$\pm$	0.90	&	0.05	\\
CH$_{3}$OCHO			&	V=0 8( 0, 8)- 7( 0, 7) E					&	90227.66	&	0.56	$\pm$	0.05	&	56.35	$\pm$	0.45	&	10.49	$\pm$	0.91	&	0.05	\\
$^{15}$NNH$^{+}$		&	1-0											&	90263.83	&	0.19	$\pm$	0.02	&	60.06	$\pm$	0.33	&	4.47	$\pm$	0.68	&	0.04	\\
CH$_{3}$CH$_{2}$CN		&	10(2,8)-9(2,7)								&	90453.35	&	0.26	$\pm$	0.05	&	55.83	$\pm$	0.78	&	8.29	$\pm$	2.12	&	0.03	\\
HC$_{5}$N				&	34-33										&	90525.89	&	0.14	$\pm$	0.02	&	58.40	$\pm$	0.42	&	3.71	$\pm$	0.87	&	0.04	\\
SO$_{2}$				&	25(3,23)-24(4,20)							&	90548.16	&	0.13	$\pm$	0.04	&	57.83	$\pm$	0.90	&	6.22	$\pm$	1.77	&	0.02	\\
HC$^{13}$CCN			&	10-9										&	90593.05	&	0.23	$\pm$	0.04	&	58.75	$\pm$	0.41	&	4.98	$\pm$	0.82	&	0.05	\\
HCC$^{13}$CN			&	10-9										&	90601.79	&	0.13	$\pm$	0.02	&	54.65	$\pm$	0.61	&	4.25	$\pm$	1.37	&	0.02	\\
HNC		&	1-0 F=0-1									&	90663.45	&	1.54	$\pm$	0.04	&	62.69	$\pm$	0.03	&	3.37	$\pm$	0.08	&	0.44	\\
HNC	&	1-0,F=2-1									&	90663.57	&	15.39	$\pm$	0.04	&	56.83	$\pm$	0.00	&	3.85	$\pm$	0.01	&	3.72	\\
CCS	&	N,J=7,7-6,6									&	90686.38	&	0.23	$\pm$	0.05	&	58.68	$\pm$	0.54	&	6.01	$\pm$	1.75	&	0.04	\\
CH$_{3}$OH				&	20(-3,17)-19(-2,17) Et						&	90812.39	&	0.09	$\pm$	0.02	&	54.72	$\pm$	0.64	&	3.97	$\pm$	1.33	&	0.02	\\
CH$_{2}$CH$^{13}$CN		&	79( 4,75)-78( 6,72)							&	90920.15	&	0.07	$\pm$	0.02	&	56.59	$\pm$	0.34	&	2.04	$\pm$	0.94	&	0.04	\\
$^{13}$C$^{34}$S		&	2-1											&	90925.99	&	0.21	$\pm$	0.05	&	57.90	$\pm$	0.57	&	5.85	$\pm$	1.60	&	0.04	\\
CH$_{3}$OCH$_{3}$		&	6(0,6)-5(1,5) AA							&	90937.50	&	0.27	$\pm$	0.06	&	54.87	$\pm$	0.93	&	8.41	$\pm$	2.62	&	0.03	\\
HCCCN					&	10-9										&	90978.98	&	6.19	$\pm$	0.04	&	58.30	$\pm$	0.01	&	4.14	$\pm$	0.03	&	1.40	\\
HDCS					&	3(1,3)-2(1,2)								&	91171.03	&	0.08	$\pm$	0.01	&	58.23	$\pm$	0.19	&	1.55	$\pm$	0.50	&	0.05	\\
H$_{2}$CN	&	13(2,11)-13(2,12),F=27/2-27/2,(n=11-11)			&	91201.96	&	0.09	$\pm$	0.02	&	62.19	$\pm$	0.42	&	2.67	$\pm$	0.79	&	0.03	\\
H$_{2}$CN				&	13(2,11)-13(2,12),F=21/2-21/2,(n=23-23)		&	91203.59	&	0.59	$\pm$	0.07	&	59.35	$\pm$	0.36	&	6.54	$\pm$	0.94	&	0.09	\\
H$_{2}$CN	&	13(2,11)-13(2,12),F=23/2-23/2,(n=22-22)		&	91209.19	&	0.23	$\pm$	0.11	&	68.64	$\pm$	1.12	&	7.30	$\pm$	3.87	&	0.03	\\
HCCCN					&	10-9 1v7 e=1f								&	91333.37	&	0.07	$\pm$	0.02	&	58.08	$\pm$	0.52	&	2.69	$\pm$	0.95	&	0.03	\\
SO$_{2}$	&	18(5,13)-19(4,16)							&	91550.44	&	0.08	$\pm$	0.02	&	58.16	$\pm$	0.54	&	3.31	$\pm$	1.09	&	0.03	\\
CH$_{3}^{13}$CN			&	5(0)-4(0)									&	91941.60	&	0.38	$\pm$	0.07	&	59.28	$\pm$	1.40	&	16.66	$\pm$	4.80	&	0.02	\\
CH$_{3}$CN	&	5(4)-4(4) F=6-5								&	91959.02	&	0.19	$\pm$	0.05	&	59.69	$\pm$	0.74	&	5.17	$\pm$	1.67	&	0.03	\\
CH$_{3}$CN				&	5(3)-4(3) F=4-3								&	91971.46	&	0.72	$\pm$	0.05	&	59.45	$\pm$	0.22	&	6.73	$\pm$	0.50	&	0.10	\\
CH$_{3}$CN	&	5(2)-4(2) F=6-5								&	91979.99	&	0.73	$\pm$	0.05	&	58.18	$\pm$	0.18	&	4.84	$\pm$	0.44	&	0.14	\\
CH$_{3}$CN	&	5(1)-4(1)									&	91985.31	&	1.34	$\pm$	0.07	&	58.39	$\pm$	0.10	&	4.47	$\pm$	0.26	&	0.28	\\
CH$_{3}$CN	&	5(0)-4(0)									&	91987.08	&	1.57	$\pm$	0.06	&	58.10	$\pm$	0.08	&	4.80	$\pm$	0.22	&	0.31	\\
H$\alpha$				&	H41$\alpha$									&	92034.43	&	2.98	$\pm$	0.08	&	54.40	$\pm$	0.37	&	27.66	$\pm$	0.90	&	0.10	\\

\hline
\end{tabular}
\end{minipage}
\end{table*}

\begin{table*}
%\begin{minipage}{200mm}
\contcaption{}
%\label{tab:G34}
\centering
\begin{minipage}{200mm}
\begin{tabular}{lrrrrrr}
 \hline 
Molecule & Transition &Frequency & $\int T_{mb}dv$	& $v_{LSR}$		&FWHM & $T_{mb}$\\
		 &			& MHz		& (K km s$^{-1}$) & (km s$^{-1}$)	&(km s$^{-1}$)    &  K\\
\hline

unidentified	&		&	143778.87	&	0.27	$\pm$	0.03	&	56.92	$\pm$	0.24	&	3.68	$\pm$	0.55	&	0.07	\\
CH$_{3}$OCHO			&	V=1 46(15,32)-67(16,29)A					&	143833.84	&	0.22	$\pm$	0.04	&	56.23	$\pm$	0.45	&	4.00	$\pm$	0.86	&	0.05	\\
unidentified	&		&	143858.80	&	0.59	$\pm$	0.06	&	57.72	$\pm$	0.27	&	5.81	$\pm$	0.67	&	0.10	\\
CH$_{3}$OH	&	3(1,3)-2(1,2) A++							&	143865.79	&	1.27	$\pm$	0.05	&	58.82	$\pm$	0.09	&	4.90	$\pm$	0.23	&	0.25	\\
DCO$^{+}$	&	2-1											&	144077.28	&	0.22	$\pm$	0.04	&	58.40	$\pm$	0.26	&	3.14	$\pm$	0.75	&	0.07	\\
$c$-C$_{3}$H$_{4}$		&	8(3,6)-8(1,7)								&	144178.68	&	0.38	$\pm$	0.05	&	59.53	$\pm$	0.32	&	4.48	$\pm$	0.82	&	0.08	\\
CCD	&	2-1 J=5/2-3/2F=5/2-3/2						&	144243.05	&	0.20	$\pm$	0.04	&	58.67	$\pm$	0.65	&	5.38	$\pm$	0.97	&	0.04	\\
H$_{2}$NCH$_{2}$CN	&	33(4,29) - 33(3,30)	&	144258.27 	&	1.76 	$\pm$	0.06 	&	61.25 	$\pm$	0.07 	&	4.39 	$\pm$	0.18 	&	0.38 	\\
CH$_{3}$OH	&	3(0,3)-2(0,2) A++  Vt=2						&	144571.97	&	0.70	$\pm$	0.06	&	60.71	$\pm$	0.50	&	8.99	$\pm$	0.97	&	0.07	\\
CH$_{3}$OH	&	3(-1,2)-2(-1,1) Et=2						&	144583.91	&	0.83	$\pm$	0.08	&	61.16	$\pm$	0.46	&	8.85	$\pm$	1.00	&	0.09	\\
CH$_{3}$OH	&	3(1,3)-2(1,2) A++  t=1						&	144589.85	&	0.83	$\pm$	0.10	&	58.00	$\pm$	0.42	&	6.83	$\pm$	0.99	&	0.12	\\
C$_{3}^{13}$CCH	&	30-29	&	144524.63	&	0.22	$\pm$	0.04	&	58.44	$\pm$	0.33	&	3.94	$\pm$	0.56	&	0.05	\\
CH$_{3}$OH	&	3(0,3)-2(0,2) A++  Vt=2		&	144571.97	&	0.16	$\pm$	0.04	&	61.39	$\pm$	1.41	&	6.67	$\pm$	2.22	&	0.03	\\
C$^{34}$S				&	3--2										&	144617.10	&	3.63	$\pm$	0.04	&	58.43	$\pm$	0.02	&	4.26	$\pm$	0.06	&	0.80	\\
CH$_{3}$C$_{5}$N		&	93(6)-92(6),F=93-92							&	144657.80	&	0.73	$\pm$	0.05	&	57.03	$\pm$	0.24	&	6.08	$\pm$	0.49	&	0.11	\\
CH$_{3}$OH	&	3(-2,1)-2(-2,0) Et=1						&	144728.77	&	1.20	$\pm$	0.08	&	59.52	$\pm$	0.20	&	6.68	$\pm$	0.56	&	0.17	\\
CH$_{3}$OH	&	3(2,2)-2(2,1) E  t=1						&	144733.24	&	2.03	$\pm$	0.05	&	55.54	$\pm$	0.12	&	8.26	$\pm$	0.29	&	0.23	\\
CH$_{3}$OH	&	3(-1,2)-2(-1,1) Et=1						&	144750.24	&	0.47	$\pm$	0.05	&	58.84	$\pm$	0.33	&	5.99	$\pm$	0.72	&	0.08	\\
CH$_{3}$OH	&	3(0,3)-2(0,2) A++  t=1						&	144768.17	&	0.37	$\pm$	0.05	&	58.39	$\pm$	0.34	&	4.33	$\pm$	0.82	&	0.08	\\
DCN	&	2-1 F1=2-2									&	144826.57	&	0.96	$\pm$	0.05	&	55.80	$\pm$	0.14	&	5.48	$\pm$	0.44	&	0.17	\\
CH$_{3}$OH				&	6(3,3)-6(2,4) EE							&	144858.98	&	0.73	$\pm$	0.08	&	57.45	$\pm$	0.55	&	10.55	$\pm$	1.37	&	0.06	\\
CH$_{3}$OH	&	3(1,2)-2(1,1) A--t=1						&	144878.57	&	0.61	$\pm$	0.06	&	58.28	$\pm$	0.51	&	8.57	$\pm$	0.99	&	0.07	\\
$c$-C$_{3}$H$_{2}$		&	3(1,2)-2(2,1)				&	145089.62	&	1.36	$\pm$	0.29	&	58.49	$\pm$	0.40	&	5.59	$\pm$	0.40	&	0.22	\\
CH$_{3}$OH	&	3(0,3)-2(0,2) E								&	145093.76	&	2.54	$\pm$	0.05	&	58.58	$\pm$	0.04	&	4.69	$\pm$	0.13	&	0.51	\\
CH$_{3}$OH	&	3(-1,3)-2(-1,2) E							&	145097.44	&	6.28	$\pm$	0.05	&	58.49	$\pm$	0.00	&	4.50	$\pm$	0.04	&	1.31	\\
CH$_{3}$OH				&	3(0,3)-2(0,2) A++							&	145103.19	&	7.19	$\pm$	0.04	&	58.55	$\pm$	0.01	&	4.44	$\pm$	0.03	&	1.52	\\
CH$_{3}$OH				&	3(2,2)-2(2,1) A--							&	145124.33	&	1.49	$\pm$	0.90	&	54.76	$\pm$	1.55	&	5.82	$\pm$	5.02	&	0.24	\\
CH$_{3}$OH				&	3(1,2)-2(1,1) E								&	145131.87	&	1.50	$\pm$	0.83	&	58.36	$\pm$	1.44	&	5.76	$\pm$	4.09	&	0.25	\\
SiC$_{2}$	&	6(2,4)-5(2,3)								&	145325.84	&	0.59	$\pm$	0.05	&	61.83	$\pm$	0.29	&	6.20	$\pm$	0.59	&	0.09	\\
c-C$_{3}$HCN	&	19(3,16)19 - 19(2,17)19	&	145373.47	&	0.13	$\pm$	0.03	&	57.05	$\pm$	0.50	&	3.56	$\pm$	1.01	&	0.03	\\
$^{13}$CH$_{2}$(OH)CHO	&	12(3,9)-11(3,8)	&	145362.26	&	0.13	$\pm$	0.03	&	58.84	$\pm$	0.37	&	3.23	$\pm$	0.75	&	0.04	\\
ONCN					&	14( 0,14)-13( 0,13)							&	145397.59	&	0.88	$\pm$	0.03	&	55.84	$\pm$	0.06	&	2.95	$\pm$	0.12	&	0.28	\\
CH$_{3}$CH$_{2}$CN		&	16(1,15)-15(1,14)							&	145418.03	&	0.31	$\pm$	0.10	&	58.01	$\pm$	0.78	&	4.62	$\pm$	1.58	&	0.06	\\
NCCONH$_{2}$	&	21(4,17)-20(5,16)	&	145550.10	&	0.43	$\pm$	0.06	&	57.75	$\pm$	0.39	&	5.95	$\pm$	0.96	&	0.07	\\
HCCCN					&	16-15 1v7 =1e								&	145560.95	&	1.75	$\pm$	0.04	&	58.60	$\pm$	0.05	&	4.80	$\pm$	0.14	&	0.34	\\
n-C$_{3}$H$_{7}$CN	&	9(7,3)-8(6,3)	&	145700.18	&	0.21	$\pm$	0.03	&	56.81	$\pm$	0.33	&	4.16	$\pm$	0.79	&	0.05	\\
unidentified	&		&	145723.50	&	0.32	$\pm$	0.04	&	57.49	$\pm$	0.50	&	7.68	$\pm$	1.15	&	0.04	\\
unidentified	&		&	145745.70	&	0.11	$\pm$	0.03	&	57.29	$\pm$	0.78	&	3.86	$\pm$	1.42	&	0.03	\\
OCS		&	12--11										&	145946.81	&	0.66	$\pm$	0.05	&	58.69	$\pm$	0.20	&	5.50	$\pm$	0.50	&	0.11	\\
CH$_{3}$CH$_{2}$CN		&	16(2,14)-15(2,13)							&	146120.04	&	0.37	$\pm$	0.05	&	56.98	$\pm$	0.34	&	4.89	$\pm$	0.73	&	0.07	\\
HCCCN					&	J=16-15, l= 1f								&	146127.53	&	0.66	$\pm$	0.06	&	58.00	$\pm$	0.38	&	7.13	$\pm$	0.90	&	0.09	\\
CH$_{3}$OH				&	20( 6, 15)- 21( 5, 16) - -					&	146286.58	&	1.04	$\pm$	0.08	&	59.60	$\pm$	0.55	&	12.92	$\pm$	1.16	&	0.08	\\
CH$_{3}$OH				&	3(1,2)-2(1,1) A--							&	146368.34	&	1.78	$\pm$	0.05	&	58.28	$\pm$	0.07	&	5.23	$\pm$	0.22	&	0.32	\\
SO$_{2}$				&	4(2,2)-4(1,3)								&	146605.51	&	0.41	$\pm$	0.05	&	57.87	$\pm$	0.29	&	4.78	$\pm$	0.60	&	0.08	\\
CH$_{3}$OH	&	14(1,14)-13(2,11) A++						&	146617.41	&	1.21	$\pm$	0.08	&	56.09	$\pm$	0.19	&	6.10	$\pm$	0.49	&	0.19	\\
H$_{2}^{13}$CO			&	2(1,1)-1(1,0)								&	146635.67	&	0.50	$\pm$	0.06	&	58.82	$\pm$	0.33	&	5.80	$\pm$	0.99	&	0.08	\\
CH$_{2}$N	&	2(0,2)-1(0,1) 7/2-5/2 9/2-7/2				&	146675.06	&	0.20	$\pm$	0.04	&	52.20	$\pm$	0.56	&	4.48	$\pm$	0.94	&	0.25	\\
CS						&	3-2											&	146969.02	&	19.62	$\pm$	0.05	&	57.63	$\pm$	0.00	&	5.55	$\pm$	0.02	&	3.32	\\

\hline
\end{tabular}
\end{minipage}
\end{table*} 

%%%%%%%%%%%%%%%%%%%%%%%%%%%%%%%%%%%%%%%%%%%%%%%%%%%%%%%%%%%%%%%%%%%%%%%%%%%%%%%%%%%%%%%%%%%%%%%%%%%

\begin{table*}
%\begin{minipage}{200mm}
\caption{S76E}
\label{tab:S76E}
\centering
\begin{minipage}{200mm}
\begin{tabular}{lrrrrrr}
 \hline 
Molecule & Transition &Frequency & $\int T_{mb}dv$	& $v_{LSR}$		&FWHM & $T_{mb}$\\
		 &			& MHz		& (K km s$^{-1}$) & (km s$^{-1}$)	&(km s$^{-1}$)    &  K\\
\hline
OCS		&	7-6								&	85139.10	&	0.15	$\pm$	0.01	&	33.69	$\pm$	0.13	&	2.40	$\pm$	0.33	&	0.06	\\
HC$^{18}$O$^{+}$	&	1-0								&	85162.22	&	0.14	$\pm$	0.01	&	33.83	$\pm$	0.14	&	2.43	$\pm$	0.30	&	0.06	\\
HC$_{5}$N	&	32-31							&	85201.34	&	0.11	$\pm$	0.01	&	33.58	$\pm$	0.31	&	2.81	$\pm$	0.71	&	0.04	\\
$c$-C$_{3}$H$_{2}$	&	2(1,2)-1(0,1)					&	85338.89	&	0.79	$\pm$	0.01	&	34.11	$\pm$	0.03	&	2.66	$\pm$	0.08	&	0.28	\\
HCS$^{+}$			&	2-1								&	85347.89	&	0.32	$\pm$	0.01	&	33.93	$\pm$	0.08	&	2.67	$\pm$	0.22	&	0.11	\\
CH$_{3}$CCH			&	5(3)-4(3)						&	85442.60	&	0.14	$\pm$	0.02	&	32.92	$\pm$	0.27	&	3.35	$\pm$	0.60	&	0.04	\\
CH$_{3}$CCH			&	5(2)-4(2)						&	85450.76	&	0.16	$\pm$	0.01	&	33.66	$\pm$	0.11	&	2.50	$\pm$	0.26	&	0.07	\\
CH$_{3}$CCH			&	5(1)-4(1)						&	85455.66	&	0.43	$\pm$	0.01	&	33.47	$\pm$	0.04	&	2.42	$\pm$	0.09	&	0.17	\\
CH$_{3}$CCH			&	5(0)-4(0)						&	85457.30	&	0.55	$\pm$	0.01	&	33.53	$\pm$	0.04	&	2.65	$\pm$	0.10	&	0.20	\\
C$_{4}$H			&	9-8 J=19/2-17/2					&	85634.01	&	0.09	$\pm$	0.01	&	34.48	$\pm$	0.27	&	2.83	$\pm$	0.59	&	0.03	\\
NH$_{2}$D			&	1(1,1)0+ - 1(0,1)0-				&	85926.27	&	0.27	$\pm$	0.02	&	33.18	$\pm$	0.22	&	4.64	$\pm$	0.66	&	0.06	\\
HC$^{15}$N			&	1-0								&	86054.96	&	0.45	$\pm$	0.02	&	33.74	$\pm$	0.07	&	2.86	$\pm$	0.17	&	0.15	\\
SO					&	N,J=2,2-1,1						&	86093.95	&	0.36	$\pm$	0.01	&	33.43	$\pm$	0.07	&	2.74	$\pm$	0.18	&	0.13	\\
H$^{13}$CN			&	1-0 F=1-1						&	86338.73	&	0.88	$\pm$	0.01	&	33.60	$\pm$	0.03	&	2.62	$\pm$	0.07	&	0.32	\\
H$^{13}$CN			&	1-0 F=2-1						&	86340.17	&	1.62	$\pm$	0.02	&	33.64	$\pm$	0.01	&	2.82	$\pm$	0.05	&	0.54	\\
H$^{13}$CN			&	1-0 F=0-1	&	86342.25	&	0.36	$\pm$	0.02	&	33.63	$\pm$	0.10	&	3.07	$\pm$	0.28	&	0.11	\\
HC13O$^{+}$			&	1-0								&	86754.28	&	1.38	$\pm$	0.01	&	33.36	$\pm$	0.02	&	2.77	$\pm$	0.04	&	0.47	\\
C$_{3}$S	&	15-14	&	86708.38	&	0.19	$\pm$	0.03	&	34.32	$\pm$	0.15	&	2.59	$\pm$	0.55	&	0.07	\\
unidentified	&		&	86777.10	&	0.18	$\pm$	0.03	&	32.87	$\pm$	0.18	&	2.93	$\pm$	0.51	&	0.06	\\
SiO		&	2-1 v=0							&	86846.96	&	0.72	$\pm$	0.04	&	33.50	$\pm$	0.13	&	5.15	$\pm$	0.40	&	0.13	\\
HN$^{13}$C			&	1-0 F=2-1						&	87090.85	&	0.81	$\pm$	0.01	&	33.38	$\pm$	0.03	&	2.81	$\pm$	0.07	&	0.27	\\
CCH					&	1-0 3/2-1/2 F=1-1				&	87284.10	&	0.57	$\pm$	0.01	&	33.51	$\pm$	0.04	&	2.58	$\pm$	0.10	&	0.21	\\
CCH					&	1-0 3/2-1/2 F=2-1				&	87316.89	&	4.08	$\pm$	0.02	&	33.64	$\pm$	0.00	&	2.50	$\pm$	0.01	&	1.53	\\
CCH					&	1-0 3/2-1/2 F=1-0				&	87328.58	&	2.26	$\pm$	0.01	&	33.58	$\pm$	0.01	&	2.53	$\pm$	0.02	&	0.84	\\
CCH					&	1-0 1/2-1/2 F=1-1				&	87401.98	&	2.26	$\pm$	0.01	&	33.60	$\pm$	0.01	&	2.51	$\pm$	0.02	&	0.85	\\
CCH					&	1-0 1/2-1/2 F=0-1				&	87407.16	&	1.09	$\pm$	0.01	&	33.55	$\pm$	0.02	&	2.59	$\pm$	0.05	&	0.40	\\
CCH					&	1-0 1/2-1/2 F=1-0				&	87446.47	&	0.59	$\pm$	0.01	&	33.52	$\pm$	0.03	&	2.39	$\pm$	0.09	&	0.23	\\
HC$_{5}$N			&	33-32							&	87863.63	&	0.25	$\pm$	0.04	&	33.29	$\pm$	0.35	&	3.02	$\pm$	0.60	&	0.03	\\
HNCO				&	4(0,4)-3(0,3)					&	87925.23	&	0.35	$\pm$	0.02	&	33.04	$\pm$	0.11	&	3.42	$\pm$	0.27	&	0.10	\\
HCN					&	1-0,F=1-1						&	88630.41	&	6.03	$\pm$	0.12	&	33.78	$\pm$	0.02	&	3.02	$\pm$	0.07	&	1.87	\\
HCN		&	1-0 F=2-1						&	88631.84	&	1.24	$\pm$	0.08	&	34.55	$\pm$	0.06	&	1.71	$\pm$	0.15	&	0.68	\\
HCN		&	1-0 F=0-1						&	88633.93	&	1.08	$\pm$	0.08	&	32.08	$\pm$	0.06	&	1.73	$\pm$	0.17	&	0.59	\\
HCO$^{+}$			&	1-0								&	89188.52	&	4.29	$\pm$	0.07	&	34.05	$\pm$	0.02	&	3.09	$\pm$	0.06	&	1.31	\\
l-C$^{13}$CC$_{2}$H$_{2}$	&	10(2,8)-9(2,7)	&	88865.38	&	0.23	$\pm$	0.03	&	31.87	$\pm$	0.14	&	2.58	$\pm$	0.31	&	0.09	\\
c-C$_{3}$D$_{2}$	&	5(4,2)-5(3,3)	&	89316.91	&	0.15	$\pm$	0.03	&	33.31	$\pm$	0.91	&	7.21	$\pm$	1.58	&	0.02	\\
unidentified	&		&	89441.30	&	0.20	$\pm$	0.04	&	32.65	$\pm$	0.70	&	6.70	$\pm$	1.54	&	0.03	\\
unidentified	&		&	89449.50	&	0.20	$\pm$	0.04	&	32.53	$\pm$	0.87	&	8.47	$\pm$	1.62	&	0.02	\\
HOCHCHCHO	&	6(4,3)-5(3,2)	&	89458.20	&	0.16	$\pm$	0.04	&	33.94	$\pm$	0.71	&	6.18	$\pm$	1.34	&	0.03	\\
HC$_{5}$N			&	34-33							&	90525.88	&	0.25	$\pm$	0.02	&	32.96	$\pm$	0.18	&	2.83	$\pm$	0.41	&	0.03	\\
HC$^{13}$CCN		&	10-9							&	90593.05	&	0.16	$\pm$	0.02	&	32.81	$\pm$	0.33	&	3.88	$\pm$	0.64	&	0.04	\\
HCC$^{13}$CN		&	10-9							&	90601.77	&	0.12	$\pm$	0.01	&	32.70	$\pm$	0.20	&	2.25	$\pm$	0.45	&	0.05	\\
HNC					&	1-0 F=2-1						&	90663.56	&	9.70	$\pm$	0.04	&	33.05	$\pm$	0.00	&	2.69	$\pm$	0.01	&	3.39	\\
HCCCN				&	10-9							&	90979.02	&	3.13	$\pm$	0.01	&	32.81	$\pm$	0.01	&	2.62	$\pm$	0.01	&	1.12	\\
HDCS	&	3(1,3)-2(1,2)					&	91171.03	&	0.13	$\pm$	0.04	&	32.09	$\pm$	0.38	&	2.85	$\pm$	0.89	&	0.03	\\
H$_{2}$CCCHCN	&	4(2,2)-2(1,3)	&	91494.56	&	0.13	$\pm$	0.03	&	33.66	$\pm$	0.29	&	3.25	$\pm$	0.99	&	0.04	\\
CH$_{3}$CN			&	5(3)-4(3)	&	91971.13	&	0.32	$\pm$	0.04	&	32.33	$\pm$	0.29	&	6.03	$\pm$	0.92	&	0.05	\\
CH$_{3}$CN			&	5(2)-4(2)	&	91979.99	&	0.38	$\pm$	0.03	&	32.42	$\pm$	0.16	&	4.40	$\pm$	0.41	&	0.08	\\
CH$_{3}$CN			&	5(1)-4(1)						&	91985.31	&	0.30	$\pm$	0.02	&	32.66	$\pm$	0.09	&	2.37	$\pm$	0.22	&	0.12	\\
CH$_{3}$CN			&	5(0)-4(0)						&	91987.08	&	0.43	$\pm$	0.02	&	32.66	$\pm$	0.09	&	3.22	$\pm$	0.21	&	0.13	\\
H$\alpha$			&	H41$\alpha$					&	92034.43	&	0.90	$\pm$	0.04	&	21.00	$\pm$	0.70	&	28.40	$\pm$	1.75	&	0.03	\\
CH$_{3}$COOH		&	v=0 18(13,6)-18(-10,8) v=0		&	143755.42	&	0.04	$\pm$	0.00	&	32.60	$\pm$	0.10	&	1.10	$\pm$	0.25	&	0.04	\\
CH$_{3}$OH	&	3(1,3)-2(1,2) A++				&	143865.79	&	0.34	$\pm$	0.01	&	32.58	$\pm$	0.08	&	3.32	$\pm$	0.24	&	0.10	\\
DCO$^{+}$	&	2-1								&	144077.28	&	0.24	$\pm$	0.03	&	32.09	$\pm$	0.34	&	6.64	$\pm$	1.05	&	0.03	\\
CH$_{3}$C$_{3}$N	&	35(16)-34(16)					&	144231.92	&	0.36	$\pm$	0.01	&	30.70	$\pm$	0.07	&	2.61	$\pm$	0.17	&	0.13	\\
CCD	&	2-1 J=5/2-3/2F=7/2-5/2			&	144241.96	&	0.06	$\pm$	0.01	&	32.38	$\pm$	0.37	&	3.08	$\pm$	0.66	&	0.02	\\
unidentified	&		&	144609.02	&	0.10	$\pm$	0.03	&	40.55	$\pm$	0.27	&	2.00	$\pm$	0.58	&	0.05	\\
C$^{34}$S	&	3-2								&	144617.10	&	0.93	$\pm$	0.01	&	33.20	$\pm$	0.02	&	2.57	$\pm$	0.05	&	0.34	\\
DCN	&	2-1 F1=2-1						&	144828.00	&	0.31	$\pm$	0.03	&	33.32	$\pm$	0.16	&	4.35	$\pm$	0.51	&	0.07	\\
$c$-C$_{3}$H$_{2}$	&	3(1,2)-2(2,1)					&	145089.62	&	0.61	$\pm$	0.07	&	32.25	$\pm$	0.15	&	6.17	$\pm$	0.77	&	0.09	\\
CH$_{3}$OH	&	3(0,3)-2(0,2) E					&	145093.76	&	0.80	$\pm$	0.05	&	32.41	$\pm$	0.10	&	3.45	$\pm$	0.28	&	0.22	\\
CH$_{3}$OH	&	3(-1,3)-2(-1,2) E				&	145097.44	&	2.33	$\pm$	0.05	&	32.76	$\pm$	0.03	&	3.05	$\pm$	0.08	&	0.72	\\
CH$_{3}$OH	&	3(0,3)-2(0,2) A++				&	145103.19	&	2.64	$\pm$	0.03	&	32.81	$\pm$	0.02	&	3.01	$\pm$	0.05	&	0.82	\\

 \hline
\end{tabular}
\end{minipage}
\end{table*}

\begin{table*}
%\begin{minipage}{200mm}
\contcaption{}
%\label{tab:G34}
\centering
\begin{minipage}{200mm}
\begin{tabular}{lrrrrrr}
 \hline 
Molecule & Transition &Frequency & $\int T_{mb}dv$	& $v_{LSR}$		&FWHM & $T_{mb}$\\
		 &			& MHz		& (K km s$^{-1}$) & (km s$^{-1}$)	&(km s$^{-1}$)    &  K\\
\hline
CH$_{3}$OH	&	3(-2,2)-2(-2,1) E				&	145126.39	&	0.36	$\pm$	0.01	&	32.97	$\pm$	0.10	&	3.30	$\pm$	0.27	&	0.10	\\
CH$_{3}$OH	&	3(1,2)-2(1,1) E					&	145131.87	&	0.37	$\pm$	0.03	&	32.49	$\pm$	0.11	&	3.82	$\pm$	0.33	&	0.09	\\
CH$_{3}$SH			&	v = 0  12( -3) - 13( -2) E		&	145386.10	&	0.78	$\pm$	0.01	&	31.98	$\pm$	0.02	&	2.67	$\pm$	0.06	&	0.28	\\
HCCCN	&	16-15							&	145560.95	&	0.80	$\pm$	0.01	&	32.98	$\pm$	0.03	&	2.53	$\pm$	0.07	&	0.30	\\
H$_{2}$CO	&	2(0,2)-1(0,1)					&	145602.95	&	4.00	$\pm$	0.03	&	33.12	$\pm$	0.01	&	2.86	$\pm$	0.02	&	1.31	\\
unidentified	&		&	145677.00	&	0.13	$\pm$	0.02	&	32.88	$\pm$	0.31	&	3.38	$\pm$	1.04	&	0.04	\\
unidentified	&		&	145692.20	&	0.06	$\pm$	0.01	&	32.88	$\pm$	0.28	&	1.77	$\pm$	0.43	&	0.03	\\
n-C$_{3}$H$_{7}$CN	&	9(7,3)-8(6,2)	&	145700.18	&	0.14	$\pm$	0.02	&	31.39	$\pm$	0.44	&	5.70	$\pm$	1.02	&	0.03	\\
t-HC(O)SH	&	15(2,13)-15(1,14)	&	145746.31	&	0.18	$\pm$	0.02	&	34.70	$\pm$	0.18	&	3.23	$\pm$	0.46	&	0.05	\\
C$^{33}$S			&	3-2 F=9/2-7/2					&	145755.62	&	0.18	$\pm$	0.01	&	32.51	$\pm$	0.17	&	3.24	$\pm$	0.48	&	0.05	\\
OCS	&	12-11							&	145946.81	&	0.10	$\pm$	0.01	&	32.46	$\pm$	0.20	&	2.23	$\pm$	0.54	&	0.04	\\
CH$_{3}$OH	&	3(1,2)-2(1,1)A--				&	146368.34	&	0.41	$\pm$	0.01	&	32.53	$\pm$	0.09	&	3.32	$\pm$	0.24	&	0.12	\\
CH$_{3}$OH	&	9(0,9)-8(1,8)A++				&	146618.83	&	0.50	$\pm$	0.03	&	32.42	$\pm$	0.07	&	3.13	$\pm$	0.20	&	0.15	\\
H$_{2} ^{13}$CO		&	2(1,1)-1(1,0)					&	146635.67	&	0.14	$\pm$	0.03	&	32.56	$\pm$	0.33	&	3.02	$\pm$	0.70	&	0.05	\\
H$_{2}$CN			&	2(0,2)-1(0,1),F=3/2-5/2			&	146640.12	&	0.41	$\pm$	0.03	&	32.30	$\pm$	0.09	&	3.07	$\pm$	0.24	&	0.13	\\
CS	&	3-2								&	146969.02	&	9.68	$\pm$	0.03	&	33.01	$\pm$	0.00	&	2.90	$\pm$	0.01	&	3.13	\\

 \hline
\end{tabular}
\end{minipage}
\end{table*}

\end{document}